\documentclass[longauth]{aa}

\usepackage{graphicx}
\usepackage{xcolor}
\usepackage{txfonts}
\usepackage{placeins}
\usepackage{hyperref}
\hypersetup{colorlinks=true,
            linkcolor=blue,
            urlcolor=blue,
            citecolor=blue,
            anchorcolor=blue}
\usepackage{orcidlink}

\begin{document} 

\def\kms{km\,s$^{-1}$}
\def\teff{$T_{\rm eff}$}
\def\logg{$\log{g}$}
\def\vmic{$\upsilon_t$}

   \title{Oxygen, sulfur, and iron radial abundance gradients of \\ classical Cepheids across the Galactic thin disk
   \thanks{
   Partly based on observations made with ESO Telescopes at the La Silla/Paranal Observatories under program IDs: 072.D-0419, 073.D-0136, and 190.D-0237 for HARPS spectra; 084.B-0029, 087.A-9013, 074.D-0008, 075.D-0676, and 60.A-9120 for FEROS spectra; 081.D-0928, 082.D-0901, 089.D-0767, and 093.D-0816 for UVES spectra.}$^,$
   \thanks{
   Partly based on data obtained with the STELLA robotic telescopes in Tenerife, a facility of The Leibniz Institute for Astrophysics Potsdam (AIP) jointly operated by the AIP and by the Instituto de Astrofisica de Canarias (IAC).}$^,$
   \thanks{
   The full versions of Tables~\ref{table:sample}, \ref{table:spectra}, and \ref{table:mean} are only available at the CDS via anonymous ftp to cdsarc.u-strasbg.fr (130.79.128.5) or via http://cdsweb.u-strasbg.fr/cgi-bin/qcat?J/A+A/.}$^,$
   \thanks{
   During the revision of this manuscript, Mario Nonino passed away. A lifelong friend and collaborator, his ideas and personality will be greatly missed.}
   }


   \author{R. da Silva \inst{1,2}
          \and
          V. D'Orazi \inst{3,4}
          \and
          M. Palla \inst{5,6}
          \and
          G. Bono \inst{3,1}
          \and
          V.F. Braga \inst{1,2}
          \and
           M. Fabrizio \inst{1,2}
          \and
          B. Lemasle\inst{7}
          \and
          E. Spitoni \inst{8}
          \and \\
          F. Matteucci \inst{9,8,10}
          \and
          H. J\"{o}nsson \inst{11}
          \and
          V. Kovtyukh \inst{12}
          \and
          L. Magrini \inst{13}
          \and
          M. Bergemann \inst{14,15}
          \and
          M. Dall'Ora \inst{16}
          \and
          I. Ferraro \inst{1}
          \and \\
          G. Fiorentino \inst{1}
          \and
          P. François \inst{17,18}
          \and
          G. Iannicola \inst{1}
          \and
          L. Inno \inst{19}
          \and
          R.-P. Kudritzki \inst{20,21}
          \and
          N. Matsunaga \inst{22,23}
          \and
          M. Monelli \inst{24,25,1}
          \and \\
          M. Nonino \inst{8}
          \and
          C. Sneden \inst{26}
          \and
          J. Storm \inst{27}
          \and
          F. Thévénin \inst{28}
          \and
          T. Tsujimoto \inst{29}
          \and
          A. Zocchi \inst{30}
          }

   \institute{
   INAF - Osservatorio Astronomico di Roma, via Frascati 33, 00078 Monte Porzio Catone, Italy
   \and
   Agenzia Spaziale Italiana, Space Science Data Center, via del Politecnico snc, 00133 Rome, Italy
   \and
   Department of Physics, University of Rome Tor Vergata, via della Ricerca Scientifica 1, 00133 Rome, Italy
   \and
   INAF - Osservatorio Astronomico di Padova, vicolo dell’Osservatorio 5, 35122 Padova, Italy
   \and
   Sterrenkundig Observatorium, Ghent University, Krĳgslaan 281 - S9, 9000 Gent, Belgium
   \and
   INAF – Osservatorio Astronomico di Bologna, Via Gobetti 93/3, 40129 Bologna, Italy
   \and
   Astronomisches Rechen-Institut, ZAH, Universit\"{a}t Heidelberg, M\"{o}nchhofstr.12-14, D-69120 Heidelberg, Germany
   \and
   INAF, Osservatorio Astronomico di Trieste, via G. B. Tiepolo 11, I-34131, Trieste, Italy
   \and
   Dipartimento di Fisica, Sezione di Astronomia, Universit\'a  degli Studi di Trieste, via G. B. Tiepolo 11, I-34131, Trieste, Italy
   \and
   INFN, Sezione di Trieste, via A. Valerio 2, I-34100, Trieste, Italy
   \and
   Materials Science and Applied Mathematics, Malm\"{o} University, SE-205 06 Malm\"{o}, Sweden
   \and
   Astronomical Observatory, Odessa National University, Shevchenko Park, UA-65014 Odessa, Ukraine
   \and
   INAF - Osservatorio Astrofisico di Arcetri, Largo E. Fermi 5, Firenze, Italy
   \and
   Max Planck Institute for Astronomy, D-69117 Heidelberg, Germany
   \and
   Niels Bohr International Academy, Niels Bohr Institute, Blegdamsvej 17, DK-2100 Copenhagen Ø, Denmark
   \and
   INAF - Osservatorio Astronomico di Capodimonte, Napoli, Italy
   \and
   GEPI, Observatoire de Paris, PSL Research University, CNRS, 61 Avenue de l'Observatoire, 75014, Paris, France
   \and
   UPJV, Université de Picardie Jules Verne, 33 rue St Leu, 80080, Amiens, France
   \and
   Science and Technology Department, Parthenope University of Naples, Naples, Italy
   \and
   LMU München, Universitätssternwarte, Scheinerstr. 1, 81679 München, Germany
   \and
   Institute for Astronomy, University of Hawaii at Manoa, 2680 Woodlawn Drive, Honolulu, HI 96822, USA
   \and
   Department of Astronomy, School of Science, The University of Tokyo, 7-3-1, Hongo, Bunkyo-ku, Tokyo 113-0033, Japan
   \and
   Laboratory of Infrared High-Resolution spectroscopy (LiH), Koyama Astronomical Observatory, Kyoto Sangyo University, Motoyama, Kamigamo, Kita-ku, Kyoto 603-8555, Japan
   \and
   Instituto de Astrofísica de Canarias (IAC), La Laguna, 38205, Spain
   \and
   Departamento de Astrofísica, Universidad de La Laguna (ULL), 38200, La Laguna, Spain
   \and
   Department of Astronomy and McDonald Observatory, The University of Texas, Austin, TX 78712, USA
   \and
   Leibniz-Institut für Astrophysik Potsdam (AIP), An der Sternwarte 16, 14482, Potsdam, Germany
   \and
   Université Côte d’Azur, Observatoire de la Côte d’Azur, CNRS, Lagrange UMR 7293, CS 34229, 06304 Nice Cedex 4, France
   \and
   National Astronomical Observatory of Japan, Mitaka, Tokyo 181-8588, Japan
   \and
   Department of Astrophysics, University of Vienna, Türkenschanzstraße 17, 1180 Vienna, Austria
   }

   \date{Received ...; accepted ...}

 
  \abstract
   {Classical Cepheids (CCs) are solid distance indicators and tracers of young stellar populations. Dating back to almost one century ago, they have been safely adopted to trace the rotation, the kinematics and the chemical enrichment history of the Galactic thin disk.}
   {The main aim of this investigation is to provide iron, oxygen, and sulfur abundances for the largest and most homogeneous sample of Galactic CCs ever analyzed (1118 spectra of 356 objects). The current sample, containing 77 CCs for which spectroscopic metal abundances are provided for the first time, covers a wide range in Galactocentric distances, pulsation modes, and pulsation periods.}
   {Optical, high-resolution, and high S/N spectra collected with different spectrographs were adopted to provide homogeneous estimates of the atmospheric parameters (effective temperature, surface gravity, microturbulent velocity) required for abundance determination. Individual distances are based either on trigonometric parallaxes by Gaia DR3 or on distances based near-infrared Period-Luminosity relations.}
   {We found that iron and $\alpha$-element radial gradients based on CCs display a well-defined change in the slope for Galactocentric distances larger than $\sim$12~kpc. We also found that logarithmic regressions take account for the variation of [X/H] abundances when moving from the inner to the outer disk. Radial gradients for the same elements, but based on open clusters covering a wide range in cluster ages, display similar trends. This means that the flattening in the outer disk is an intrinsic feature of the radial gradients, since it is independent of age. Empirical evidence indicates that the S radial gradient is steeper than the Fe radial gradient. The difference in the slope is a factor of two in the linear fit ($-$0.081 vs. $-$0.041~dex\,kpc$^1$) and changes from $-$1.62 to $-$0.91 in the logarithmic distance. Moreover, we found that S (explosive nucleosynthesis) is, on average, under-abundant when compared with O (hydrostatic nucleosynthesis). The difference becomes clearer in the metal-poor regime and in dealing with the [O/Fe] and [S/Fe] abundance ratios. We performed a detailed comparison with Galactic chemical evolution models and we found that a constant Star Formation Efficiency for Galactocentric distances larger than 12~kpc takes account for the flattening observed in both iron and $\alpha$-elements. To further constrain the impact that predicted S yields for massive stars have on radial gradients, we adopted a "toy model" and we found that the flattening in the outermost regions requires a decrease of a factor of four in the current S predictions.} 
   {CCs are solid beacons to trace the recent chemical enrichment of young stellar populations. Sulfur photospheric abundances, when compared with other $\alpha$-elements, have the key advantage of being a volatile element. Therefore, stellar S abundances can be directly compared with nebular sulfur abundances in external galaxies.}


   \maketitle
%
\section{Introduction}
\label{sec:intro}

Radial and azimuthal metallicity gradients (with respect to the galactic center and to the galactic plane, respectively) are crucial diagnostics to trace the chemical enrichment history of individual galactic components \citep{Lemasleetal2013,Lemasleetal2018,Genovalietal2014,Genovalietal2015,daSilvaetal2016,daSilvaetal2022}. The radial variation of different heavy elements, produced either by SN~Ia or by massive and AGB stars or by more exotic objects (e.g., neutron star mergers), provides fundamental constraints on chemical evolution models \citep{Cavichiaetal2014,SchonrichMcMillan2017,Grisonietal2018,Prantzosetal2018,Matteuccietal2020,Spitonietal2022,Tsujimoto2023}. Moreover, the use of stellar tracers ({\it old:} RR Lyrae, blue Horizontal Branch, globular clusters; {\it intermediate-age:} red clump, anomalous Cepheids; {\it young:} red and blue supergiants, classical Cepheids -- CCs, open clusters -- OCs) covering a broad range in stellar ages allows us to investigate the role that galaxy mergers, stellar migrations, and kinematics plays in the formation and evolution of galaxies. In addition to these indisputable advantages, radial metallicity gradients allow us to trace the coupling among star-forming regions \citep{Genovalietal2014}, spiral structure \citep{Lemasleetal2022}, and geometrical complexity \citep[warps, flares, streams;][]{Feastetal2014,Matsunagaetal2018,Chenetal2019,Skowronetal2019,Dehnenetal2023}.

\setlength{\tabcolsep}{3.5pt}

\begin{table*}
\centering
\caption{Excerpt from the list of our sample of 379 Galactic classical Cepheids.}
\label{table:sample}
\begin{tabular}{l c c r@{}l c r@{ }l c c c c c c c}
\noalign{\smallskip}\hline\hline\noalign{\smallskip}
Name & $\alpha_{\rm ICRS}$ & $\delta_{\rm ICRS}$ &
\multicolumn{2}{c}{\parbox[c]{1.0cm}{\centering Period [days]}} &
Mode &
\multicolumn{2}{c}{${\rm [Fe/H]}_{\rm lit}$ $\pm$ $\sigma$} &
Ref.\tablefootmark{a} &
\parbox[c]{0.6cm}{\centering $X$ [pc]} &
\parbox[c]{0.6cm}{\centering $Y$ [pc]} &
\parbox[c]{0.6cm}{\centering $Z$ [pc]} &
\parbox[c]{0.6cm}{\centering $R_{\rm H}$ [pc]} &
\parbox[c]{0.6cm}{\centering $R_{\rm G}$ [pc]} &
\parbox[c]{0.7cm}{\centering $\sigma(R)$ [pc]} \\
\noalign{\smallskip}\hline\noalign{\smallskip}
\object{AA\,Gem}      & 06:06:34.946 &   +26:19:45.191 & 11&.3128451 &   0 & $-$0.08 & $\pm$ 0.05 &   1 & $-$11353 & $-$259 &   182 & 3245 & 11356 & 184 \\
\object{AA\,Ser}      & 18:41:21.761 & $-$01:06:40.442 & 17&.1424446 &   0 &    0.38 & $\pm$ 0.20 &   1 &  $-$5015 &   1851 &   122 & 3618 &  5346 & 419 \\
...                   & ...          & ...             &   & ...     & ... &         &        ... & ... &      ... &    ... &   ... &  ... &   ... & ... \\
\object{l\,Car}       & 09:45:14.782 & $-$62:30:28.323 & 35&.5580599 &   0 &    0.24 & $\pm$ 0.10 &   1 &  $-$8013 & $-$465 & $-$38 &  481 &  8027 &  12 \\
\object{$\zeta$\,Gem} & 07:04:06.522 &   +20:34:13.059 & 10&.1485988 &  0B &    0.01 & $\pm$ 0.06 &   2 &  $-$8484 & $-$102 &   101 &  385 &  8485 &   4 \\
\hline
\end{tabular}
\tablefoot{The first five columns give the star name, the right ascension and declination, the pulsation period, and the pulsation mode (0: fundamental; 0B: fundamental with bump; 1: first overtone; 2: multi-mode; $-$1: no mode identification available). Columns (6) and (7) give the iron abundance from literature and the corresponding references. The last six columns list the Heliocentric $X, Y, Z$ projected distances and the Heliocentric and Galactocentric radial distances together with their errors.
\tablebib{
\tablefoottext{a}{
1: \citet{Genovalietal2014} -- for some of the stars we assumed a typical error of 0.1~dex since they were not provided by the original authors;
2: \citet{daSilvaetal2022} -- the uncertainty adopted is the largest value between $\sigma$ and std. The complete table is available at the CDS.
}}}
\end{table*}

Dating back to more than half a century ago \citep{Kraft1966}, CCs have been widely used to trace the variation of iron as a function of the Galactocentric distance. The key advantage of using CCs to trace young stellar populations is manifold: {\em i)} CCs are very solid primary distance indicators and their individual distances can be estimated with accuracy better than 3\%, on average; {\em ii)} They are associated with central helium burning phases (blue loop) of intermediate-mass (from $\sim$3 to $\sim$10~$M_\odot$) stars. This means that they are quite common and ubiquitous across the Galactic thin disk; {\em iii)} CCs have low surface gravities and their spectra are quite rich in absorption lines, with long-period Cepheids being characterized by a large presence of molecular bands. The main drawback is that the variation of the physical properties along the pulsation cycle requires a solid approach to estimate the effective temperature, the surface gravity, and the microturbulent velocity. Moreover, different lines might also experience a significant variation along the pulsation cycle and, therefore, they need to be properly identified. The reader interested in a detailed and more quantitative discussion concerning these issues is referred to \citet{daSilvaetal2022}. 

The empirical scenario emerging from the recent paramount spectroscopic effort on CCs is that all the investigated elements (iron peak, $\alpha$, neutron capture) display a well-defined radial gradient, with the exception of barium \citep{Andrievskyetal2013}. In spite of this global agreement, some problems emerge. The current estimates of the global metallicity gradient very often differ in zero point, as expected, due to different assumptions on the solar abundances and on the approach adopted to estimate the atmospheric parameters. Moreover, a difference is also quite often seen in the slope of the radial metallicity gradient, which is commonly associated with the adopted line list. Several atomic lines, from both neutral and ionized species, are affected by non Local Thermodynamical Equilibrium (NLTE) effects and this dependence changes as a function of surface gravity, effective temperature, and iron abundance. This means that in the estimate of the mean abundance based on different lines, we are summing up a significant fraction of possible systematic errors.

Finally, it is worth mentioning that the slope is also tightly correlated with the rate at which the individual elements are enriched. This is the main reason why the iron radial gradient is typically the steepest one, not only among iron-peak elements but also among $\alpha$ and neutron-capture elements. Indeed, iron is produced by both SN~Ia and massive stars, with the former enriching the interstellar medium on longer timescales (of the order of Gyr). In a recent investigation based on more than 400 high-resolution and high signal-to-noise ratio (S/N) spectra for two dozen calibrating Cepheids \citep{daSilvaetal2022}, we found preliminary evidence that the slope of the sulfur radial gradient is steeper than that found for iron abundances. To further investigate this crucial issue, we performed the same measurements plus another $\alpha$-element (oxygen) over a significantly larger sample of CCs (more than 350 variables). To provide a comprehensive analysis of Galactic CCs, we also performed a re-analysis of literature data. In passing, we also mention that recent findings indicate that stellar metallicities measured in extragalactic systems agree, in general, with the nebular abundances based on the analysis of the auroral lines \citep{Bresolinetal2009,Bresolinetal2022,Gazaketal2015,Liuetal2022}. Studies on H\,II regions and luminous young stars in local disk galaxies show a significant gas metallicity gradient ($-$0.04 to $-$0.06~dex\,kpc$^{-1}$), that is, a strong decrease in the abundances when moving from the center to the outskirts of the galaxy \citep{Zaritskyetal1994,SanchezBlazquezetal2014,Bresolinetal2012,Bresolinetal2016,KudritzkiUrbaneja2018}. The same outcome applies to the $\alpha$-element abundance gradients \citep{Urbanejaetal2005}. However, many disks are surrounded by huge areas of neutral hydrogen with constant low metallicity and low star formation rate \citep[][and references therein]{Bresolinetal2012,Kudritzkietal2014}.

The main aim here is to determine and compare iron, sulfur, and oxygen radial abundance gradients across the Galactic thin disk. The paper is therefore organized as follows. In Sect.~\ref{sec:observ} we introduce the spectroscopic data and the properties of our sample of classical Cepheids and open clusters. In Sect.~\ref{sec:atmpar_ab} we describe how we derive the atmospheric parameters and the abundances. Section~\ref{sec:gradients} shows the Galactic radial gradients that we obtain for different abundance ratios together with their dependence on the stellar age. In Sects.~\ref{sec:comparison_observations} and \ref{sec:comparison_theory} we compare our results with those from literature derived from both observations and theory. A summary of our results together with some final considerations are provided in Sect.~\ref{sec:summary}.

\section{Observations and sample properties}
\label{sec:observ}



\subsection{The sample of classical Cepheids}

The sample of classical Cepheids investigated in the current study is the result of a compilation of several subsamples for which high-resolution and high S/N spectra are available in different databases. Such spectra were collected using four high-resolution spectrographs: the High Accuracy Radial velocity Planet Searcher spectrograph \citep[HARPS;][]{Mayoretal2003} mounted at the 3.6~m telescope of the European Southern Observatory (ESO) at La Silla (Chile), the Fiber-fed Extended Range Optical Spectrograph \citep[FEROS;][]{Kauferetal1999} installed at the 2.2~m MPG/ESO, the Ultraviolet and Visual Echelle Spectrograph \citep[UVES;][]{Dekkeretal2000} at the Very Large Telescope of ESO at Paranal (Chile), and the STELLA Echelle Spectrograph \citep[SES;][]{Strassmeieretal2004,Strassmeieretal2010} located at the Iz\~ana Observatory on Tenerife in the Canary islands. The spectral resolution achieved for the instrument settings used are R$\sim$40\,000 for UVES spectra, R$\sim$115\,000 for HARPS, R$\sim$48\,000 for FEROS, and R$\sim$55\,000 for STELLA. For details on the wavelength ranges, we refer the reader to \citet{Proxaufetal2018} and \citet{Crestanietal2021}.

\begin{figure*}
\centering
\resizebox{0.8\hsize}{!}{\includegraphics{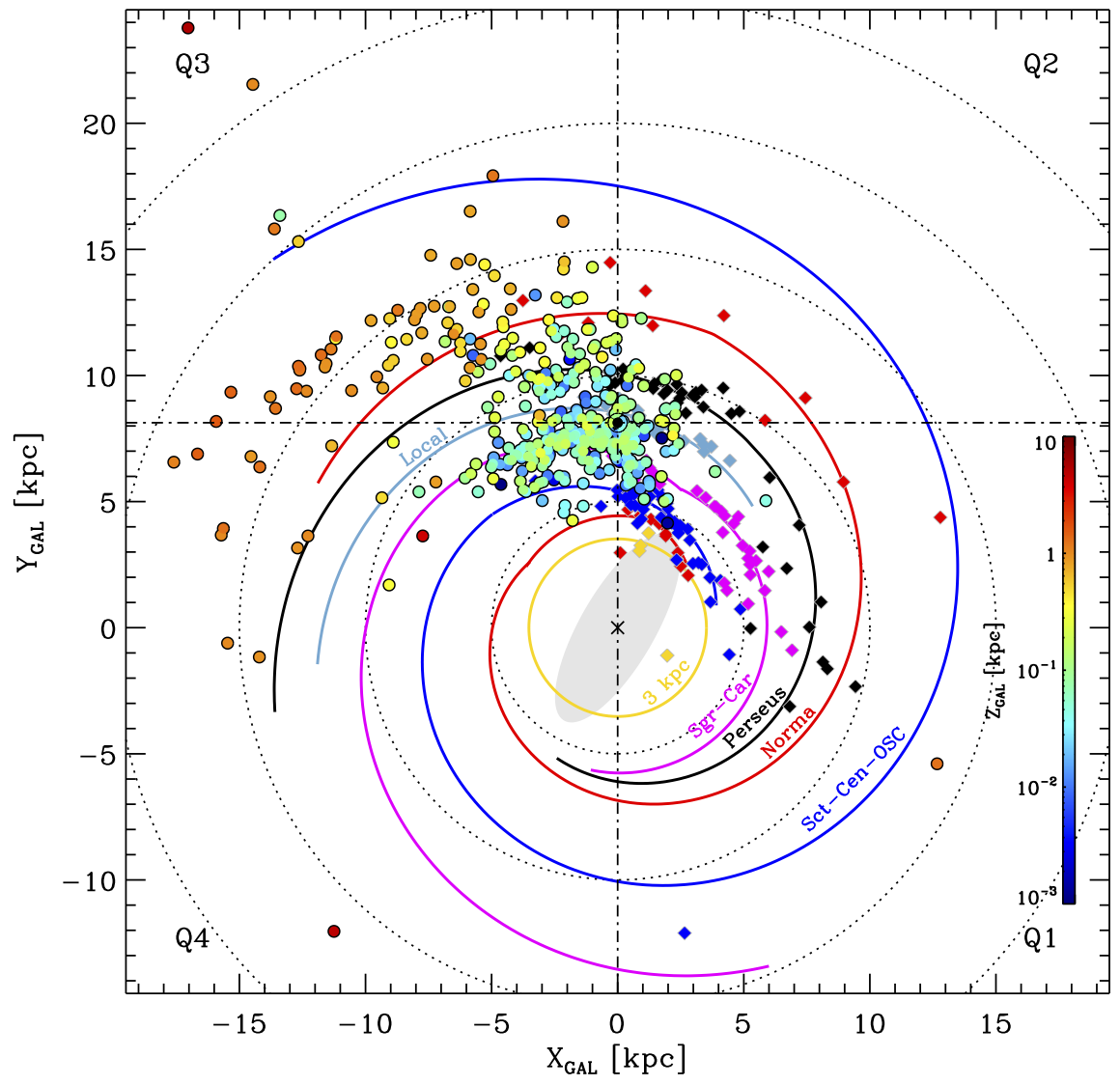}}
\caption{Radial distribution of the current Cepheid sample (colored circles) projected onto the Galactic plane. The symbols are color-coded (see color bar on the right side) according to the distance from the Galactic plane. The dotted annuli display distances from the Galactic center multiples of 5~kpc. The position adopted for the Sun is at $X=0$ and $Y=8.127$~kpc \citep{Gravity2018}. The colored diamonds show the locations of high-mass star-forming regions used by \citet{Reidetal2019} to fit the spiral arms. They are color-coded according to the spiral arm association. The "long" bar of the inner part of the Galaxy is indicated with a shaded ellipse \citep{Weggetal2015}. The solid curved lines trace the Milky Way spiral arms from \citet{Reidetal2019} and from \citet[Outer-Scutum-Centaurus]{Sunetal2015}.}
\label{figure:xyz_galactic}
\end{figure*}

Our spectroscopic sample includes proprietary data and spectra downloaded from the ESO and the STELLA archives: 183 HARPS spectra of 10 stars, 339 FEROS spectra of 161 stars, 363 UVES spectra of 215 stars, and 400 STELLA spectra of 64 stars (some of the stars have spectra collected with more than one instrument), for a total of 1285 spectra of 379 stars. The S/N is of at least 100 for more than 80\% of these spectra. For 1118 spectra of 356 stars we were able to derive all the stellar atmospheric parameters required for abundance determination (effective temperature, surface gravity, microturbulent velocity, and metallicity). For reasons discussed in the next section, only the effective temperature was derived for 167 spectra of 23 stars. In this context, it is worth mentioning that the abundance estimates based on high-resolution spectra for 77 out of the 356 CCs are provided here for the first time. The complete sample of 379 stars is listed in Table~\ref{table:sample} and details on the number of spectra are given in Table~\ref{table:mean}. A more detailed discussion concerning the classification of our sample as classical Cepheids is provided in the Appendix~\ref{appendix:bailey_fourier}.

\subsection{The sample of open clusters}

We complemented our sample of classical Cepheids with a sample of open clusters from \citet{Randichetal2022}, which includes 62 open clusters observed by the Gaia-ESO Survey (GES) plus 18 retrieved from the ESO archive. We refer the reader to \citet{Viscasillasetal2022} for the general characteristics of the sample and the selection criteria of members in each cluster. Briefly, the sample covers a range in age from 1.4~Myr to 6.8~Gyr, in Galactocentric distances from 5.8 to 20.6~kpc, and in metallicity from $-$0.45 to 0.27~dex. 

\begin{table*}
\centering
\caption{Excerpt from the list of atmospheric parameters, Fe abundances, and S abundances for each spectrum in our sample.}
\label{table:spectra}
{\small
\begin{tabular}{l c c r@{ }l c c r@{ }l c r@{ }l c r@{ }l r@{ }l}
\noalign{\smallskip}\hline\hline\noalign{\smallskip}
Name & Dataset &
\parbox[c]{0.7cm}{\centering MJD [d]} &
\multicolumn{2}{c}{\parbox[c]{1.1cm}{\centering \teff\ $\pm$ $\sigma$ [K]}} &
\logg\ &
\parbox[c]{1.1cm}{\centering \vmic\ [\kms]} &
\multicolumn{2}{c}{\ion{Fe}{i}  $\pm$ $\sigma$} & $N_{\rm{\ion{Fe}{i}}}$ &
\multicolumn{2}{c}{\ion{Fe}{ii} $\pm$ $\sigma$} & $N_{\rm{\ion{Fe}{ii}}}$ &
\multicolumn{2}{c}{\ion{O}{i} $\pm$ $\sigma$} &
\multicolumn{2}{c}{\ion{S}{i} $\pm$ $\sigma$} \\
\noalign{\smallskip}\hline\noalign{\smallskip}
\object{AA\,Gem}      &   UVES & 54846.1489855 & 5577 & $\pm$ 132 & 1.1 & 3.5 & $-$0.18 & $\pm$ 0.10 &  67 & $-$0.17 & $\pm$ 0.11 &  12 & $-$0.24 & $\pm$ 0.08 & $-$0.34 & $\pm$ 0.11 \\
\object{AA\,Ser}      &   UVES & 54708.0398980 & 4814 & $\pm$  88 & ... & ... &         &        ... & ... &         &        ... & ... &		  & 	   ... &		 &  	  ... \\
...                   &    ... &           ... &      &       ... & ... & ... &         &        ... & ... &         &        ... & ... &		  & 	   ... &		 &  	  ... \\
\object{$\zeta$\,Gem} & STELLA & 57683.1372787 & 5347 & $\pm$  93 & 1.1 & 3.1 &    0.00 & $\pm$ 0.09 & 122 &	0.00 & $\pm$ 0.11 &  16 &	 0.23 & $\pm$ 0.13 & $-$0.08 & $\pm$ 0.08 \\
\object{$\zeta$\,Gem} & STELLA & 57703.2482580 & 5352 & $\pm$  97 & 1.1 & 3.1 & $-$0.01 & $\pm$ 0.09 & 124 & $-$0.02 & $\pm$ 0.11 &  19 &	 0.15 & $\pm$ 0.14 & $-$0.03 & $\pm$ 0.13 \\
\hline
\end{tabular}}
\tablefoot{The first three columns give the target name, spectroscopic dataset, and Modified Julian Date at which each spectrum was collected. Columns (4), (5), and (6) give, respectively, the effective temperature and its standard deviation, the surface gravity, and the microturbulent velocity. The uncertainties in \logg\ and \vmic\ were assumed to be 0.3~dex and 0.5~\kms, respectively \citep[see discussion in][]{Genovalietal2014}. Columns (7)-(8) and (9)-(10) list the \ion{Fe}{i} and \ion{Fe}{ii} abundances derived from individual lines together with the standard deviations and the number of lines used (given the small number of \ion{Fe}{ii} lines, a typical value of 0.11~dex was adopted as a minimal uncertainty for the abundances from this specie). The last two columns give the \ion{O}{i} and the \ion{S}{i} abundances with their uncertainties, estimated as described in Sect.~\ref{sec:atmpar_ab}. The complete table is available at the CDS.}
\end{table*}

\begin{table*}
\caption{Excerpt from the list of mean atmospheric parameters, Fe abundances, and S abundances derived for each star in our sample.}
\label{table:mean}
\centering
{\small
\begin{tabular}{l r@{ }l r@{ }l r@{ }l c r@{ }l c r@{ }l c c c c c}
\noalign{\smallskip}\hline\hline\noalign{\smallskip}
Name &
\multicolumn{2}{c}{[\ion{Fe}{i}/H]  $\pm$ $\sigma$} &
\multicolumn{2}{c}{[\ion{Fe}{ii}/H] $\pm$ $\sigma$} &
\multicolumn{2}{c}{[Fe/H] $\pm$ $\sigma$ (std)} &
$N$ &
\multicolumn{2}{c}{[O/H] $\pm$ $\sigma$ (std)} &
$N$ &
\multicolumn{2}{c}{[S/H] $\pm$ $\sigma$ (std)} &
$N$ &
$N_{\rm F}$ &
$N_{\rm H}$ &
$N_{\rm U}$ &
$N_{\rm S}$ \\
\noalign{\smallskip}\hline\noalign{\smallskip}
\object{AA\,Gem}      & $-$0.18 & $\pm$ 0.10 & $-$0.17 & $\pm$ 0.11 & $-$0.18 & $\pm$ 0.10        &   1 & $-$0.24 & $\pm$ 0.08        &   1 & $-$0.34 & $\pm$ 0.11        &   1 & ... & ... &   1 & ... \\
\object{AA\,Ser}      &         &        ... &         &        ... &         &        ...        &   0 &         &        ...        &   0 &         &        ...        &   0 & ... & ... & ... & ... \\
...                   &         &        ... &         &        ... &         &        ...        & ... &         &        ...        & ... &         &        ...        & ... & ... & ... & ... & ... \\
\object{l\,Car}       & $-$0.04 & $\pm$ 0.11 & $-$0.04 & $\pm$ 0.16 & $-$0.04 & $\pm$ 0.11        &   1 &         &        ...        &   0 & $-$0.08 & $\pm$ 0.13        &   1 &   1 & ... & ... & ... \\
\object{$\zeta$\,Gem} &    0.01 & $\pm$ 0.01 &    0.01 & $\pm$ 0.01 &    0.01 & $\pm$ 0.01 (0.06) & 131 &    0.03 & $\pm$ 0.01 (0.09) & 129 & $-$0.03 & $\pm$ 0.01 (0.05) & 131 & ... &  47 & ... &  84 \\
\hline
\end{tabular}}
\tablefoot{The first three columns give the target name and the iron abundances from neutral and ionized lines, which are either the same values from Table~\ref{table:spectra} for stars with single spectrum or the weighted mean and standard errors computed using the abundances from multiple spectra. Column (4) lists the iron abundances and their uncertainties, either calculated as a weighted mean from Cols. (2) and (3) or adopted from Col. (2) if only one spectra is available. For stars with multiple spectra, the standard deviation is also shown within parentheses, which was calculated using individual abundances from \ion{Fe}{i} and \ion{Fe}{ii} lines together. Column (5) gives the number of spectra for which we were able to derive the iron abundances and the atmospheric parameters. Columns from (6) to (9) list the oxygen and sulfur abundances (again, either the same values from Table~\ref{table:spectra} for stars with single spectrum or the weighted mean and standard errors computed using the abundances from multiple spectra), the standard deviations calculated for stars with multiple spectra, and the number of spectra used. The four last columns show the number of optical spectra available from each spectrograph: $N_{\rm F}$: FEROS; $N_{\rm H}$: HARPS; $N_{\rm U}$: UVES; $N_{\rm S}$: STELLA.} 
\end{table*}

For the sample clusters, we adopted the [O/H] ratios from \citet{Magrinietal2023}. The abundance of O is computed using the forbidden [\ion{O}{i}] line at 6300.3~\AA, in the LTE approximation \citep[see][for details]{Tautvaivsieneetal2015}. For S, to be consistent with the spectral analysis done on the CCs, we re-measured the abundance using only the \ion{S}{i} triplet at 6757~\AA. We then computed the mean [S/H] for each cluster. 

\subsection{Spatial distribution across the thin disk}

The heliocentric distances of our Cepheids were estimated based on four different diagnostics, listed here in priority order: 1) Gaia EDR3 distances from parallaxes within the $external.gaiaedr3\_distance$ table \citep{BailerJonesetal2021}; 2) W1-band Period-Luminosity relation \citep[PL, calibrated from Galactic CCs by][]{Wangetal2018}; 3) K-band PL \citep[calibrated on Galactic CCs by][]{Ripepietal2020}; IV) J-band PL \citep[calibrated from LMC CCs by][]{Ripepietal2022}. We adopted the Gaia distances only for Cepheids with $parallax\_over\_error > 10$ and $ruwe < 1.4$. The first condition is self-explanatory, whereas the second was included because $ruwe$ is an estimate of the goodness of the astrometric solution for single stars. A value of 1.4 is a typical threshold adopted to separate single detections from multiple detections candidates (or problematic solutions). For the other Cepheids, we adopted the PL relations using 2MASS and WISE magnitudes. Notice that we have ruled out all the magnitudes with bad photometric quality (X, U or E), from both surveys. The apparent magnitudes were unreddened by using E(B-V) from \citet{SchlaflyFinkbeiner2011} and the reddening law by \citet{Cardellietal1989}, extended to the MIR by \citet{Madoreetal2013}. We ended up with 282 CCs with distances from Gaia and 115/12/3 from W1-/K-/J-band PLs.

After deriving the heliocentric distances, we used the Gaia coordinates to estimate the Galactocentric Cartesian coordinates ($X$, $Y$, $Z$) and the Galactocentric distance on the plane ($R_{\rm G} = \sqrt(X^2 + Y^2)$). For these calculation, a distance of 8.127~kpc from the Galactic center was assumed \citep{Gravity2018}. Figure~\ref{figure:xyz_galactic} shows the radial distribution of the entire spectroscopic sample projected onto the Galactic plane. The symbols are color coded according to the distance from the Galactic plane (the absolute value of $Z$). Data plotted in this figure display several interesting findings worth being discussed in more detail:

{\em i) Galactocentric distances} -- The Cepheid Galactocentric distances cover almost 25~kpc, ranging from about 5~kpc for stars in the inner disk to almost 30~kpc for stars in the outer disk. The current sample includes, for the first time, eight Cepheids with Galactocentric distances larger than 18~kpc;

{\em ii) Height from the Galactic plane} -- The bulk of the spectroscopic sample has distances from the Galactic plane smaller than 500~pc, as expected for typical disk stellar populations. Interestingly enough, Cepheids located at Galactocentric distances larger than 12~kpc display a systematic drift towards negative $Z$ values and approach $Z$ = $-$1.5~kpc for $R_{\rm G}$$\sim$17-18~kpc. This trend and the increase at even larger Galactocentric distances are associated with the Milky Way (MW) warp. There is only one exception: the Cepheid \object{V1253\,Cen} ($X=-7.73$, $Y=3.63$~kpc) is located across the solar circle, but its height above the Galactic plane is $\sim$4~kpc. This circumstantial evidence indicates that this variable does not appear to be a typical disk star, and in turn a typical CCs. In passing, we notice that, in a recent investigation, \citet{Gaia2023} called attention to the fact that a large sample of Galactic field stars, included in the Gaia Data Release 3 (Gaia DR3), display a strong vertical asymmetry of the thin disk. The conclusions of this investigation are independent of the inclusion of this object and it was neglected. However, it is worth keeping this Cepheid under special surveillance. 

{\em iii) Distribution across the disk} -- The current sample covers the four quadrants and, in particular, the 3rd and the 4th quadrant. Interestingly enough, the spatial distribution agrees quite well with high-mass star forming regions observed by \citet{Reidetal2019}. The agreement is further supported by the comparison with the spiral arms identified by those authors, and it applies not only to the main arms (Sgr-Car, Perseus, Norma, Sct-Cen) but to the Local arm as well. However, there is evidence that the Norma-Outer arm has a larger pitch angle in the 3rd quadrant. The separation among the different arms is not very sharp due to the presence of several inter-arm objects, but the over-densities trace quite well the main arms.

{\em iv) Orbital properties} -- We also investigated the orbital properties of the current sample of Galactic Cepheids. In particular, we computed the circularity of the orbits, defined as the angular momentum $J_Z$ around the short $Z$-axis, normalized by the maximum angular momentum of a circular orbit with the same binding energy $E$: $\lambda_Z = J_Z/J_{\rm max}(E)$. The orbits were split into four components \citep{Zhuetal2018,Santuccietal2023}: a cold component with near circular orbits ($\lambda_Z > 0.8$) typical of the thin disk; a hot component with near radial orbits ($-0.25 < \lambda_Z < 0.25$) typical of the bulge; a warm component ($0.25 < \lambda_Z < 0.8$) typical of the thick disk, and a counter-rotating component ($\lambda_Z < -0.25$). To integrate the orbits we used the \textit{galpy} code \citep{Bovy2015}, modeling the potential of the Galaxy by means of the \textit{MWPotential2014} implementation, which is composed of a power-law with an exponential cut-off for the bulge, a \citet{MiyamotoNagai1975} disk, and a \citet{Navarroetal1996} halo\footnote{We assumed the position of the Sun in the Galaxy adopted by \citet{Gaia2018}: height above the Galactic plane $Z_\sun = 27$~pc \citep{Chenetal2001}, distance from the Galactic center $R_{\rm G}^\sun = 8.34$~kpc, and circular velocity at the solar radius $V_{\rm c}$ = 240~\kms\ \citep{Reidetal2014}. We also adopted the solar reflex motion components from \citet{Schonrichetal2010}, that is ($U_\sun$, $V_\sun$, $W_\sun$) = (11.1, 12.24, 7.25)~\kms.}. To perform the orbit integration we used the \citet{DormandPrince1980} integration method (by setting the option $method = dopr54\_c$ in \textit{galpy}). We integrated the orbits for 10~Gyr, starting from the current position and velocity of the variables. We found that the bulk of the current spectroscopic sample has cold orbits typical of thin disk stars. The same outcome applies to the Cepheids with larger heights above the disk. The fraction of objects with warm orbits typical of the thick disk is of the order of 1\%. The orbital and kinematic properties of Galactic Cepheids will be addressed in a forthcoming investigation (Zocchi et al., in prep.).

\section{Atmospheric parameters and abundances}
\label{sec:atmpar_ab}

The effective temperature (\teff), surface gravity (\logg), microturbulent velocity (\vmic), and metallicity ([Fe/H]) for our sample of classical Cepheids were derived using pyMOOGi\,\footnote{https://github.com/madamow/pymoogi}, a Python version of the MOOG code \citep{Sneden2002}. The model atmospheres, interpolated in the grid of \citet{CastelliKurucz2004}, and the equivalent widths (EWs), measured using ARES \citep{Sousaetal2007,Sousaetal2015}, are taken as input by the code. The atomic data for the \ion{Fe}{i} and \ion{Fe}{ii} lines are the same adopted by \citet{daSilvaetal2022}, in which a clean and homogeneous line list with revised atomic parameters were created. The procedure is the same adopted in our previous works, which is described in detail in \citet{Proxaufetal2018}. In that procedure, first we derive the \teff\ for each spectrum using the line depth ratios (LDR) method \citep[][and references therein]{Kovtyukh2007}. Then, keeping \teff\ fixed, the three other parameters are iteratively changed until convergence. A solution for the surface gravity is accepted only if the ionization equilibrium of \ion{Fe}{i} and \ion{Fe}{ii} lines is achieved. An accepted value for the microturbulent velocity is obtained only if the abundances from \ion{Fe}{i} lines do not depend, within the errors, on their EWs. The metallicity passed as input to our algorithm is constantly updated, and the final iron abundance adopted is the mean value computed from individual \ion{Fe}{i} lines. Table~\ref{table:spectra} lists the atmospheric parameters derived for each spectrum of each star in our sample.

Once the stellar parameters were computed, we derived S abundances by performing line profile fitting of the \ion{S}{i} triplet at 6757~\AA, adopting the same model atmospheres. This line is proven to be unblended and relatively strong in our sample stars \citep{Duffauetal2017}. We computed synthetic profiles using the Python version of SME (Spectroscopy Made Easy, \citealt{PiskunovValenti2017}), made available by A. Wehrhahn\footnote{\url{https://github.com/AWehrhahn}} \citep{Wehrhahn2021}. The adopted line list, covering 15 \AA\ around the \ion{S}{i} triplet, was retrieved from VALD\footnote{\url{http://vald.astro.uu.se}} using the "extract stellar" option and including hyperfine/isotopic splitting information when available. We updated atomic parameters for the line under scrutiny in this study following \cite{Duffauetal2017}. The reader is referred to that paper for further details and an extensive discussion concerning the 3D/NLTE effects of sulfur lines. Here we briefly recall that the lines in Multiplet 8 exhibit negligible departures (up to 0.1~dex) from the LTE approximation (see also \citealt{Takedaetal2005,Korotin2009}). We have also adopted the same solar abundance of A(S)$_\odot$ = 7.16~dex as \cite{Duffauetal2017}. To evaluate systematic effects due to the different code and line list with respect to \cite{daSilvaetal2022}, we reanalyzed the 20 calibrating Cepheids published in our previous work and found a mean difference of +0.03\,$\pm$\,0.02~dex, which is well within the observational uncertainties.

Oxygen abundances were derived via spectral synthesis of the [O~{\sc i}] forbidden line at 6300.3~\AA, for which NLTE departures are negligible. We analyzed an UVES solar spectrum, finding A(O)$_\odot$ = 8.78, which is the value we adopted as reference abundance throughout this work. Atomic parameters for the blending Ni~{\sc i} line at 6300.37 were adopted as such that $EP$ = 4.266~eV, and $\log{gf}$ = $-$2.11 (see also \citealt{Caffauetal2015}). Telluric lines affecting this spectral region around the oxygen line were removed using a synthetic template computed with the updated version of the {\sc TelFit} code \citep{Gulliksonetal2014}. 

Internal errors were computed considering the best-fit uncertainties as given by SME (see \citealt{PiskunovValenti2017} for further details) and errors in atmospheric stellar parameters. These have been estimated in the standard way, that is by changing one parameter at a time and computing the corresponding variations in abundance. The different contributions were then added in quadrature (we refer the reader to our previous works, e.g., \citealt{Dorazietal2020}).

\begin{table}
\centering
\caption{Slopes and zero-points of the abundance gradients as a function of the Galactocentric distance.}
\label{table:slopes}
\begin{tabular}{c r@{ }l r@{ }l c c}
\noalign{\smallskip}\hline\hline\noalign{\smallskip}
\parbox[c]{1.6cm}{\centering Abundance ratio} &
\multicolumn{2}{c}{\parbox[c]{1.6cm}{\centering Slope [dex\,kpc$^{-1}$]}} &
\multicolumn{2}{c}{\parbox[c]{1.6cm}{\centering Zero-point [dex]}} &
Fit &
Sample \\
\noalign{\smallskip}\hline\noalign{\smallskip}
 {[Fe/H]} & $-$0.041 & $\pm$ 0.003 &    0.32 & $\pm$ 0.02 & linear & CCs \\
 {[Fe/H]} & $-$0.907 & $\pm$ 0.046 &    0.81 & $\pm$ 0.04 & log    & CCs \\
 {[Fe/H]} & $-$0.055 & $\pm$ 0.004 &    0.48 & $\pm$ 0.04 & linear & OCs \\[0.1cm]
 {[O/H]}  & $-$0.029 & $\pm$ 0.006 &    0.21 & $\pm$ 0.05 & linear & CCs \\
 {[O/H]}  & $-$0.564 & $\pm$ 0.104 &    0.48 & $\pm$ 0.09 & log    & CCs \\
 {[O/H]}  & $-$0.040 & $\pm$ 0.009 &    0.30 & $\pm$ 0.07 & linear & OCs \\[0.1cm]
 {[S/H]}  & $-$0.081 & $\pm$ 0.004 &    0.59 & $\pm$ 0.03 & linear & CCs \\
 {[S/H]}  & $-$1.624 & $\pm$ 0.053 &    1.40 & $\pm$ 0.05 & log    & CCs \\
 {[S/H]}  & $-$0.086 & $\pm$ 0.009 &    0.68 & $\pm$ 0.08 & linear & OCs \\[0.1cm]
 {[O/Fe]} &    0.003 & $\pm$ 0.007 & $-$0.04 & $\pm$ 0.05 & linear & CCs \\[0.1cm]
 {[S/Fe]} & $-$0.037 & $\pm$ 0.003 &    0.25 & $\pm$ 0.03 & linear & CCs \\
 {[S/Fe]} & $-$0.606 & $\pm$ 0.047 &    0.50 & $\pm$ 0.04 & log    & CCs \\
\hline
\end{tabular}
\tablefoot{Based on the data plotted in Figs.~\ref{figure:xh_rgal_open_cluster} and \ref{figure:xfe_rgal_open_cluster}. The linear regressions for open clusters are not shown in the quoted figures.}
\end{table}

\begin{figure*}
\centering
\begin{minipage}[t]{0.8\textwidth}
\centering
\resizebox{\hsize}{!}{\includegraphics{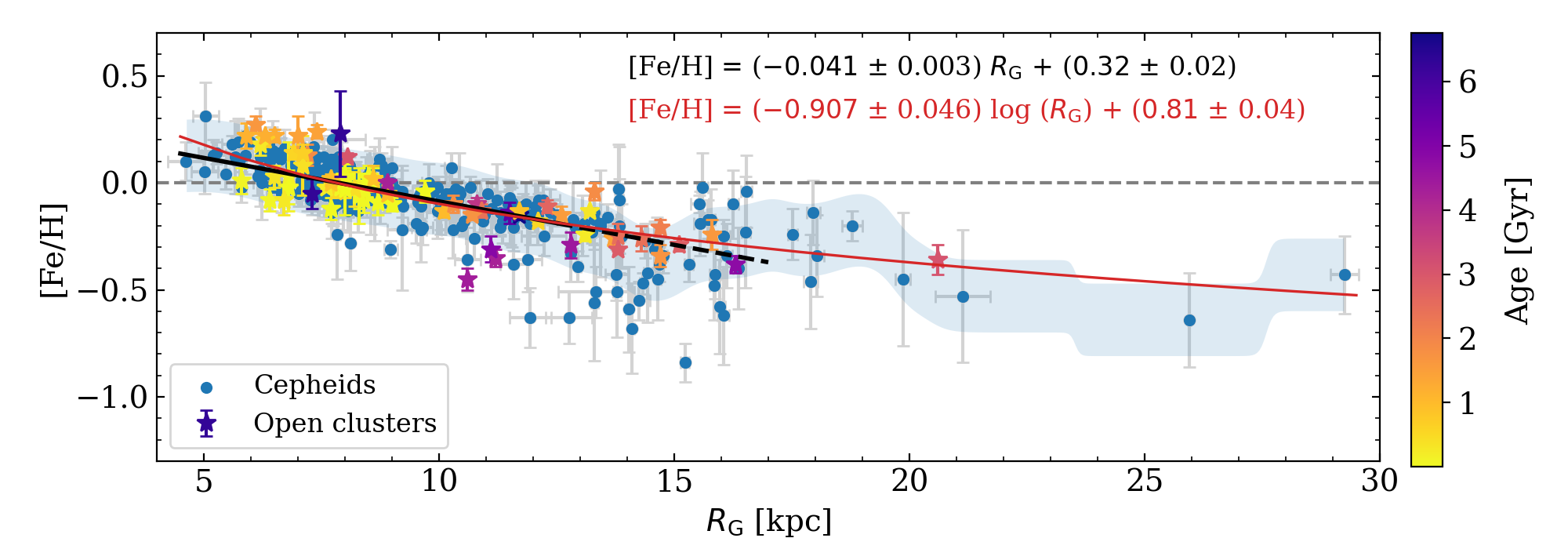}}
\end{minipage} \\
\begin{minipage}[t]{0.8\textwidth}
\centering
\resizebox{\hsize}{!}{\includegraphics{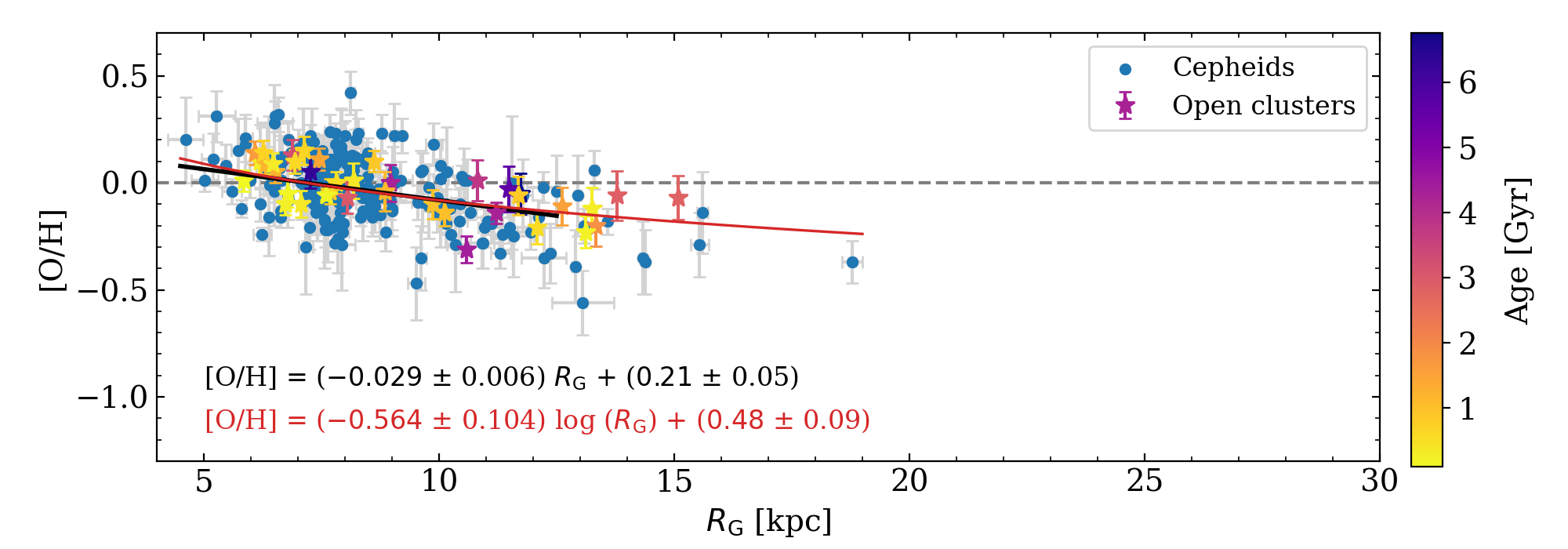}}
\end{minipage} \\
\begin{minipage}[t]{0.8\textwidth}
\centering
\resizebox{\hsize}{!}{\includegraphics{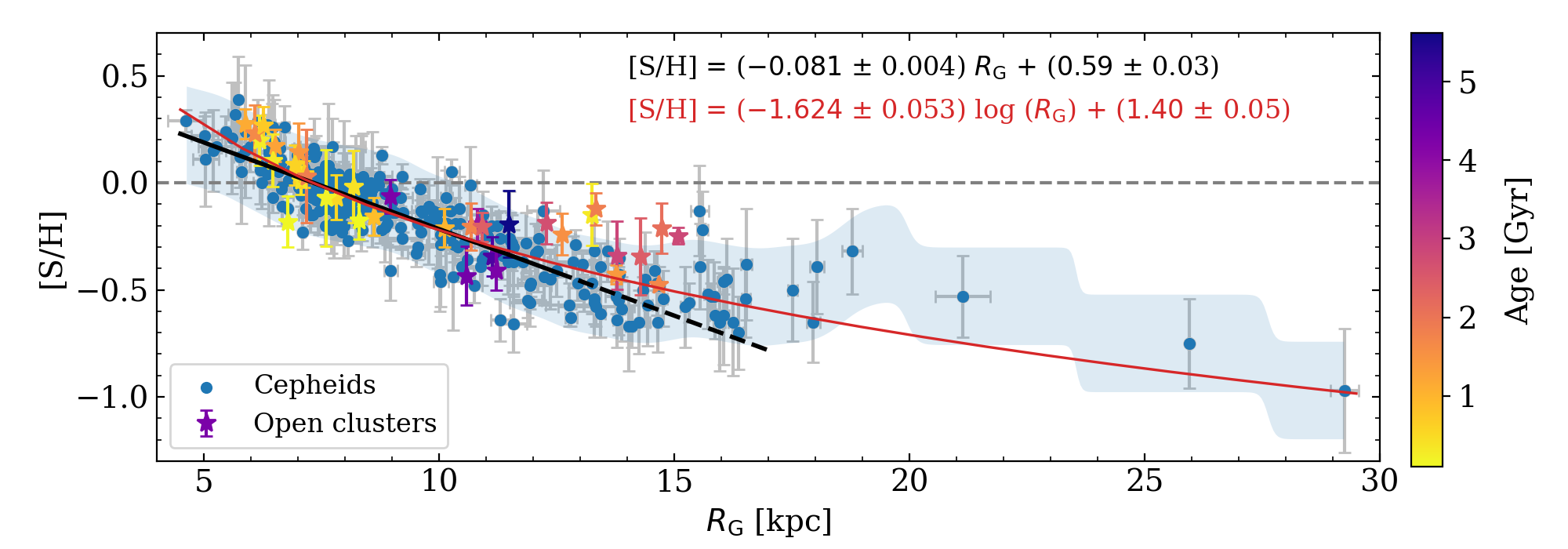}}
\end{minipage}
\caption{Abundance ratios as a function of the Galactocentric distance. Our sample of Cepheids (blue circles) is compared with our sample of open clusters (symbols color-coded according to the stellar age). The Fe and O abundance ratios are from \citet{Magrinietal2023} whereas the S abundances are from the present work. The black line displays a linear regression fitted to Cepheids in the range of Galactocentric distance from 4 to 12.5~kpc, while the dashed line is the extrapolation up to 17~kpc. The red line shows a logarithmic fit over the whole range of Galactocentric distances. The corresponding equations are also shown. The shaded area shows one standard deviation around the running mean, weighted with a Gaussian function taking the errors into account.}
\label{figure:xh_rgal_open_cluster}
\end{figure*}

Figure~\ref{figure:examples_synth} shows examples of spectral synthesis fit around the sulfur line at 6757~\AA\ for HARPS, FEROS, UVES, and STELLA spectra. Figure~\ref{figure:paratm_ab_literature} shows a comparison of the atmospheric parameters and abundances derived in the present work with the data available in the literature for a set of stars in common. Figure D.1 shows O and S abundances as a function of the pulsation phase for a sub-sample of CCs with multiple spectra collected along the pulsation cycle \citep[see][]{daSilvaetal2022}. Similarly to our results previously published for iron and alpha elements, this figure clearly shows that there is no significant correlation of the [O/H] and [S/H] abundance ratios with the pulsation cycle.

\section{Galactic radial Gradients}\label{sec:rad_grad}
\label{sec:gradients}

\subsection{[Fe/H] and [$\alpha$/H] radial gradients} 
\label{ss:FeH_alphaH_gradients}

Figure~\ref{figure:xh_rgal_open_cluster} displays the radial gradients for derived iron, oxygen, and sulfur. The corresponding coefficients of both linear (for $R_{\rm G}$ within 12.5~kpc) and logarithmic (for the whole range of $R_{\rm G}$) regressions are listed in Table~\ref{table:slopes}. A glance at the data plotted in this figure discloses several interesting new findings:

{\em i) Slope of the iron radial gradient} --
Dating back to more than twenty years ago, it has been suggested since then that the radial gradient, based either on classical Cepheids or on open clusters, shows a change in the slope across the solar circle, despite of the few data available for radii above $\sim$12~kpc \citep[see, e.g.,][]{Magrinietal2017}. To support this preliminary evidence, a few authors suggested that this variation could be associated with a resonance in the spiral arm \citep{Lepineetal2011,Lepineetal2017}. However, from a more detailed analysis of the radial distribution of Galactic Cepheids across the thin disk, and using a more detailed algorithm to constrain the clustering of such stars, \cite{Genovalietal2014} instead found a linear radial gradient when moving from the inner to the outer disk. Moreover, they also suggested that the dispersion, at fixed Galactocentric distance, is mainly caused by a spread in chemical composition inside the same Cepheid group. This finding was soundly and independently supported in a recent and more sophisticated approach done by \citet{Lemasleetal2022} by taking the disk flare into account.

Data plotted in the top panel of Fig.~\ref{figure:xh_rgal_open_cluster} show that the metallicity radial gradient is almost linear for Galactocentric distances smaller than $\sim$12~kpc, but it becomes significantly flatter for $R_{\rm G}$ larger than 15~kpc. By performing a number of numerical fits using different analytical functions, we found that a logarithmic fit can take account for the steepening of the gradient when approaching the inner disk and for the flattening when approaching the outer disk. We also performed a linear fit for Cepheids with Galactocentric distances smaller than 12.5~kpc. The current estimate agrees quite well with similar estimates available in the literature \citep{Magrinietal2023}. The flattening in the outer disk fully supports preliminary results based on both CCs and OCs \citep{Carraroetal2007,Yongetal2012,Donoretal2020}. In passing, we notice that the current empirical evidence concerning the linearity is based on the largest and most homogeneous sample of Galactic Cepheids.

On the other hand, \citet{Twarogetal1997}, by using a large sample of open clusters, found evidence of a metallicity discontinuity located at $R_{\rm G}$$\sim$10~kpc and suggested a two-zone model for the chemical enrichment of the Galactic thin disk. This working hypothesis was also supported by \citet{Caputoetal2001} by using photometric metallicities for a large sample of Galactic CCs. This empirical evidence was interpreted by \citet{Lepineetal2017} as the possible presence of the co-rotation resonance of the Galactic thin disk. It was also questioned by \citet{Genovalietal2014}, since the residuals of the iron gradients appeared to be correlated with the location of the spiral arms. Indeed, these authors suggested that the evidence of a possible change in the slope and the dip in the rotation curve of the Galactic thin disk between 9 and 10~kpc \citep{Sofue2013} were associated with the Perseus arm. More recently, by analyzing a larger sample of CCs, \citet{Trentinetal2023} also found some evidence of a possible change in the slope at $R_{\rm G}$$\sim$9.25~kpc. They performed linear regressions to the entire dataset and to the data obtained by binning the CCs in twelve radial intervals of 1.33~kpc each. Similar fits to individual objects and to the binned data were also performed for Galactocentric distances smaller and larger than 9.25~kpc. They found that one single fit to the entire sample can account for both inner and outer disk Cepheids. However, the possible presence of a break in the radial gradient could not be excluded on the basis of their dataset. The current data indicate a departure from linearity for Galactocentric distances larger than 12-14~kpc, but a more quantitative analysis requires a larger sample and a more detailed statistical approach (Lemasle et al., in prep.), in order to confirm whether the radial gradient shows either a break at a given Galactocentric distance or a smooth change with distance, as suggested by the logarithmic fit.

{\em ii) Comparison with the iron gradient of open clusters} --
To investigate the impact that age has on the radial metallicity gradient, we performed a detailed comparison with our sample of Galactic open clusters. CCs cover a limited range in age, from a few tens of Myr (long period) to a few hundreds of Myr (short period). Therefore, we took advantage of the homogeneous abundance estimates recently provided by the last GES data release. The key advantage of this sample is that, in addition to the homogeneous abundances, it also has homogeneous estimates of individual distances and ages from isochrone fitting of their members detected in Gaia DR2 \citep{Cantatetal2020}. Furthermore, the GES sample of OCs covers a broad range in Galactocentric distances ($R_{\rm G}$$\sim$6-20~kpc) and their cluster ages range from a few Myr (i.e., even younger than the youngest CCs) to almost 7~Gyr. Data plotted in Fig.~\ref{figure:xh_rgal_open_cluster} (see also Table~\ref{table:slopes}) show that CCs and OCs display a similar iron radial gradient when moving from the inner to the outer disk. Moreover, the OCs do not show any clear variation both in the zero-point and in the slope as a function of the cluster age. The cluster ages of the GES sample changes by more than one order of magnitude, but their distribution is similar, within the errors. This is in agreement with what was found in \citet{Magrinietal2023} for the radial gradient of several elements belonging to different nucleosynthesis channels, whereby a limited temporal evolution is observed in the age range covered by open clusters. The similarity of CCs and OCs iron radial gradients is soundly supported by the similarity of both zero-point and slope of the linear fits listed in Table~\ref{table:slopes}. However, it is worth noting that only few of the clusters have ages larger than 3~Gyr ($\sim$10 clusters), therefore limiting the considerations we can draw concerning the old age tail. The flattening of the radial gradient in the outermost disk regions is only partially supported by OCs, since most of them have Galactocentric distance smaller than $\sim$15~kpc and only one cluster is located at 20~kpc.

\begin{figure*}
\centering
\resizebox{0.8\hsize}{!}{\includegraphics{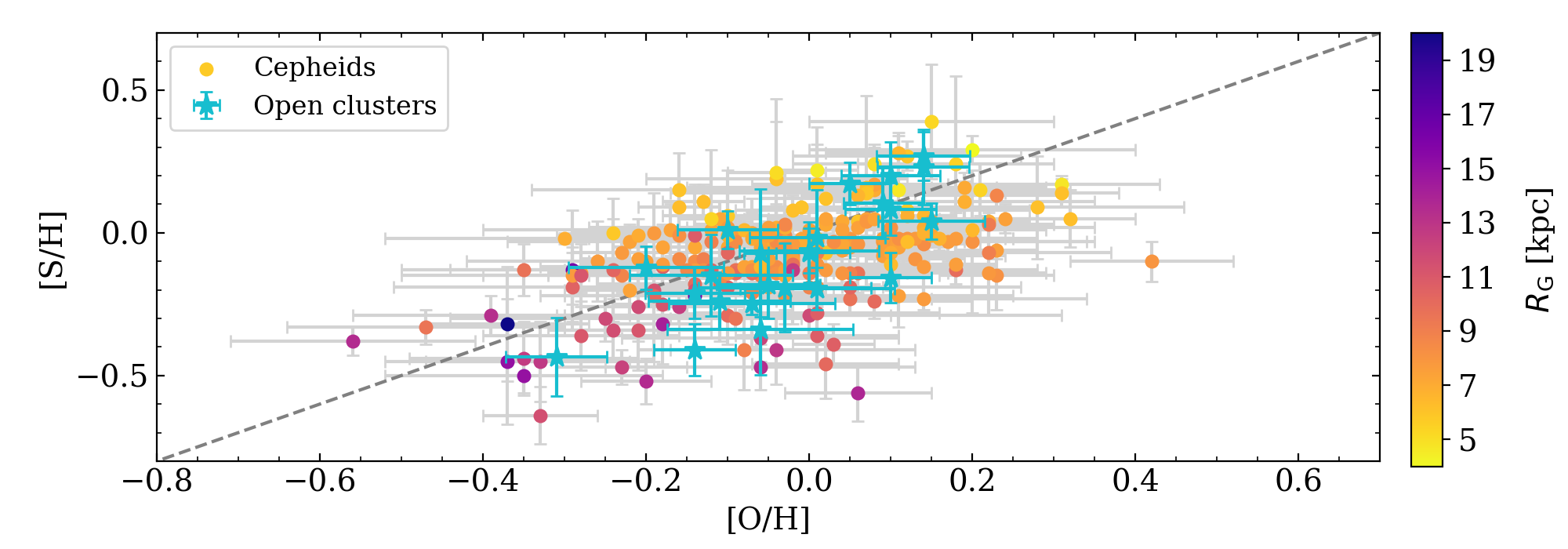}}
\caption{[S/H] as a function of the [O/H] abundances. 
Our sample of Cepheids, color-coded according to the distance from the Galactic center (see the color bar on the right side), is compared with the same open clusters plotted in Fig.~\ref{figure:xh_rgal_open_cluster}. The grey dashed line is the 1:1 linear relationship.}
\label{figure:sh_oh_rgal}
\end{figure*}

{\em iii) Slope of the oxygen radial gradient} --
Oxygen abundances plotted in Fig.~\ref{figure:xh_rgal_open_cluster} display quite clearly that the oxygen radial gradient is shallower than the iron gradient. The slope of the linear fit is 25\% shallower whereas the logarithmic fit can be barely compared since oxygen abundances in the outermost disk regions are missing. CCs and OCs display, once again, very similar radial trends with no clear dependence on the cluster age. The comparison with similar estimates available in the literature based on CCs indicates that the current slope is shallower. In particular, it is shallower than the slope obtained by \citet{Trentinetal2023} using a sample of 65 CCs plus literature data ($-$0.029\,$\pm$\,0.006 vs. $-$0.046\,$\pm$\,0.002~dex\,kpc$^{-1}$) and by \citet{LuckLambert2011} using a sample of 313 CCs ($-$0.056\,$\pm$\,0.003~dex\,kpc$^{-1}$). This comparison should be cautiously treated since it might be related to the different lines exploited for measuring oxygen abundances, and more in general to the details of the spectroscopic analysis (such as code, model atmospheres, and atmospheric parameters). Concerning OCs, by using data provided by \cite{Magrinietal2023}, we performed an independent fit following the same approach adopted for CCs. We found that the two gradients are, within the errors, in agreement with each other ($-$0.029\,$\pm$\,0.006 vs. $-$0.040\,$\pm$\,0.009~dex\,kpc$^{-1}$), thus supporting a similarity of the oxygen radial gradients between CCs and OCs.

The oxygen abundances provided by \cite{LuckLambert2011} and by \cite{Trentinetal2023} are based on a combination of the 6156$-$8 triplet and the [\ion{O}{i}] forbidden line at 6300.3~\AA, whereas cluster abundances \citep{Magrinietal2023,Randichetal2022} are only based on the line at 6300.3~\AA. Similarly, we only adopted the forbidden line at 6300.3~\AA\ because it is stronger than the 6156$-$8 triplet in all our sample stars.

{\em iv) Slope of the sulfur radial gradient} -- Data plotted in the bottom panel of Fig.~\ref{figure:xh_rgal_open_cluster} bring forward the key role that sulfur can play and that can help us to better understand the chemical enrichment of the Galactic thin disk:

\begin{figure*}
\centering
\begin{minipage}[t]{0.8\textwidth}
\centering
\resizebox{\hsize}{!}{\includegraphics{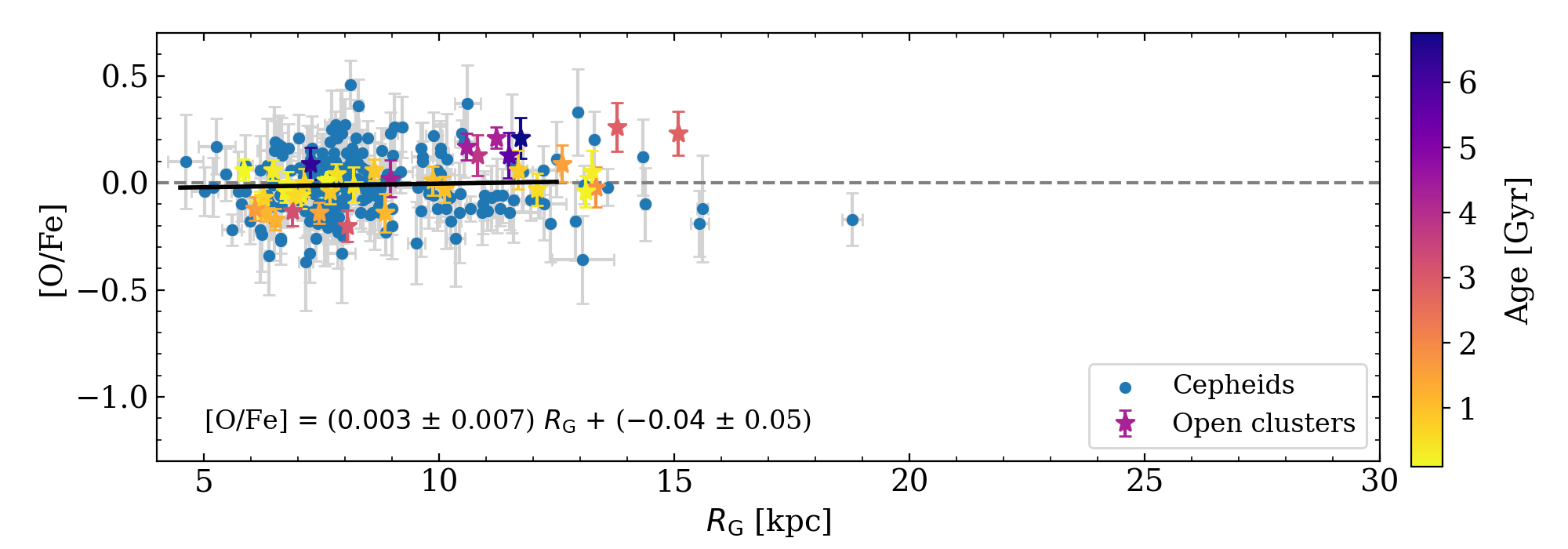}}
\end{minipage}
\begin{minipage}[t]{0.8\textwidth}
\centering
\resizebox{\hsize}{!}{\includegraphics{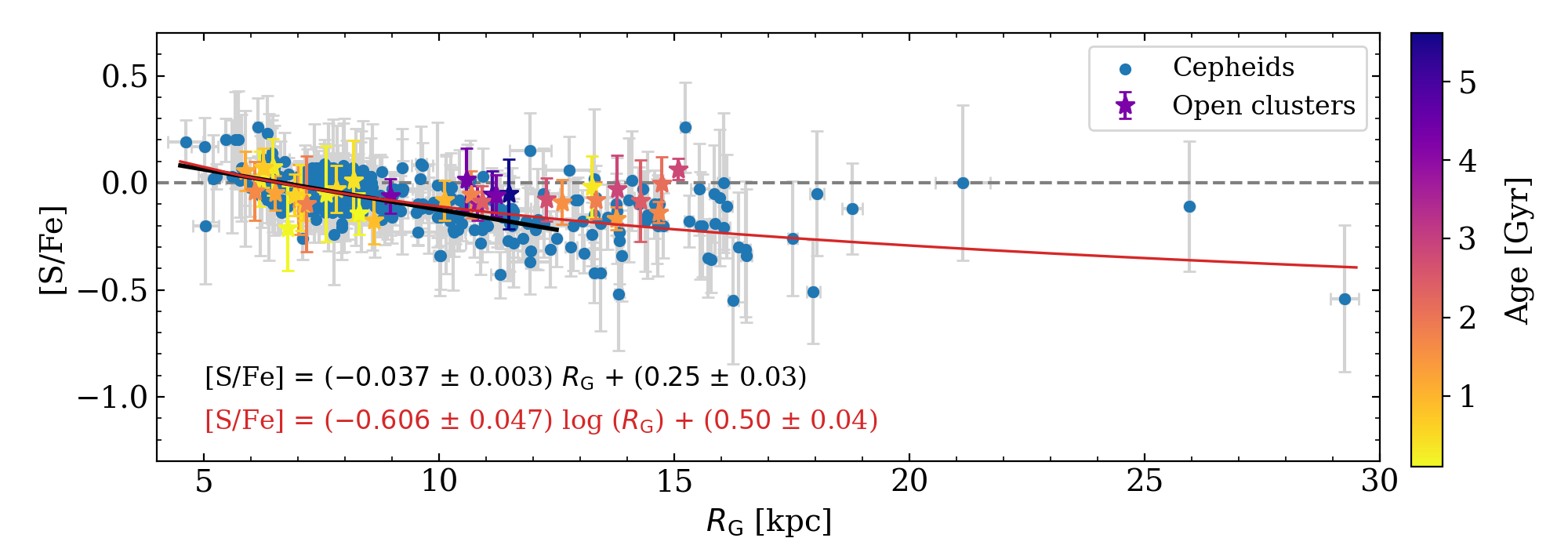}}
\end{minipage}
\caption{Abundance ratios as a function of the Galactocentric distance. Same as in Fig.~\ref{figure:xh_rgal_open_cluster} but showing the [O/Fe] 
and the [S/Fe] abundance ratios.}
\label{figure:xfe_rgal_open_cluster}
\end{figure*}

a) The slope of the sulfur radial gradient based on the linear fit is a factor of two steeper than the slope of the iron gradient. The difference in the logarithmic fit is of the order of 0.7~dex\,($\log{\rm kpc})^{-1}$. Notice that the latter fit ranges from the inner disk to the outskirts of the thin disk. This finding supports preliminary results obtained by \citet{daSilvaetal2022}, although the significance of those results was hampered by the sample size. Notice that the sulfur radial gradients for both CCs and OCs agree quite well. Indeed, the zero-points and the slopes of the linear fits attain, within the errors, very similar values (see Table~\ref{table:slopes}). It is also worth mentioning that the current data do not show any evidence of an abrupt change in the slope when moving from the inner to the outer disk.

b) The standard deviation of sulfur abundances are, at fixed Galactocentric distance, smaller than the dispersion of oxygen and iron abundances, thus suggesting that the spectroscopic diagnostic we adopted is also minimally affected by possible changes in effective temperature, surface gravity, or microturbulent velocity. To our knowledge, this is the very first time in which there is solid evidence of an $\alpha$-element gradient that is steeper than the iron gradient, and paves the way to a new empirical framework concerning the role played by massive stars in the chemical enrichment of both iron and sulfur. 

\begin{figure*}
\centering
\begin{minipage}[t]{0.497\textwidth}
\centering
\resizebox{\hsize}{!}{\includegraphics{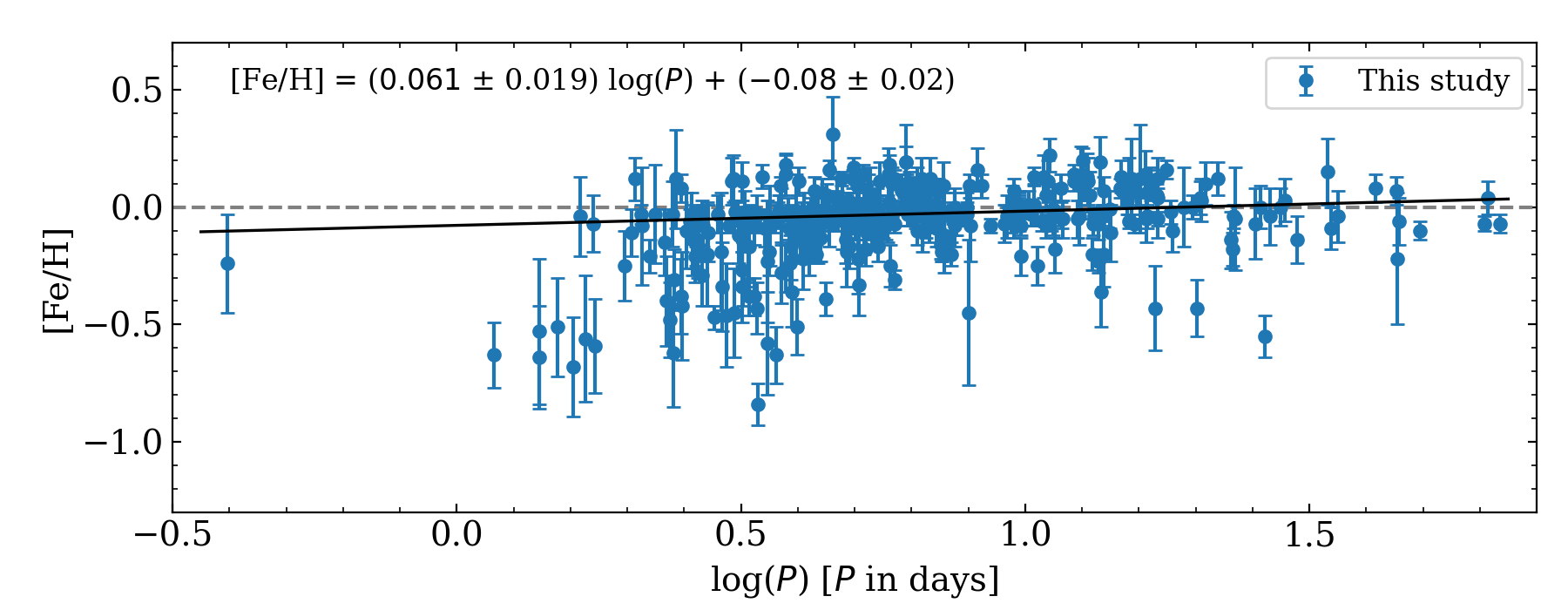}}
\end{minipage}
\begin{minipage}[t]{0.497\textwidth}
\centering
\resizebox{\hsize}{!}{\includegraphics{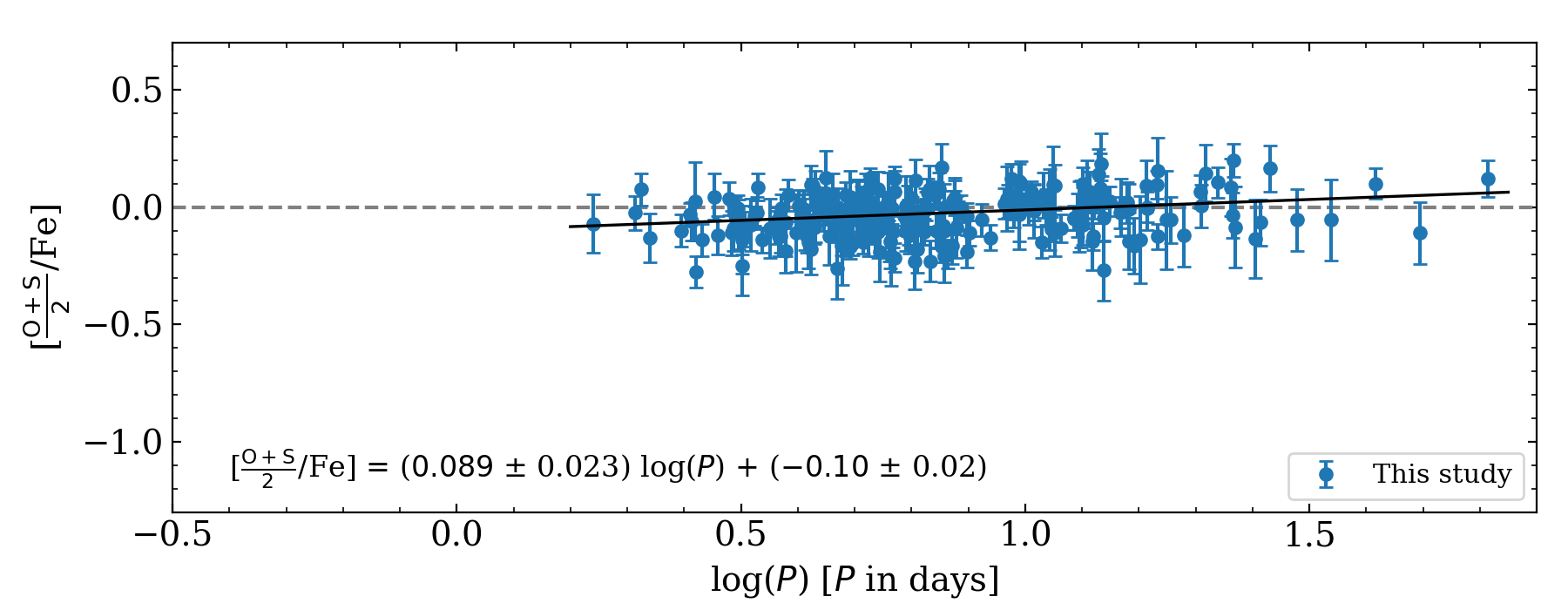}}
\end{minipage} \\
\begin{minipage}[t]{0.497\textwidth}
\centering
\resizebox{\hsize}{!}{\includegraphics{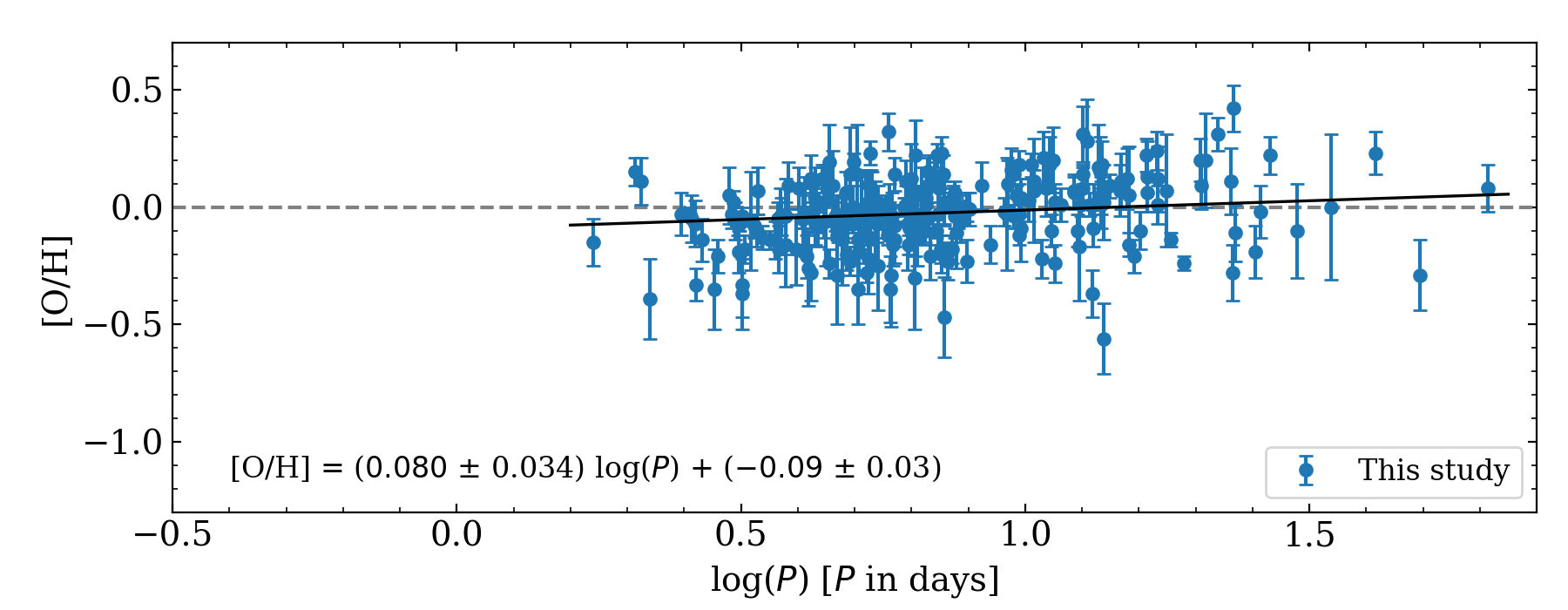}}
\end{minipage}
\begin{minipage}[t]{0.497\textwidth}
\centering
\resizebox{\hsize}{!}{\includegraphics{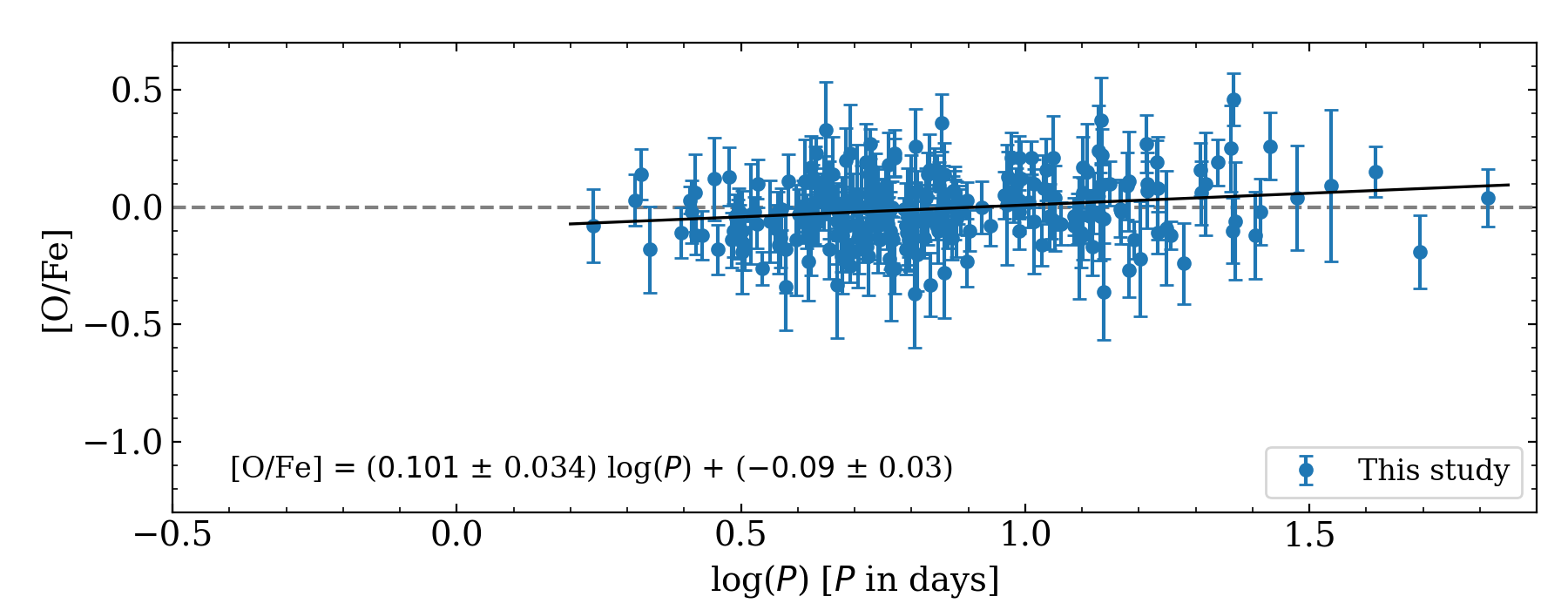}}
\end{minipage} \\
\begin{minipage}[t]{0.497\textwidth}
\centering
\resizebox{\hsize}{!}{\includegraphics{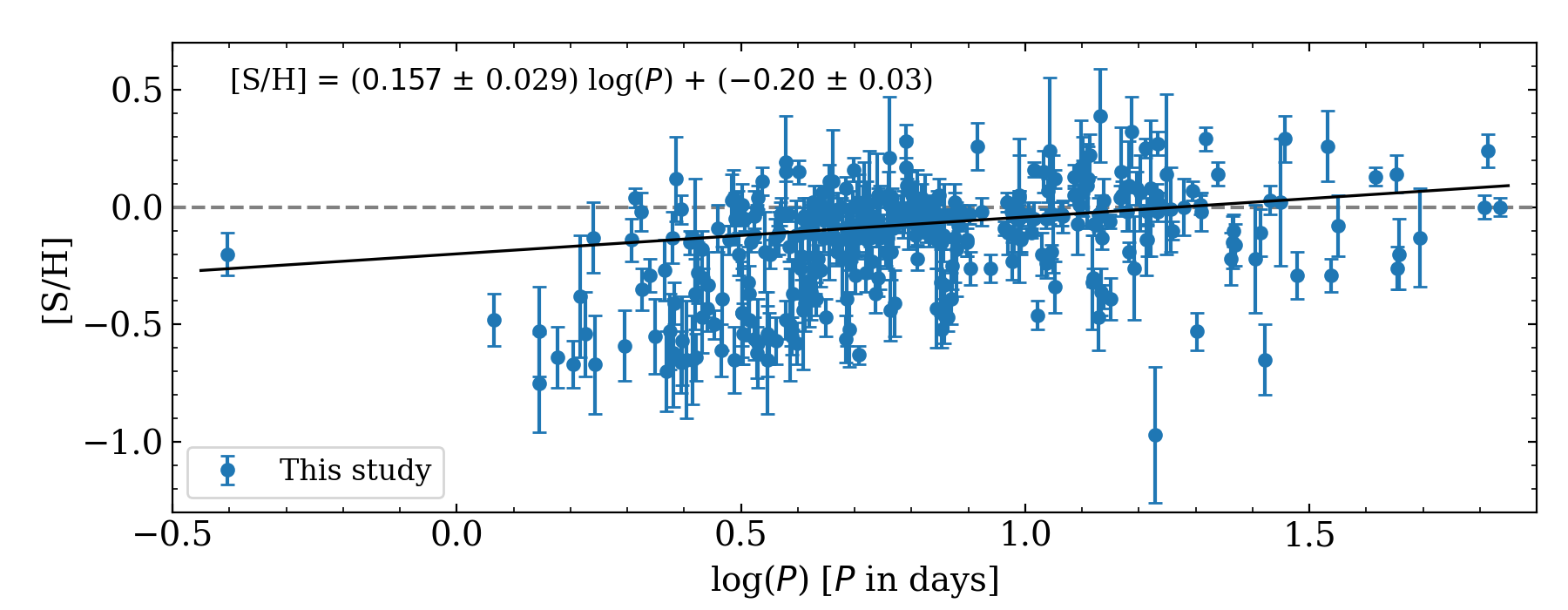}}
\end{minipage}
\begin{minipage}[t]{0.497\textwidth}
\centering
\resizebox{\hsize}{!}{\includegraphics{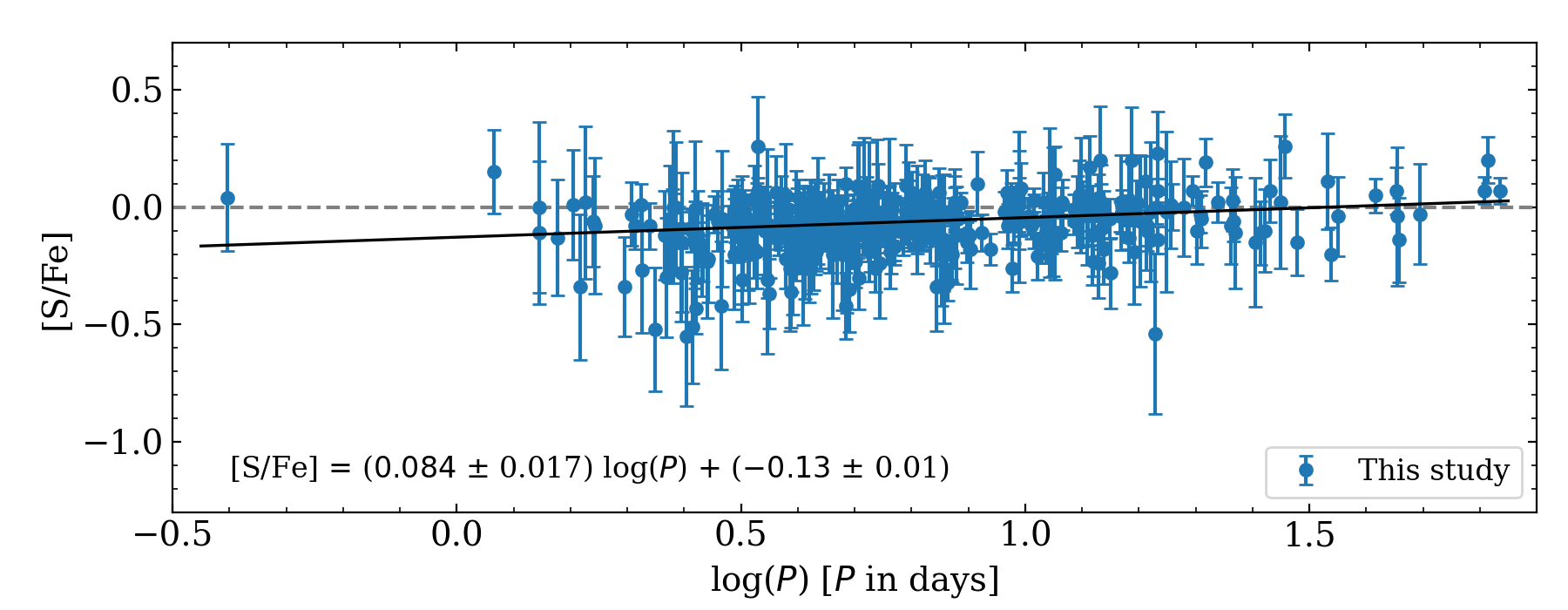}}
\end{minipage}
\caption{Abundance ratios as a function of the logarithmic pulsation period. The black line shows a linear regression fitted to the current Cepheid sample. The corresponding equation is also shown.}
\label{figure:xh_logP}
\end{figure*}

It is worth noting that the slope of the sulfur gradient provided by \cite{Trentinetal2023} is about 25\% shallower ($-$0.060\,$\pm$\,0.006~dex\,kpc$^{-1}$), but it is based on three \ion{S}{i} lines, namely $\lambda$ = 6743.585, 6748.153, 6757.153~\AA. However, we cannot ascertain whether each line of the multiplet has been measured in all their Cepheids. On the other hand, in all our sample stars only the strongest line at 6757~\AA\ can be useful for abundance determination (see a similar discussion in \citealt{Duffauetal2017}). The slope provided by \cite{Luck2018} is slightly shallower ($-$0.0693\,$\pm$\,0.0035~dex\,kpc$^{-1}$), but there is missing information on which lines they used.

In passing, we also notice that the current slope for the [S/H] radial gradient is similar, within the errors, to the estimate provided by \citet[][$-$0.050\,$\pm$\,0.025~dex\,kpc$^{-1}$]{Perdigonetal2021} based on a large sample of field stars with Galactocentric distances between 6 and 10~kpc, and steeper than the radial gradient provided by \citet[][$-$0.035\,$\pm\,$0.006~dex\,kpc$^{-1}$]{Arellanoetal2020} using H\,II regions with Galactocentric distances ranging from 7 to 14~kpc. Note that the comparison with similar estimates available in the literature should be considered as a global validation test, since CCs, when compared with field stars, cover a well defined range in age and they are associated to a specific 
evolutionary channel.

To further investigate the difference between the two $\alpha$ elements studied in the present work, Fig.~\ref{figure:sh_oh_rgal} shows the S-O relationship for both classical Cepheids and open clusters. Data plotted in this figure indicate that in the metal-poorer regime S and O display, within the errors, the same trend. Notice that these objects are located, as expected, in the outer disk (see the color bar on the right). Interestingly enough, the main finding emerging from the current analysis is that CCs and OCs are, at fixed sulfur abundance, more oxygen enhanced. Indeed, the distribution across the one-to-one liner relationship is asymmetrical, and a good fraction of CCs and the large majority of OCs are located below the dashed line. The trend becomes even clearer for sulfur abundances close to solar, where the oxygen abundances are on average 0.1-0.2~dex more enhanced. This may be explained by the fact that oxygen is only produced by massive stars \citep[see, e.g.,][]{Matteucci2021}, while a non-negligible fraction of sulfur is also produced on longer timescales by SNe Ia \citep[e.g.,][]{LeungNomoto2018,LeungNomoto2020}.

\subsection{[$\alpha$/Fe] radial gradients} 
\label{ss:alphaFe_gradients}

Figure~\ref{figure:xfe_rgal_open_cluster} shows the $\alpha$-to-iron abundance ratio as a function of the Galactocentric distance. In the top panel, the oxygen-to-iron abundance ratio ([O/Fe]) is, as expected, almost constant over the entire range of distances. The slope is quite flat and agrees, within the errors, with similar estimates of the [$\alpha$/Fe] radial gradients available in the literature. In a recent investigation, \citet{SantosPeral2021} found a [Mg/Fe] radial gradient of 0.025\,$\pm$\,0.009~dex\,kpc$^{-1}$ using field stars with Galactocentric distances between 6 and 11~kpc.

In the bottom panel of the same figure, the sulfur-to-iron abundance ratio ([S/Fe]) shows a well-defined negative gradient, supported by both the linear and the logarithmic regressions. The quoted evidence is somehow puzzling, since a similar estimate done by \citet{Perdigonetal2021}, based on a large sample of field stars, indicates a flat [S/Fe] radial gradient between 6 and 10~kpc (0.004\,$\pm$\,0.006~dex\,kpc$^{-1}$). The comparison with similar estimates available in the literature should be cautiously treated, since field stars cover a broad range in age and in evolutionary phases.

Oxygen and magnesium are mainly produced in hydrostatic nucleosynthesis of massive stars \citep{McWilliametal2008,Kobayashietal2020}. Sulfur has a mixed origin since it is produced during the final evolutionary phases of massive ($M\,\ge\,20$~M$_\odot$) stars, but it is also produced during type II SNe explosions \citep{LimongiChieffi2003}. Sulfur has, therefore, a key advantage when compared with other $\alpha$-elements.

The range in iron abundance covered by the current sample of CCs, and for which we were able to measure the oxygen abundance, is of the order of 0.7~dex. In this range, the [O/Fe] abundance ratio is constant whereas the [S/Fe] ratio shows a steady decrease when moving from the inner (more metal-rich) to the outer (more metal-poor) disk. Notice that the depletion in sulfur is already of the order of 0.2~dex at $R_{\rm G}$$\sim$12.5~kpc and becomes of the order of 0.5~dex in the outermost disk regions.
 
The physical reasons driving the difference in the abundance ratios of these two $\alpha$-elements are not clear. It would be quite interesting to investigate the abundance ratios of the other explosive $\alpha$-elements (Si, Ca, Ti) to provide a more complete analysis of their enrichment in the Milky Way. Sulfur can play a key role in this context, since it is a moderately volatile element and, therefore, it is not blocked into the dust grains of the interstellar medium. For this reason, S abundance in stars can be directly compared with abundances in H\,II regions, supernovae remnants, P\,Ne \citep{Henryetal2004}, damped Ly-$\alpha$ systems and, in general, high-redshift galaxies \citep{Vladiloetal2018,Dessaugesetal2007}, and in active galactic nuclei \citep{Liuetal2015,Mizumotoetal2023}.

\subsection{Dependence on age of classical Cepheids abundances} 

To further investigate the difference in the slope between iron and $\alpha$-elements, the left panels of Fig.~\ref{figure:xh_logP} show the same abundance ratios plotted in Fig.~\ref{figure:xh_rgal_open_cluster}, but as a function of the logarithmic pulsation period. As already mentioned in Sects.~\ref{sec:intro} and \ref{sec:rad_grad}, the pulsation period is a solid diagnostic for the individual age of CCs \citep{Bonoetal2005}. The abundances plotted in the top and in the bottom panels show, for the first time, that CCs with periods shorter than $\sim$4~days ($\log{P}$$\lesssim$\,0.6, older ages) are also iron- and sulfur-poorer. In the middle panel, we do not see the same trend, but only because oxygen abundances are missing in a good fraction of the metal-poor CCs. This is a spectroscopic validation of an intrinsic property of CCs. Their period distribution systematically shifts towards shorter periods when moving into the metal-poor regime. Up to now, this circumstantial evidence was only based on the difference in the period distribution of CCs in the Small Magellanic Cloud (SMC) and the Large Magellanic Cloud (LMC) \citep{Soszynskietal2017,Pietrukowiczetal2021,Bonoetal2023}. This is the very first time that the drift toward shorter periods is observed in metal-poor Galactic CCs.

The right panels of Fig.~\ref{figure:xh_logP} display the element-to-iron abundance ratios as a function of the logarithmic pulsation period. The linear regressions show that the slope of the [O/Fe] ratio is steeper when compared with the [S/Fe] and [((O+S)/2)/Fe] slopes. Moreover, there is a clear increase of the spread in the range of short periods.

\begin{figure*}
\centering
\begin{minipage}[t]{0.52\textwidth}
\centering
\resizebox{\hsize}{!}{\includegraphics{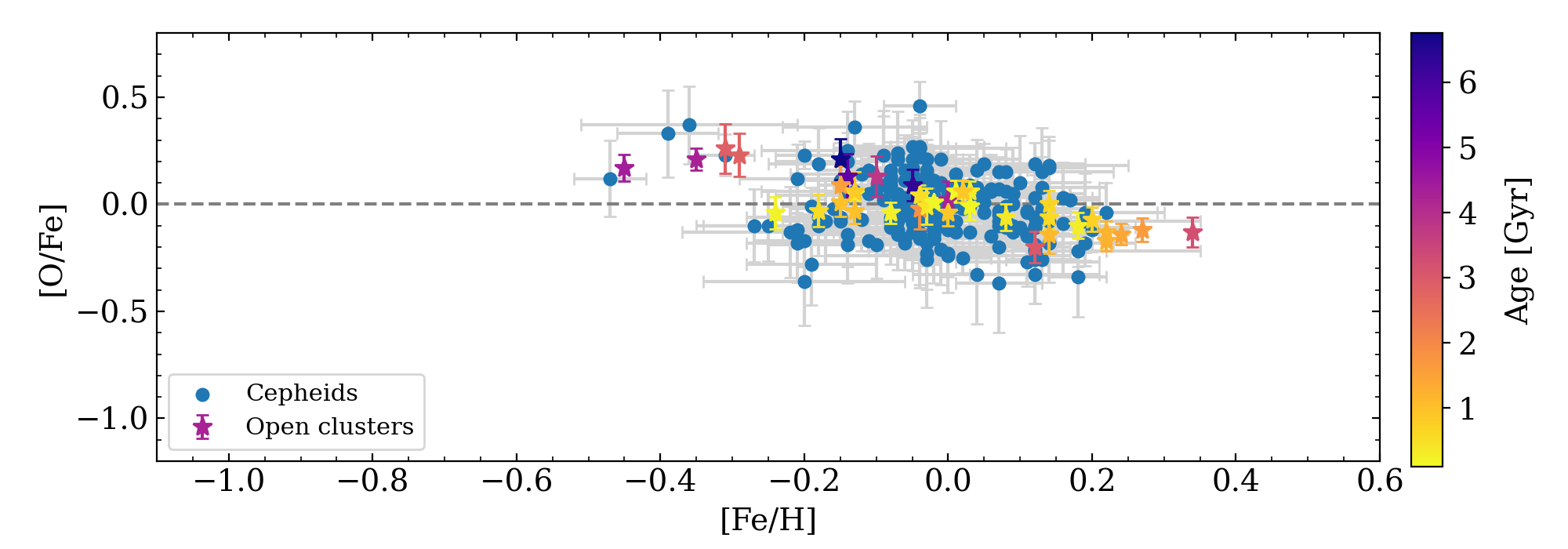}}
\end{minipage}
\begin{minipage}[t]{0.47\textwidth}
\centering
\resizebox{\hsize}{!}{\includegraphics{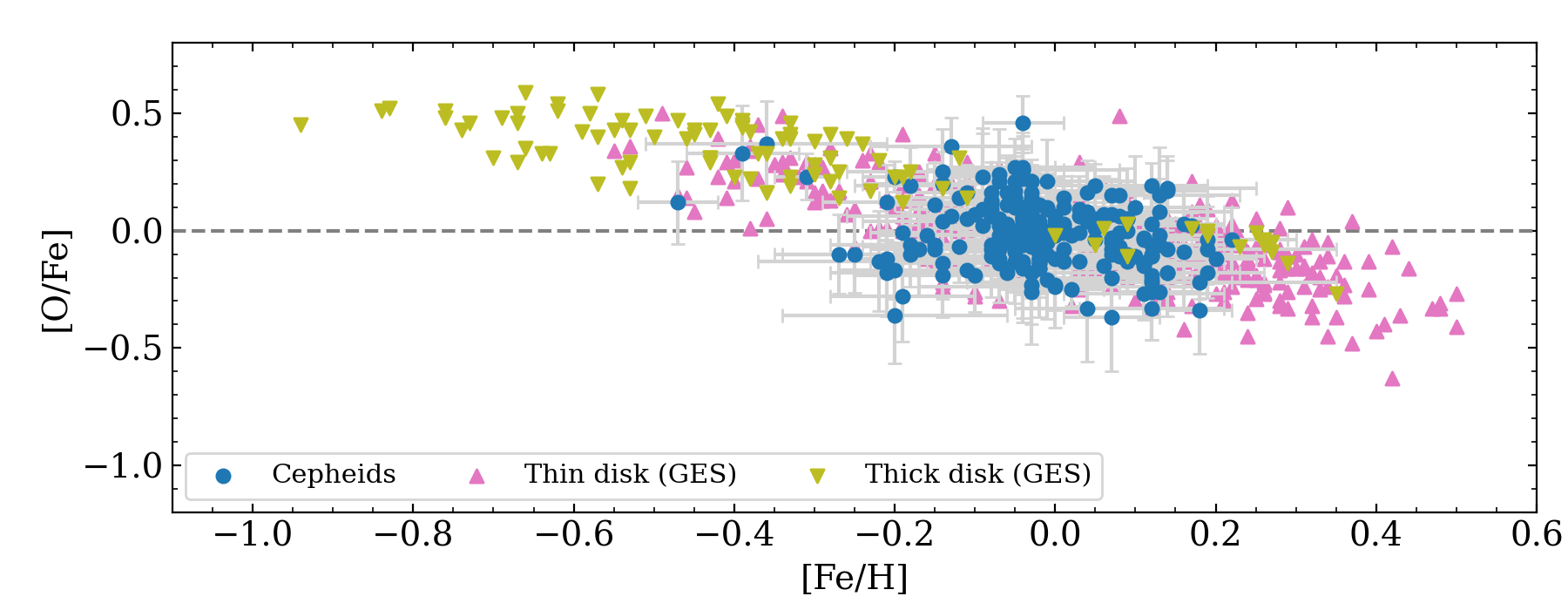}}
\end{minipage} \\
\begin{minipage}[t]{0.52\textwidth}
\centering
\resizebox{\hsize}{!}{\includegraphics{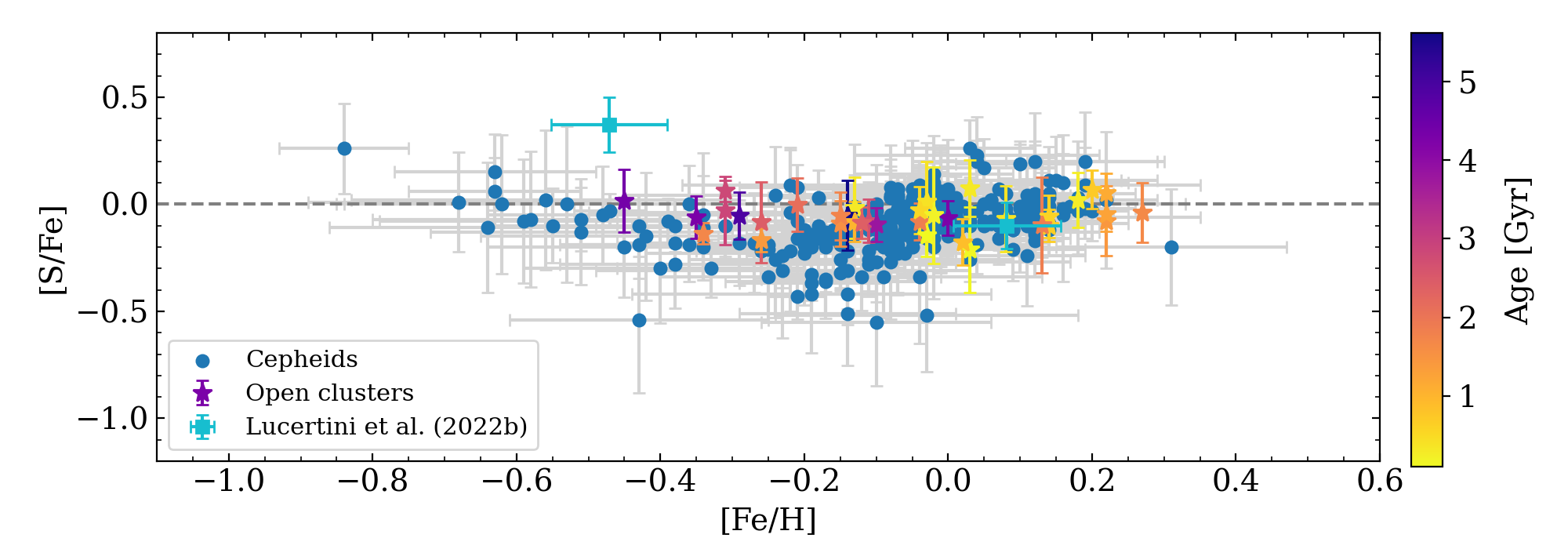}}
\end{minipage}
\begin{minipage}[t]{0.47\textwidth}
\centering
\resizebox{\hsize}{!}{\includegraphics{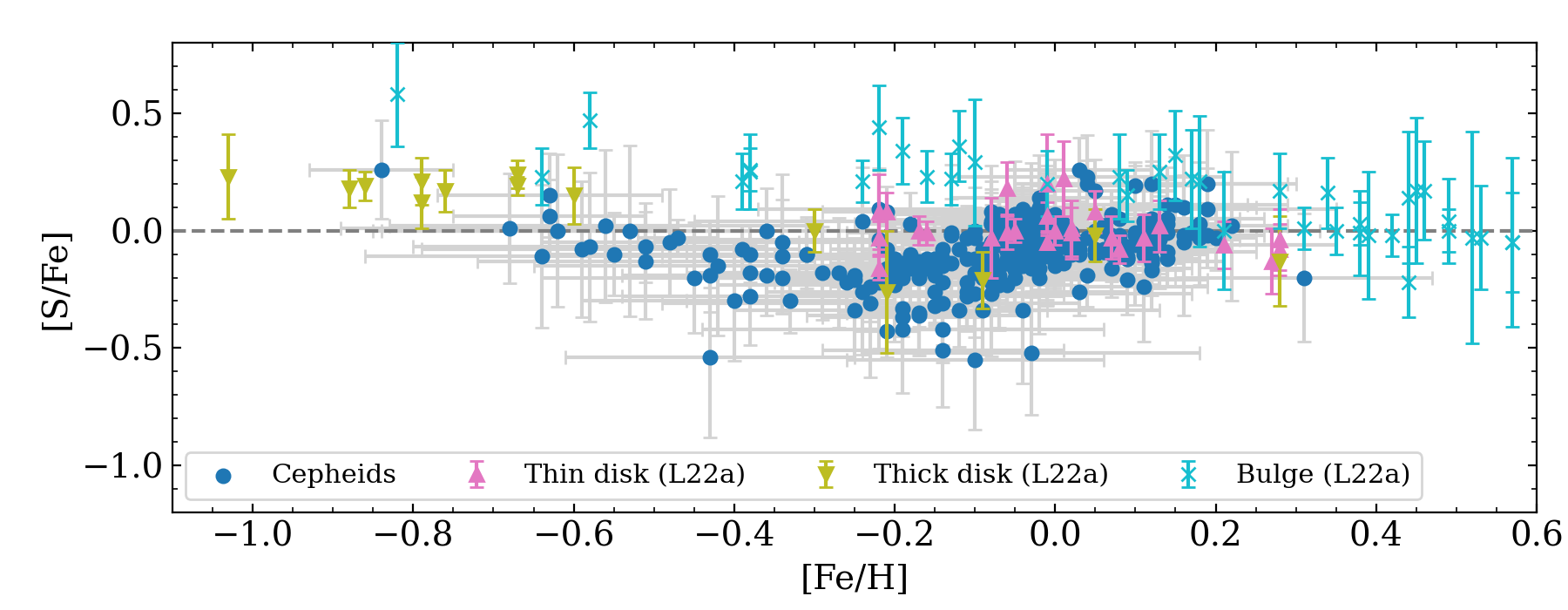}}
\end{minipage}
\caption{Abundance ratios as a function of the iron abundance.
{\it Left panels:} The data are the same as in Fig.~\ref{figure:xh_rgal_open_cluster}, but comparing the [O/Fe] (top) and [S/Fe] (bottom) ratios for our samples of classical Cepheids and open clusters. The bottom panel also shows the abundance ratios for the open clusters (\object{Trumpler\,5 and Trumpler\,20}) provided by \citet{Lucertinietal2022b}.
{\it Right panels:} The [O/Fe] and [S/Fe] ratios for Galactic Cepheids are compared with thin and thick disk stars provided by the GES (DR5.0) collaboration and with thin disk, thick disk, and bulge field stars provided by \citet[][L22a]{Lucertinietal2022a}.}
\label{figure:ofe_sfe_feh_clusters_field}
\end{figure*}

\begin{figure*}
\centering
\begin{minipage}[t]{0.497\textwidth}
\centering
\resizebox{\hsize}{!}{\includegraphics{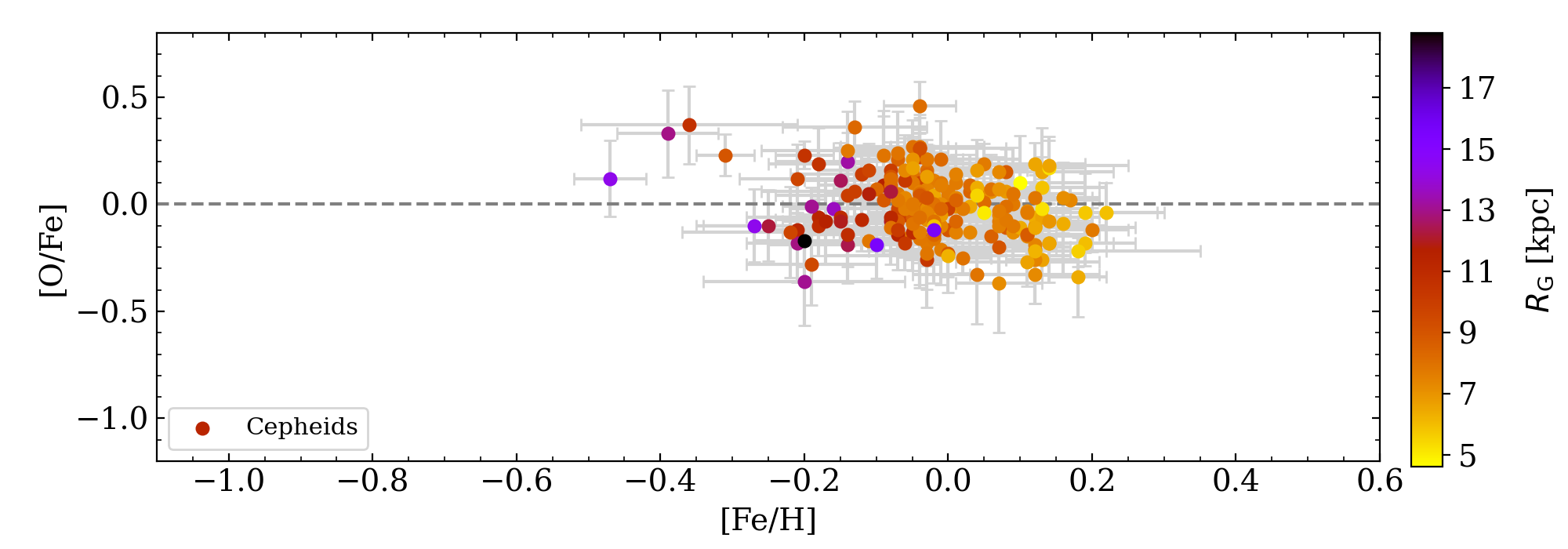}}
\end{minipage}
\begin{minipage}[t]{0.497\textwidth}
\centering
\resizebox{\hsize}{!}{\includegraphics{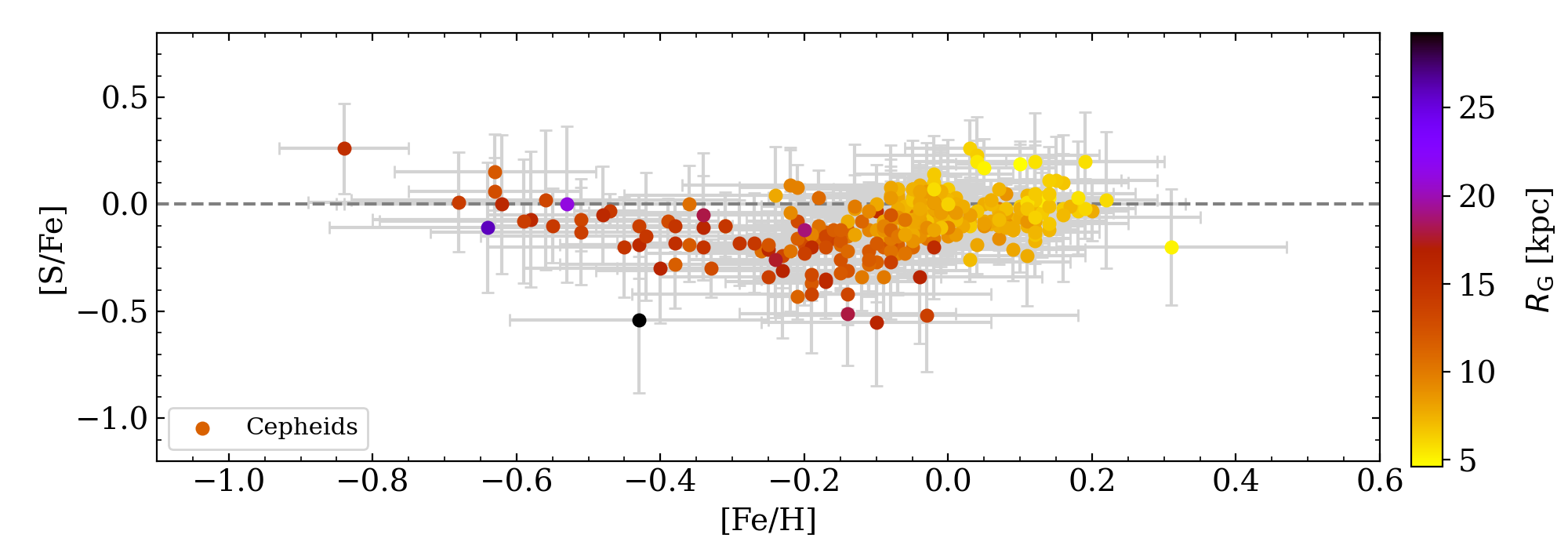}}
\end{minipage}
\caption{Abundance ratios as a function of the iron abundance. The [O/Fe] and [S/Fe] ratios for the current CC sample are color-coded according to their Galactocentric distance.}
\label{figure:ofe_sfe_feh_RG}
\end{figure*}

This empirical evidence indicates that both O and S are overabundant in very young (long period) Cepheids. However, this plain physical argument is hampered by the fact that the $\alpha$-elements are typically produced on a time scale of the order of tens of Myr whereas iron-peak elements, if we assume that they are mainly produced by Type Ia SNe, are produced on a timescale of the order of a few Gyr. 

\section{Comparison with literature data}
\label{sec:comparison_observations}

The top left panel of Fig.~\ref{figure:ofe_sfe_feh_clusters_field} shows the [O/Fe] abundance ratio as a function the iron abundance for both the CC and OC samples. Data plotted in this figure display that, for Fe abundances more metal-poor than solar, the O abundance is, on average, overabundant. In particular, CCs and OCs for [Fe/H]$\lesssim$$-$0.3 show a well defined O enhancement. The data plotted in the bottom panel display that S abundance is, on average, under-abundant. The abundance ratio attains solar values only in the metal rich regime ([Fe/H] $\ge$ 0). \object{Trumpler\,20}, one of the open clusters observed by \citet{Lucertinietal2022b}, agrees quite well with the current sample of OCs and CCs, whereas \object{Trumpler\,5} shows a well defined enhancement in sulfur. In passing, we also notice that the abundance ratios are independent of the age, since CCs and OCs display similar variations. 

The right panels of the Fig.~\ref{figure:ofe_sfe_feh_clusters_field} show the same comparison, but with thin and thick disk field stars observed by GES and with bulge field stars observed by \citet{Lucertinietal2022b}. The [O/Fe] abundance ratio is, as expected, enhanced in the metal-poor regime and becomes under-abundant in the metal-rich regime. CCs display a remarkable agreement when compared with thin and thick disk stars. This suggests that O abundance of metal-poor CCs is expected to be enhanced in the metal-poor regime and depleted in the metal-rich regime \citep{Romanielloetal2022}. 

The variation of the [S/Fe] as a function of the iron abundance for CCs is far from being linear. The ratio attains solar values in the metal-poor ([Fe/H]$\sim$$-$0.6) and in the metal-rich ([Fe/H]$\ge$0) regimes, but shows a parabolic trend for intermediate iron abundances. The parabolic trend is less clear in OCs, since the current sample only includes a few OCs located in the outer disk and with accurate S abundances.

Thin disk stars agree quite well with CCs, but thick disk stars are systematically enhanced when compared with classical Cepheids. The bulge stars attain solar ratios only in the very metal-rich ([Fe/H]$\ge$0.2) regime, whereas in the more metal-poor regime it is systematically enhanced, suggesting a different chemical enrichment compared with young thin disk stars. However, it is worth noting that while thick, thin, and bulge stellar abundances indicate evolutionary trends of individual MW components \citep[see time-delay model, e.g.,][]{Matteucci2021}, CCs display present-day frames of different regions with diverse histories of star formation. Therefore, cautions must be taken when comparing these different datasets.

To further constrain the variation in the abundance ratios, Fig.~\ref{figure:ofe_sfe_feh_RG} shows the same data for CCs plotted in Fig.~\ref{figure:ofe_sfe_feh_clusters_field}, but color-coded according to the Galactocentric distance. Data plotted in the left panel show that the [O/Fe] is, as expected, solar in the solar neighborhood and in the inner disk, whereas it becomes enhanced in the more metal-poor regions of the outer disk. The transition from inner to outer disk is even more clearly traced by the [S/Fe] abundance ratios plotted in the right panel of the same figure. Indeed, CCs located inside the solar circle display solar values whereas those located at larger Galactocentric distances appear to be under-abundant, and approach solar values once again in the outermost regions. The number of stars approaching solar abundance at low metallicities is quite small to draw firm conclusions, but the data plotted on the bottom panel of Fig.~\ref{figure:xfe_rgal_open_cluster} and on the right panel of Fig.~\ref{figure:ofe_sfe_feh_RG} indicate that sulfur abundances do not follow iron abundances for Galactocentric distances larger than $\sim$12-13~kpc. Such a behavior could be supported by a S production that is variable with distance. This working hypothesis is discussed on a more quantitative basis in the next section.

\section{Comparison between theory and observations}
\label{sec:comparison_theory}

To investigate on a more quantitative basis the impact of the observed CCs abundance gradients on the Milky Way disk evolution, we compare them with the results of detailed Galactic chemical evolution models. The starting point is the best fit model recently provided by \citet{Pallaetal2020}. The model is a revised version of the two-infall paradigm (see also \citealt{Spitonietal2019}) in which two consecutive gas accretion episodes, separated by an age gap of at least 3~Gyr, form the so-called high-$\alpha$ and low-$\alpha$ sequences observed in the Galactic disk. For the second gas infall episode forming the low-$\alpha$ sequence, the timescale for gas accretion increases with radius according to the inside-out scenario. In addition, inward radial gas flows with constant velocity (of 1~\kms) and a star formation efficiency (SFE) for the Schmidt-Kenicutt law\footnote{We adopt $\Sigma_{\rm SFR}=\nu \Sigma_{\rm gas}^k$ (\citealt{Kennicutt1998}), with $k=1.5$} variable with radius (with values between $\nu=5$ and 0.1~Gyr$^{-1}$ ), act together with the inside-out mechanism. The reader interested in a more detailed discussion concerning the physical assumptions adopted in the model is referred to \citet{Pallaetal2020}.

This model takes account for radial abundance gradients based either on CCs or on OCs observational programs (\citealt{LuckLambert2011,Genovalietal2015,Magrinietal2018}) as well as gas, stellar, and SFR density gradients (see, e.g., \citealt{NakanishiSofue2003,NakanishiSofue2006,StahlerPalla2005,Green2014}) in the Galactic thin disk. Moreover, the model setup allows to reproduce the so-called $\alpha$-dichotomy/bimodality in [$\alpha$/Fe] vs. [Fe/H] trends by APOGEE survey at different Galactocentric distances (see, e.g., \citealt{Queirozetal2020}).

\begin{figure*}
\centering
\includegraphics[width=0.5\textwidth]{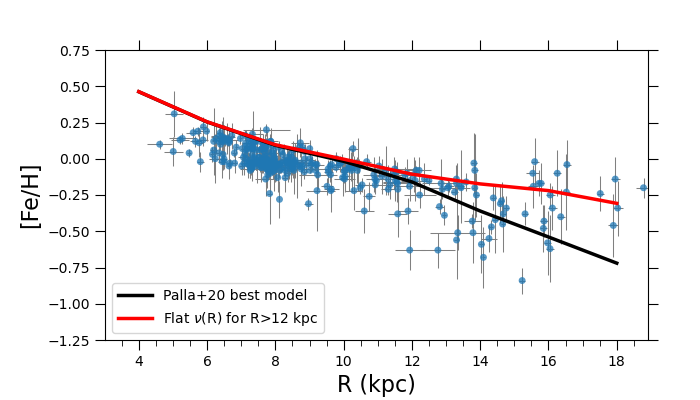}\\
\includegraphics[width=0.95\textwidth]{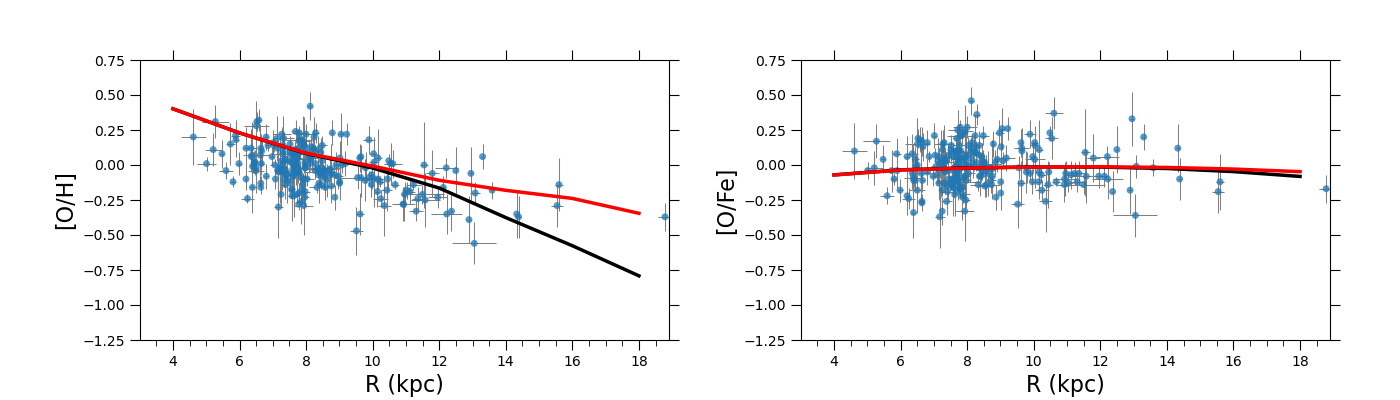}\\
\includegraphics[width=0.95\textwidth]{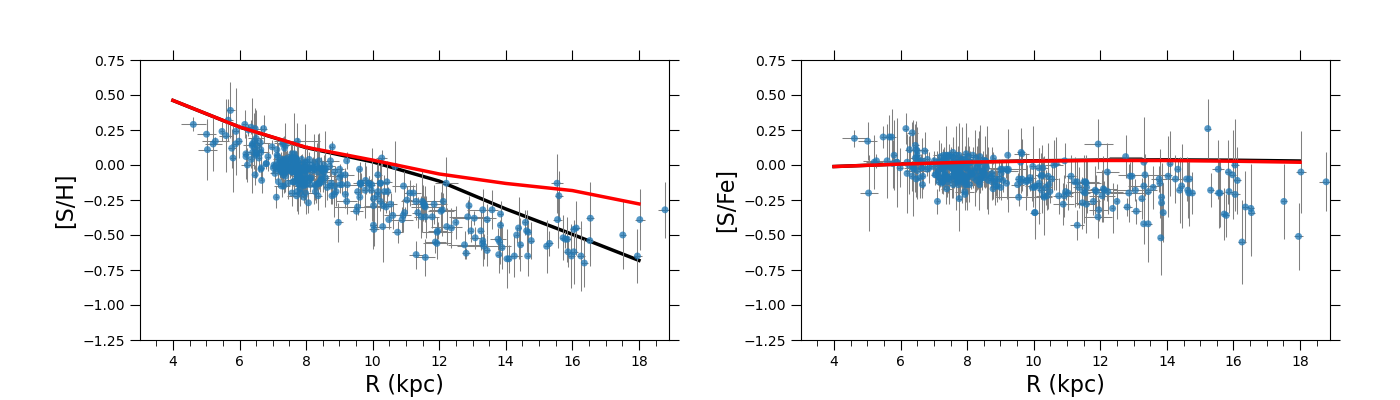}
\caption{Predicted [X/H] and [X/Fe] radial gradients compared with the Cepheid data presented in this work (blue points with error bars). Black lines are from the best model presented by \citet{Pallaetal2020}, whereas red lines are from a model with analogue setup but with flattened profile for the efficiency of star formation $\nu$ in the outer Galactic regions ($R_{\rm G}>12$~kpc).}
\label{fig:model_gradients}
\end{figure*}

Figure~\ref{fig:model_gradients} shows the comparison between the observed abundance gradients for Fe, O, and S with the best fit model of \citet{Pallaetal2020}. Starting from Fe, the model by \citet{Pallaetal2020} underestimates the present-day [Fe/H] abundance in the outermost radii ($R_{\rm G}$$\gtrsim$12~kpc). The same seems to happen for the [O/H] gradient, even though the number of stars in outer regions ($\sim$10 with $R_{\rm G}$$>$13~kpc) is limited to reach firm conclusions. To reproduce the radial trends, we run another chemical evolution model with modified prescriptions relative to the best fit model by \citet{Pallaetal2020}. In particular, we relax the condition of a variable SFE for $R_{\rm G}$$>$12~kpc by fixing its value to the one at $R_{\rm G}$ = 12~kpc ($\nu=0.5$~Gyr$^{-1}$). This new model agrees quite well with the observed abundance plateau at large Galactocentric distances and at the same time it does not significantly affect the predicted [$\alpha$/Fe] radial trends, as can be seen in the right panels for both O and S.

Concerning sulfur, the bending of the slope (see also Fig.~\ref{figure:xh_rgal_open_cluster}) at large radii should also agree with our modified model showing a plateau at $R_{\rm G}$$>$12~kpc. However, the lower panels of Fig.~\ref{fig:model_gradients} highlight an overestimation of the observed [S/H] and [S/Fe] by the “modified model” in the outer regions, whereas present-day abundances within the solar ring ($R_{\rm G}$ $\lesssim$ 8~kpc) are generally reproduced. This indicates that the larger steepness of the sulfur gradient (see the slope coefficients in Figs.~\ref{figure:xh_rgal_open_cluster} and \ref{figure:xfe_rgal_open_cluster} or in Table~\ref{table:slopes}) is not captured by the current models.

\begin{figure*}
\centering
\includegraphics[width=0.95\textwidth]{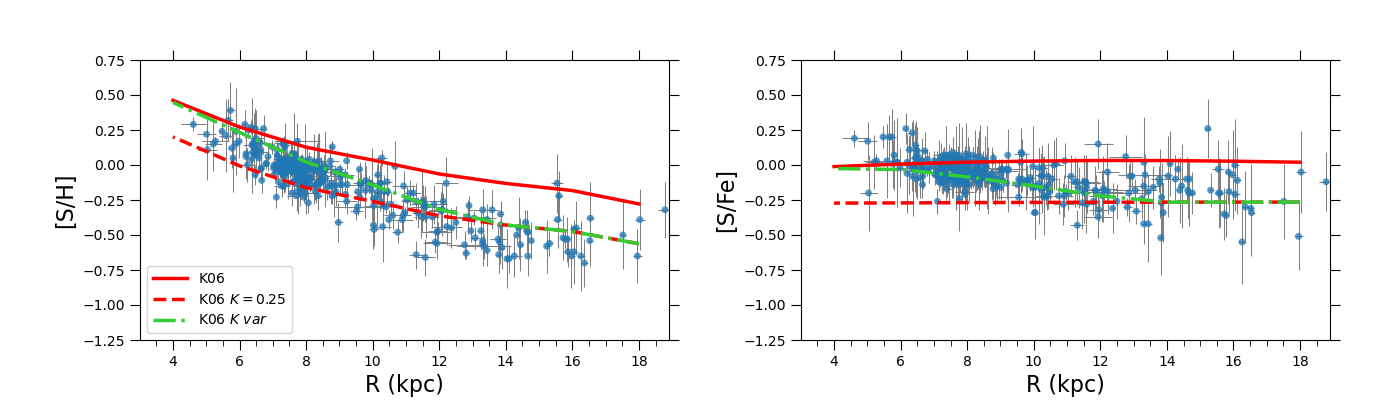}
\caption{Effect of the S yield for massive stars on the [S/H] and [S/Fe] radial gradients for the model with flattened star formation efficiency profile for outer regions. The red solid line shows the model with standard yield prescriptions (from \citealt{Kobayashietal2006,Kobayashietal2011}); the red dashed line shows the model with the S yield reduced by a factor 4; the green dash-dotted line displays a model with scaling factor for the S yields variable with radius. Symbols are the same as in Fig.~\ref{fig:model_gradients}. }
\label{fig:model_Svariation}
\end{figure*}

Such a discrepancy may be searched in the nucleosynthesis prescriptions for sulfur. Indeed, this element is also synthesized by Type II SN explosive nucleosynthesis (see also \ref{ss:alphaFe_gradients}), and it has a more uncertain outcome than those of purely hydrostatic elements, such as oxygen (\citealt{WoosleyWeaver1995,Nomotoetal2013,LimongiChieffi2018}). Moreover, the comparison with stellar abundances in the [S/Fe] vs. [Fe/H] diagram is often hampered by the difficulty in getting reliable and homogeneous stellar abundances over a broad range in iron abundances (see \citealt{Romanoetal2010,Kobayashietal2020}, and references therein). Therefore, we decided to change the stellar yield prescriptions adopted in \citet{Pallaetal2020}. In particular, we allow the CC-SN yields by \citet{Kobayashietal2006,Kobayashietal2011} to be multiplied by a scaling factor $K$. It is worth noting that, despite of being also not negligibly produced by Type Ia SNe, we keep fixed the S yields prescriptions by \citet{Iwamotoetal1999}. In fact, the variations in [S/Fe] that can be obtained by changing the plethora of Type Ia SN models available in the literature is of the order of $\sim$0.1~dex (see \citealt{Palla2021}), much smaller than the difference in [S/Fe] at different Galactocentric distances ($\sim$0.3~dex).

Figure~\ref{fig:model_Svariation} shows the results of this test. Together with the "modified model" presented in Fig. \ref{fig:model_gradients}, we also plot an analogue model with CC-SN sulfur yield multiplied by a factor $K=0.25$. We can clearly see that such a model recover the agreement with the data in the outer regions for both the [S/H] and [S/Fe] but, as expected, underestimate the observed values for radii within the solar ring. For this reason, in Fig.~\ref{fig:model_Svariation} we show also a third model for which the $K$ factor changes with radius, from $K=1$ at 4~kpc to $K=0.25$ from $12$~kpc onwards. This "toy model", as we called it, indeed reproduces very well both the observed trends for both [S/H] and [S/Fe] and highlights the necessity of a source of production for sulfur that varies in its contribution with the Galactocentric radius.

Whether it could be the contribution by stellar populations with different typical rotational velocity (e.g., \citealt{LimongiChieffi2018}), different binary fraction (\citealt{Farmeretal2023}), or combined effects (including the contribution by different Type Ia SNe progenitors) is beyond the scope of this work. A much larger number of chemical elements are needed to perform such a study.

The new spectroscopic sample for Galactic CCs presented in this work points toward a revision in the prescriptions adopted by chemical evolution models aiming at reproducing Galactic radial gradients (e.g., \citealt{Grisonietal2018,Pallaetal2020}). This is mainly driven by the much larger statistical significance of the data for the outermost regions of the Galactic thin disk. Several previous OCs catalogues also suggested the possibility of a change in the slope between inner and outer disk (e.g., \citealt{Yongetal2012,Magrinietal2023}), however, uncertainties affecting individual distances, the limited number of objects located in the outermost regions, and the possibility that the change in the slope was driven by an age effect hampered firm conclusions. In passing, we also notice that the different scenario in the modeling of radial gradients have, within the two-infall paradigm, a marginal impact on the fit of the observed $\alpha$-bimodality (e.g., \citealt{Spitonietal2019,Spitonietal2021,Pallaetal2022}). The stellar data adopted for such analyses have Galactocentric distances smaller than 12-14~kpc, as those from outer regions are significant severely limited in number (see, e.g., \citealt{Spitonietal2021}). Next generation spectroscopic surveys focusing on the outer Galaxy (e.g., WEAVE, \citealt{Daltonetal2020}, 4MOST, \citealt{deJongetal2019}) may help in significantly increasing the number of field stars with abundance ratios in outer regions and thus in having simultaneous constraints to probe Galactic evolutionary models.

Finally, the detection of significant difference between sulfur gradient and those from other elements underline the importance of getting data for the largest number of chemical elements, as it highlights features of nucleosynthetic processes that are hidden if looking at oxygen and iron gradients only. Therefore, we plan to extend the analysis for other chemical elements in order to unveil the causes of the different gradient behaviors.

\section{Summary and final remarks}
\label{sec:summary}

We present the largest (1285 spectra) and most homogeneous spectroscopic sample for Galactic classical Cepheids (379 objects). For a significant fraction of them (1118 spectra, 356 objects) we measured the abundances of iron and of two $\alpha$-elements (O, S). The new abundances are based on optical and high S/N spectra collected with a variety of high resolution spectrographs. The current sample, when compared with similar estimates available in the literature, brings forward several key advantages:
{\em new abundances} -- accurate abundances are provided, for the first time, for 77 CCs in our sample;
{\em Galactocentric distances} -- the current CC sample ranges from the inner ($R_{\rm G}$$\sim$5~kpc) to the outer ($R_{\rm G}$$\sim$25-30~kpc) disk and they are distributed across the four quadrants;
{\em period distribution} -- the spectroscopic sample includes both short (old) and long (young) period CCs;
{\em mode distribution} -- the sample includes CCs pulsating in the fundamental mode, in first overtone, in mixed mode, and in double mode;
{\em Sample selection} -- a number of acid tests have been performed to remove possible contamination either from old (Type II Cepheids) or intermediate (anomalous Cepheids) age tracers or from geometrical (binaries) variables.
We also investigated the orbital properties of the current spectroscopic sample and we found that the bulk of them have orbits typical of thin disk stars.

Since we are dealing with objects that experience, due to radial oscillations, steady variations in their physical properties, special care was paid in the estimate of the atmospheric parameters (effective temperature, surface gravity, microturbulent velocity). These were often double checked using different approaches and/or diagnostics. The internal consistency and the accuracy of the atmospheric parameters and of the adopted line list were investigated in a previous study \citep{daSilvaetal2022} by using two dozen calibrating Cepheids for which high-resolution spectra cover the entire pulsation cycle.

Classical Cepheids are excellent standard candles. The individual distances that we estimated are based either on Gaia DR3 trigonometric parallaxes or on NIR/MIR PL relations \citep{Ripepietal2022}. The latter were estimated by using individual color excesses from \citet{SchlaflyFinkbeiner2011} and the reddening law by \citet{Cardellietal1989} and by \citet[][MIR]{Madoreetal2013}.

In order to constrain the sensitivity of the radial gradients to the age, we took advantage of a sample of 62 open clusters collected by the Gaia-ESO survey \citep{Randichetal2022} and complemented with 18 OCs available in the ESO archive. The key advantage of this sample is that homogeneous estimates of their abundances (Fe, O) were already available together with homogeneous cluster parameters (cluster age, distance, reddening). Furthermore, we measured S abundances by using the same diagnostics. All in all, the current OC sample covers Galactocentric distances similar to the CC sample, but the age distribution ranges from a few Myr to about 7~Gyr. 

The main results concerning the radial abundance gradients are 
the following:

{\em Slope of the radial gradients} -- Iron and sulfur radial gradients based on both CCs and OCs display a well defined departure from linearity when moving from the inner to the outer disk. The oxygen radial gradient shows a similar trend, though the flattening at large Galactocentric distance is partially hampered by the lack of measurements for $R_{\rm G}$$\ge$19~kpc. We found that logarithmic regressions take account for the variation of [X/H] abundances when moving from the inner to the outer disk. This is a solid finding because we individually checked CCs located at large Galactocentric distances, by using pulsation observables ($I$-band amplitude, Fourier parameters) and the height above the Galactic plane. 

{\em Difference in the slope between iron and $\alpha$-elements} -- The slope of the S radial gradient is steeper than the Fe slope. The difference is approximately a factor of two concerning the linear fit ($-$0.081 vs. $-$0.041~dex\,kpc$^{-1}$) and changes from $-$1.62 to $-$0.91 in the logarithmic distance. The OCs fully support the difference brought forward by CCs. It is worth mentioning that the current finding is minimally affected by the spread, at fixed Galactocentric distance, in iron and in sulfur abundances. The uncertainty on individual measurements is significantly smaller. The linear fit to the oxygen radial gradient is similar, as expected, to the iron radial gradient, whereas the logarithmic fit is slightly shallower.

{\em Difference between O and S} -- Empirical evidence indicates that S (explosive nucleosynthesis) is, on average, under-abundant when compared with O (hydrostatic nucleosynthesis). The difference becomes clearer in the metal-poor regime, and even clearer if dealing with the [O/Fe] and [S/Fe] abundance ratios. Indeed, the former one shows either a flat distribution or a tiny positive slope. On the other hand, the [S/Fe] abundance ratio displays a well defined negative gradient, suggesting that sulfur is under-abundant compared with iron over a substantial fraction of the Galactic thin disk. In this context, it is worth mentioning that OCs display either a flat distribution or a mild difference when compared with CCs. Indeed, OCs for Galactocentric distances larger than 10~kpc attain solar [S/Fe] abundance ratios.

{\em Age effects} -- 
1) The current abundances and the element-to-iron abundance ratios display positive gradients when plotted as a function of the logarithmic period. More metal-poor CCs display a systematic drift toward shorter periods, as expected on the basis of theoretical and empirical evidence. Circumstantial evidence indicates that very young CCs are overabundant in O and in S. The difference is mainly a consequence of the fast chemical enrichment of $\alpha$-elements when compared with iron-peak elements;
2) Radial gradients and abundance ratios do not appear to be correlated with age. CCs together with young and old open clusters display similar variations when moving from the inner to the outer disk and from more metal-poor to more metal-rich regimes. In this context, it is worth mentioning that, dating back to \citet{Carraroetal2007} and to \citet{Frieletal2010}, a flattening of the radial gradients based on OCs has already been suggested in the literature. However, this variation was interpreted as an age effect in the sense that an increase in cluster age caused a flattening of the radial gradients. CCs, being systematically younger than 250~Myr, are actually tracing the present-day radial gradient. This means that, if we assume that the average Galactocentric velocity is of the order of a few km/sec \citep{Tianetal2017,LopezCorredoiraetal2019}, a CC that was born at a Galactocentric distance of 10~kpc moved, after 200~Myr, by at most 1~kpc. The consequence of this plain evidence is that the flattening we are dealing with is associated with the physical properties of the outer disk.

To investigate on a more quantitative basis the impact that the current radial abundance gradients have on the chemical evolution of the Galactic thin disk, we also performed a detailed comparison between theory and observations. We adopted the best-fit model recently provided by \citet{Pallaetal2020} and we found that a constant Star Formation Efficiency for Galactocentric distances larger than 12~kpc can take account for the flattening observed in iron and in 
$\alpha$-elements.

Moreover, we developed a "toy model" to investigate the impact that current predictions concerning S yields for massive stars have on the radial abundance gradients. We found that inside the solar circle the current [S/H] and [S/Fe] gradients are well reproduced by canonical yield prescriptions by \citet{Kobayashietal2006, Kobayashietal2011}. The same comparison indicates that the flattening in the outermost regions requires a decrease of a factor of four in the current S predictions. This indicates a substantial decrease in the S yields in the more metal poor regime.

The current findings open the path to a few key pieces of evidence worth being discussed in more detail. 

Biochemistry mainly relies on six crucial elements: carbon, hydrogen, oxygen, nitrogen, phosphorus, and sulfur (CHONPS). This means that molecules containing these key elements played a key role in prebiotic chemistry and on the origins of life on Earth. The overabundance of sulfur across the solar circle and in the inner disk seems a good viaticum for the identification of prebiotic molecules. 

Moreover, sulfur in contrast with other $\alpha$-elements is a volatile element. This means that stellar sulfur abundances can be directly compared with nebular sulfur abundances in external galaxies. There is mounting evidence that both barred and unbarred galaxies display a flattening in the metallicity gradients at large radial distances, as discussed in, for instance, \citet{SanchezMenguianoetal2016}. These authors also found that the outer flattening regions contain a relatively high metallicity, which can be barely explained in the framework of a canonical inside-out scenario with a constant SFR. Many mechanisms have been suggested to explain this trend and, in particular, the role played by a bar and/or spiral arms \citep{Minchevetal2011,Spitonietal2019}, minor mergers of satellites \citep{Quillenetal2009}, and "wind recycling" accretions \citep{OppenheimerDave2008}. The preliminary findings brought forward by this investigation indicates that the MW radial gradients traced by young stellar populations show a similar trend. This means that they are a fundamental laboratory to investigate here and now the physical mechanisms driving the flattening.

Finally, let us mention that the possibility to compare sulfur abundances in external galaxies appears even more promising since, having strong emission lines, the S abundances can be measured even at large redshifts in Seyfert galaxies \citep[see][and references therein]{Mizumotoetal2023,Dorsetal2023} and damped Lyman alpha systems \citep[see][and references therein]{Gioanninietal2017}.

\begin{acknowledgements}
We thank the reviewer for their positive words and constructive suggestions concerning an earlier version of the present paper, which improved its content and readability.
We acknowledge financial support from the ASI-INAF agreement n. 2022-14-HH.0.
V.K. is grateful to the Vector-Stiftung at Stuttgart, Germany, for support within the program "2022--Immediate help for Ukrainian refugee scientists" under grant P2022-0064.
M.B. is supported through the Lise Meitner grant from the Max Planck Society. We acknowledge support by the Collaborative Research Centre SFB 881 (projects A5, A10), Heidelberg University, of the Deutsche Forschungsgemeinschaft (DFG, German Research Foundation) and by the European Research Council (ERC) under the European Union’s Horizon 2020 research and innovation programme (grant agreement 949173).
M.P. acknowledges funding support from ERC starting grant 851622 Dust Origin.
M.M. thanks financial support from the Spanish Ministry of Science and Innovation (MICINN) through the Spanish State Research Agency, under Severo Ochoa Programe 2020-2023 (\mbox{CEX2019-000920-S}), and from the Agencia Estatal de Investigaci\'on del Ministerio de Ciencia e Innovaci\'on (MCINN/AEI) under the grant "RR Lyrae stars, a lighthouse to distant galaxies and early galaxy evolution" and the European Regional Development Fun (ERDF) with reference PID2021-127042OB-I00.
R.P.K. acknowledges support by the Munich Excellence Cluster Origins, funded by the Deutsche Forschungsgemeinschaft (DFG; German Research Foundation) under Germany’s Excellence Strategy EXC-2094 390783311.

\end{acknowledgements}
\bibliographystyle{aa}
\bibliography{daSilvaetal2022.bib}

\begin{thebibliography}{148}
\expandafter\ifx\csname natexlab\endcsname\relax\def\natexlab#1{#1}\fi

\bibitem[{{Andrievsky} {et~al.}(2013){Andrievsky}, {L{\'e}pine}, {Korotin},
  {Luck}, {Kovtyukh}, \& {Maciel}}]{Andrievskyetal2013}
{Andrievsky}, S.~M., {L{\'e}pine}, J.~R.~D., {Korotin}, S.~A., {et~al.} 2013,
  \mnras, 428, 3252

\bibitem[{{Arellano-C{\'o}rdova} {et~al.}(2020){Arellano-C{\'o}rdova},
  {Esteban}, {Garc{\'\i}a-Rojas}, \& {M{\'e}ndez-Delgado}}]{Arellanoetal2020}
{Arellano-C{\'o}rdova}, K.~Z., {Esteban}, C., {Garc{\'\i}a-Rojas}, J., \&
  {M{\'e}ndez-Delgado}, J.~E. 2020, \mnras, 496, 1051

\bibitem[{{Bailer-Jones} {et~al.}(2021){Bailer-Jones}, {Rybizki}, {Fouesneau},
  {Demleitner}, \& {Andrae}}]{BailerJonesetal2021}
{Bailer-Jones}, C.~A.~L., {Rybizki}, J., {Fouesneau}, M., {Demleitner}, M., \&
  {Andrae}, R. 2021, \aj, 161, 147

\bibitem[{{Bono} {et~al.}(2023){Bono}, {Braga}, {Pietrinferni}, {}, {}, \&
  {}}]{Bonoetal2023}
{Bono}, G., {Braga}, V., {Pietrinferni}, A., {et~al.} 2023, TAAR, submitted

\bibitem[{{Bono} {et~al.}(2005){Bono}, {Marconi}, {Cassisi}, {Caputo},
  {Gieren}, \& {Pietrzynski}}]{Bonoetal2005}
{Bono}, G., {Marconi}, M., {Cassisi}, S., {et~al.} 2005, \apj, 621, 966

\bibitem[{{Bovy}(2015)}]{Bovy2015}
{Bovy}, J. 2015, \apjs, 216, 29

\bibitem[{{Bresolin} {et~al.}(2009){Bresolin}, {Gieren}, {Kudritzki},
  {Pietrzy{\'n}ski}, {Urbaneja}, \& {Carraro}}]{Bresolinetal2009}
{Bresolin}, F., {Gieren}, W., {Kudritzki}, R.-P., {et~al.} 2009, \apj, 700, 309

\bibitem[{{Bresolin} {et~al.}(2012){Bresolin}, {Kennicutt}, \&
  {Ryan-Weber}}]{Bresolinetal2012}
{Bresolin}, F., {Kennicutt}, R.~C., \& {Ryan-Weber}, E. 2012, \apj, 750, 122

\bibitem[{{Bresolin} {et~al.}(2022){Bresolin}, {Kudritzki}, \&
  {Urbaneja}}]{Bresolinetal2022}
{Bresolin}, F., {Kudritzki}, R.-P., \& {Urbaneja}, M.~A. 2022, \apj, 940, 32

\bibitem[{{Bresolin} {et~al.}(2016){Bresolin}, {Kudritzki}, {Urbaneja},
  {Gieren}, {Ho}, \& {Pietrzy{\'n}ski}}]{Bresolinetal2016}
{Bresolin}, F., {Kudritzki}, R.-P., {Urbaneja}, M.~A., {et~al.} 2016, \apj,
  830, 64

\bibitem[{{Caffau} {et~al.}(2015){Caffau}, {Ludwig}, {Steffen}, {Livingston},
  {Bonifacio}, {Malherbe}, {Doerr}, \& {Schmidt}}]{Caffauetal2015}
{Caffau}, E., {Ludwig}, H.~G., {Steffen}, M., {et~al.} 2015, \aap, 579, A88

\bibitem[{{Cantat-Gaudin} {et~al.}(2020){Cantat-Gaudin}, {Anders},
  {Castro-Ginard}, {Jordi}, {Romero-G{\'o}mez}, {Soubiran}, {Casamiquela},
  {Tarricq}, {Moitinho}, {Vallenari}, {Bragaglia}, {Krone-Martins}, \&
  {Kounkel}}]{Cantatetal2020}
{Cantat-Gaudin}, T., {Anders}, F., {Castro-Ginard}, A., {et~al.} 2020, \aap,
  640, A1

\bibitem[{{Caputo} {et~al.}(2001){Caputo}, {Marconi}, {Musella}, \&
  {Pont}}]{Caputoetal2001}
{Caputo}, F., {Marconi}, M., {Musella}, I., \& {Pont}, F. 2001, \aap, 372, 544

\bibitem[{{Cardelli} {et~al.}(1989){Cardelli}, {Clayton}, \&
  {Mathis}}]{Cardellietal1989}
{Cardelli}, J.~A., {Clayton}, G.~C., \& {Mathis}, J.~S. 1989, \apj, 345, 245

\bibitem[{{Carraro} {et~al.}(2007){Carraro}, {Geisler}, {Villanova},
  {Frinchaboy}, \& {Majewski}}]{Carraroetal2007}
{Carraro}, G., {Geisler}, D., {Villanova}, S., {Frinchaboy}, P.~M., \&
  {Majewski}, S.~R. 2007, \aap, 476, 217

\bibitem[{{Castelli} \& {Kurucz}(2004)}]{CastelliKurucz2004}
{Castelli}, F. \& {Kurucz}, R.~L. 2004, ArXiv Astrophysics e-prints
  [\eprint{astro-ph/0405087}]

\bibitem[{{Cavichia} {et~al.}(2014){Cavichia}, {Moll{\'a}}, {Costa}, \&
  {Maciel}}]{Cavichiaetal2014}
{Cavichia}, O., {Moll{\'a}}, M., {Costa}, R.~D.~D., \& {Maciel}, W.~J. 2014,
  \mnras, 437, 3688

\bibitem[{{Chen} {et~al.}(2001){Chen}, {Stoughton}, {Smith}, {Uomoto}, {Pier},
  {Yanny}, {Ivezi{\'c}}, {York}, {Anderson}, {Annis}, {Brinkmann}, {Csabai},
  {Fukugita}, {Hindsley}, {Lupton}, {Munn}, \& {SDSS
  Collaboration}}]{Chenetal2001}
{Chen}, B., {Stoughton}, C., {Smith}, J.~A., {et~al.} 2001, \apj, 553, 184

\bibitem[{{Chen} {et~al.}(2019){Chen}, {Wang}, {Deng}, {de Grijs}, {Liu}, \&
  {Tian}}]{Chenetal2019}
{Chen}, X., {Wang}, S., {Deng}, L., {et~al.} 2019, Nature Astronomy, 3, 320

\bibitem[{{Crestani} {et~al.}(2021){Crestani}, {Braga}, {Fabrizio}, {Bono},
  {Sneden}, {Preston}, {Ferraro}, {Iannicola}, {Nonino}, {Fiorentino},
  {Th{\'e}venin}, {Lemasle}, {Prudil}, {Alves-Brito}, {Altavilla}, {Chaboyer},
  {Dall'Ora}, {D'Orazi}, {Gilligan}, {Grebel}, {Koch-Hansen}, {Lala},
  {Marengo}, {Marinoni}, {Marrese}, {Mart{\'\i}nez-V{\'a}zquez}, {Matsunaga},
  {Monelli}, {Mullen}, {Neeley}, {da Silva}, {Stetson}, {Salaris}, {Storm},
  {Valenti}, \& {Zoccali}}]{Crestanietal2021}
{Crestani}, J., {Braga}, V.~F., {Fabrizio}, M., {et~al.} 2021, \apj, 914, 10

\bibitem[{{da Silva} {et~al.}(2022){da Silva}, {Crestani}, {Bono}, {Braga},
  {D'Orazi}, {Lemasle}, {Bergemann}, {Dall'Ora}, {Fiorentino},
  {Fran{\c{c}}ois}, {Groenewegen}, {Inno}, {Kovtyukh}, {Kudritzki},
  {Matsunaga}, {Monelli}, {Pietrinferni}, {Porcelli}, {Storm}, {Tantalo}, \&
  {Th{\'e}v{\'e}nin}}]{daSilvaetal2022}
{da Silva}, R., {Crestani}, J., {Bono}, G., {et~al.} 2022, \aap, 661, A104

\bibitem[{{da Silva} {et~al.}(2016){da Silva}, {Lemasle}, {Bono}, {Genovali},
  {McWilliam}, {Cristallo}, {Bergemann}, {Buonanno}, {Fabrizio}, {Ferraro},
  {Fran{\c c}ois}, {Iannicola}, {Inno}, {Laney}, {Kudritzki}, {Matsunaga},
  {Nonino}, {Primas}, {Przybilla}, {Romaniello}, {Th{\'e}venin}, \&
  {Urbaneja}}]{daSilvaetal2016}
{da Silva}, R., {Lemasle}, B., {Bono}, G., {et~al.} 2016, \aap, 586, A125

\bibitem[{{Dalton} {et~al.}(2020){Dalton}, {Trager}, {Abrams}, {Bonifacio},
  {Aguerri}, {Vallenari}, {Bishop}, {Middleton}, {Benn}, {Dee}, {Mignot},
  {Lewis}, {Pragt}, {Pico}, {Walton}, {Rey}, {Allende Prieto}, {Lhom{\'e}},
  {Balcells}, {Terrett}, {Brock}, {Ridings}, {Skvar{\v{c}}}, {Verheijen},
  {Steele}, {Stuik}, {Kroes}, {Tromp}, {Kragt}, {Lesman}, {Mottram}, {Bates},
  {Gribbin}, {Burgal}, {Herreros}, {Delgado}, {Martin}, {Cano}, {Navarro},
  {Irwin}, {Peralta de Arriba}, {O'Mahoney}, {Bianco}, {Moleinezhad}, {ter
  Horst}, {Molinari}, {Lodi}, {Guerra}, {Baruffalo}, {Carrasco}, {Farcas},
  {Schallig}, {Hughes}, {Hill}, {Smith}, {Drew}, {Poggianti}, {Iovino},
  {Pieri}, {Jin}, {Dominguez Palmero}, {Fari{\~n}a}, {Mart{\'\i}n}, {Worley},
  {Murphy}, {Guest}, {Morris}, {Elswijk}, {de Haan}, {Hanenburg}, {Salasnich},
  {Mayya}, {Izazaga-P{\'e}rez}, {Gafton}, {Caffau}, {Horville}, {Paz
  Chinch{\'o}n}, {Falcon-Barosso}, {G{\"a}nsicke}, {San Juan}, \&
  {Hernandez}}]{Daltonetal2020}
{Dalton}, G., {Trager}, S., {Abrams}, D.~C., {et~al.} 2020, in Society of
  Photo-Optical Instrumentation Engineers (SPIE) Conference Series, Vol. 11447,
  Ground-based and Airborne Instrumentation for Astronomy VIII, ed. C.~J.
  {Evans}, J.~J. {Bryant}, \& K.~{Motohara}, 1144714

\bibitem[{{de Jong} {et~al.}(2019){de Jong}, {Agertz}, {Berbel}, {Aird},
  {Alexander}, {Amarsi}, {Anders}, {Andrae}, {Ansarinejad}, {Ansorge},
  {Antilogus}, {Anwand-Heerwart}, {Arentsen}, {Arnadottir}, {Asplund}, {Auger},
  {Azais}, {Baade}, {Baker}, {Baker}, {Balbinot}, {Baldry}, {Banerji},
  {Barden}, {Barklem}, {Barth{\'e}l{\'e}my-Mazot}, {Battistini}, {Bauer},
  {Bell}, {Bellido-Tirado}, {Bellstedt}, {Belokurov}, {Bensby}, {Bergemann},
  {Bestenlehner}, {Bielby}, {Bilicki}, {Blake}, {Bland-Hawthorn}, {Boeche},
  {Boland}, {Boller}, {Bongard}, {Bongiorno}, {Bonifacio}, {Boudon}, {Brooks},
  {Brown}, {Brown}, {Br{\"u}ggen}, {Brynnel}, {Brzeski}, {Buchert},
  {Buschkamp}, {Caffau}, {Caillier}, {Carrick}, {Casagrande}, {Case}, {Casey},
  {Cesarini}, {Cescutti}, {Chapuis}, {Chiappini}, {Childress}, {Christlieb},
  {Church}, {Cioni}, {Cluver}, {Colless}, {Collett}, {Comparat}, {Cooper},
  {Couch}, {Courbin}, {Croom}, {Croton}, {Daguis{\'e}}, {Dalton}, {Davies},
  {Davis}, {de Laverny}, {Deason}, {Dionies}, {Disseau}, {Doel}, {D{\"o}scher},
  {Driver}, {Dwelly}, {Eckert}, {Edge}, {Edvardsson}, {Youssoufi}, {Elhaddad},
  {Enke}, {Erfanianfar}, {Farrell}, {Fechner}, {Feiz}, {Feltzing}, {Ferreras},
  {Feuerstein}, {Feuillet}, {Finoguenov}, {Ford}, {Fotopoulou}, {Fouesneau},
  {Frenk}, {Frey}, {Gaessler}, {Geier}, {Gentile Fusillo}, {Gerhard},
  {Giannantonio}, {Giannone}, {Gibson}, {Gillingham},
  {Gonz{\'a}lez-Fern{\'a}ndez}, {Gonzalez-Solares}, {Gottloeber}, {Gould},
  {Grebel}, {Gueguen}, {Guiglion}, {Haehnelt}, {Hahn}, {Hansen}, {Hartman},
  {Hauptner}, {Hawkins}, {Haynes}, {Haynes}, {Heiter}, {Helmi}, {Aguayo},
  {Hewett}, {Hinton}, {Hobbs}, {Hoenig}, {Hofman}, {Hook}, {Hopgood},
  {Hopkins}, {Hourihane}, {Howes}, {Howlett}, {Huet}, {Irwin}, {Iwert},
  {Jablonka}, {Jahn}, {Jahnke}, {Jarno}, {Jin}, {Jofre}, {Johl}, {Jones},
  {J{\"o}nsson}, {Jordan}, {Karovicova}, {Khalatyan}, {Kelz}, {Kennicutt},
  {King}, {Kitaura}, {Klar}, {Klauser}, {Kneib}, {Koch}, {Koposov},
  {Kordopatis}, {Korn}, {Kosmalski}, {Kotak}, {Kovalev}, {Kreckel}, {Kripak},
  {Krumpe}, {Kuijken}, {Kunder}, {Kushniruk}, {Lam}, {Lamer}, {Laurent},
  {Lawrence}, {Lehmitz}, {Lemasle}, {Lewis}, {Li}, {Lidman}, {Lind}, {Liske},
  {Lizon}, {Loveday}, {Ludwig}, {McDermid}, {Maguire}, {Mainieri}, {Mali},
  {Mandel}, {Mandel}, {Mannering}, {Martell}, {Martinez Delgado}, {Matijevic},
  {McGregor}, {McMahon}, {McMillan}, {Mena}, {Merloni}, {Meyer}, {Michel},
  {Micheva}, {Migniau}, {Minchev}, {Monari}, {Muller}, {Murphy},
  {Muthukrishna}, {Nandra}, {Navarro}, {Ness}, {Nichani}, {Nichol}, {Nicklas},
  {Niederhofer}, {Norberg}, {Obreschkow}, {Oliver}, {Owers}, {Pai},
  {Pankratow}, {Parkinson}, {Paschke}, {Paterson}, {Pecontal}, {Parry},
  {Phillips}, {Pillepich}, {Pinard}, {Pirard}, {Piskunov}, {Plank},
  {Pl{\"u}schke}, {Pons}, {Popesso}, {Power}, {Pragt}, {Pramskiy}, {Pryer},
  {Quattri}, {Queiroz}, {Quirrenbach}, {Rahurkar}, {Raichoor}, {Ramstedt},
  {Rau}, {Recio-Blanco}, {Reiss}, {Renaud}, {Revaz}, {Rhode}, {Richard},
  {Richter}, {Rix}, {Robotham}, {Roelfsema}, {Romaniello}, {Rosario},
  {Rothmaier}, {Roukema}, {Ruchti}, {Rupprecht}, {Rybizki}, {Ryde}, {Saar},
  {Sadler}, {Sahl{\'e}n}, {Salvato}, {Sassolas}, {Saunders}, {Saviauk},
  {Sbordone}, {Schmidt}, {Schnurr}, {Scholz}, {Schwope}, {Seifert}, {Shanks},
  {Sheinis}, {Sivov}, {Sk{\'u}lad{\'o}ttir}, {Smartt}, {Smedley}, {Smith},
  {Smith}, {Sorce}, {Spitler}, {Starkenburg}, {Steinmetz}, {Stilz}, {Storm},
  {Sullivan}, {Sutherland}, {Swann}, {Tamone}, {Taylor}, {Teillon}, {Tempel},
  {ter Horst}, {Thi}, {Tolstoy}, {Trager}, {Traven}, {Tremblay}, {Tresse},
  {Valentini}, {van de Weygaert}, {van den Ancker}, {Veljanoski}, {Venkatesan},
  {Wagner}, {Wagner}, {Walcher}, {Waller}, {Walton}, {Wang}, {Winkler},
  {Wisotzki}, {Worley}, {Worseck}, {Xiang}, {Xu}, {Yong}, {Zhao}, {Zheng},
  {Zscheyge}, \& {Zucker}}]{deJongetal2019}
{de Jong}, R.~S., {Agertz}, O., {Berbel}, A.~A., {et~al.} 2019, The Messenger,
  175, 3

\bibitem[{{Dehnen} {et~al.}(2023){Dehnen}, {Semczuk}, \&
  {Sch{\"o}nrich}}]{Dehnenetal2023}
{Dehnen}, W., {Semczuk}, M., \& {Sch{\"o}nrich}, R. 2023, \mnras, 523, 1556

\bibitem[{{Dekker} {et~al.}(2000){Dekker}, {D'Odorico}, {Kaufer}, {Delabre}, \&
  {Kotzlowski}}]{Dekkeretal2000}
{Dekker}, H., {D'Odorico}, S., {Kaufer}, A., {Delabre}, B., \& {Kotzlowski}, H.
  2000, in \procspie, Vol. 4008, Optical and IR Telescope Instrumentation and
  Detectors, ed. M.~{Iye} \& A.~F. {Moorwood}, 534--545

\bibitem[{{Dessauges-Zavadsky} {et~al.}(2007){Dessauges-Zavadsky}, {Calura},
  {Prochaska}, {D'Odorico}, \& {Matteucci}}]{Dessaugesetal2007}
{Dessauges-Zavadsky}, M., {Calura}, F., {Prochaska}, J.~X., {D'Odorico}, S., \&
  {Matteucci}, F. 2007, \aap, 470, 431

\bibitem[{{Donor} {et~al.}(2020){Donor}, {Frinchaboy}, {Cunha}, {O'Connell},
  {Allende Prieto}, {Almeida}, {Anders}, {Beaton}, {Bizyaev}, {Brownstein},
  {Carrera}, {Chiappini}, {Cohen}, {Garc{\'\i}a-Hern{\'a}ndez}, {Geisler},
  {Hasselquist}, {J{\"o}nsson}, {Lane}, {Majewski}, {Minniti}, {Bidin}, {Pan},
  {Roman-Lopes}, {Sobeck}, \& {Zasowski}}]{Donoretal2020}
{Donor}, J., {Frinchaboy}, P.~M., {Cunha}, K., {et~al.} 2020, \aj, 159, 199

\bibitem[{{D'Orazi} {et~al.}(2020){D'Orazi}, {Oliva}, {Bragaglia}, {Frasca},
  {Sanna}, {Biazzo}, {Casali}, {Desidera}, {Lucatello}, {Magrini}, \&
  {Origlia}}]{Dorazietal2020}
{D'Orazi}, V., {Oliva}, E., {Bragaglia}, A., {et~al.} 2020, \aap, 633, A38

\bibitem[{{Dormand} \& {Prince}(1980)}]{DormandPrince1980}
{Dormand}, J. \& {Prince}, P. 1980, Journal of Computational and Applied
  Mathematics, 6, 19

\bibitem[{{Dors} {et~al.}(2023){Dors}, {Valerdi}, {Riffel}, {Riffel},
  {Cardaci}, {H{\"a}gele}, {Armah}, {Revalski}, {Flury}, {Freitas-Lemes},
  {Am{\^o}res}, {Krabbe}, {Binette}, {Feltre}, \&
  {Storchi-Bergmann}}]{Dorsetal2023}
{Dors}, O.~L., {Valerdi}, M., {Riffel}, R.~A., {et~al.} 2023, \mnras, 521, 1969

\bibitem[{{Duffau} {et~al.}(2017){Duffau}, {Caffau}, {Sbordone}, {Bonifacio},
  {Andrievsky}, {Korotin}, {Babusiaux}, {Salvadori}, {Monaco},
  {Fran{\c{c}}ois}, {Sk{\'u}lad{\'o}ttir}, {Bragaglia}, {Donati}, {Spina},
  {Gallagher}, {Ludwig}, {Christlieb}, {Hansen}, {Mott}, {Steffen}, {Zaggia},
  {Blanco-Cuaresma}, {Calura}, {Friel}, {Jim{\'e}nez-Esteban}, {Koch},
  {Magrini}, {Pancino}, {Tang}, {Tautvai{\v{s}}ien{\.{e}}}, {Vallenari},
  {Hawkins}, {Gilmore}, {Randich}, {Feltzing}, {Bensby}, {Flaccomio},
  {Smiljanic}, {Bayo}, {Carraro}, {Casey}, {Costado}, {Damiani}, {Franciosini},
  {Hourihane}, {Jofr{\'e}}, {Lardo}, {Lewis}, {Morbidelli}, {Sousa}, \&
  {Worley}}]{Duffauetal2017}
{Duffau}, S., {Caffau}, E., {Sbordone}, L., {et~al.} 2017, \aap, 604, A128

\bibitem[{{Farmer} {et~al.}(2023){Farmer}, {Laplace}, {Ma}, {de Mink}, \&
  {Justham}}]{Farmeretal2023}
{Farmer}, R., {Laplace}, E., {Ma}, J.-z., {de Mink}, S.~E., \& {Justham}, S.
  2023, \apj, 948, 111

\bibitem[{{Feast} {et~al.}(2014){Feast}, {Menzies}, {Matsunaga}, \&
  {Whitelock}}]{Feastetal2014}
{Feast}, M.~W., {Menzies}, J.~W., {Matsunaga}, N., \& {Whitelock}, P.~A. 2014,
  \nat, 509, 342

\bibitem[{{Friel} {et~al.}(2010){Friel}, {Jacobson}, \&
  {Pilachowski}}]{Frieletal2010}
{Friel}, E.~D., {Jacobson}, H.~R., \& {Pilachowski}, C.~A. 2010, \aj, 139, 1942

\bibitem[{{Gaia Collaboration} {et~al.}(2018){Gaia Collaboration}, {Katz},
  {Antoja}, {Romero-G{\'o}mez}, {Drimmel}, {Reyl{\'e}}, {Seabroke}, {Soubiran},
  {Babusiaux}, {Di Matteo}, {Figueras}, {Poggio}, {Robin}, {Evans}, {Brown},
  {Vallenari}, {Prusti}, {de Bruijne}, {Bailer-Jones}, {Biermann}, {Eyer},
  {Jansen}, {Jordi}, {Klioner}, {Lammers}, {Lindegren}, {Luri}, {Mignard},
  {Panem}, {Pourbaix}, {Randich}, {Sartoretti}, {Siddiqui}, {van Leeuwen},
  {Walton}, {Arenou}, {Bastian}, {Cropper}, {Lattanzi}, {Bakker}, {Cacciari},
  {Casta n}, {Chaoul}, {Cheek}, {De Angeli}, {Fabricius}, {Guerra}, {Holl},
  {Masana}, {Messineo}, {Mowlavi}, {Nienartowicz}, {Panuzzo}, {Portell},
  {Riello}, {Tanga}, {Th{\'e}venin}, {Gracia-Abril}, {Comoretto},
  {Garcia-Reinaldos}, {Teyssier}, {Altmann}, {Andrae}, {Audard},
  {Bellas-Velidis}, {Benson}, {Berthier}, {Blomme}, {Burgess}, {Busso},
  {Carry}, {Cellino}, {Clementini}, {Clotet}, {Creevey}, {Davidson}, {De
  Ridder}, {Delchambre}, {Dell'Oro}, {Ducourant},
  {Fern{\'a}ndez-Hern{\'a}ndez}, {Fouesneau}, {Fr{\'e}mat}, {Galluccio},
  {Garc{\'\i}a-Torres}, {Gonz{\'a}lez-N{\'u}{\~n}ez}, {Gonz{\'a}lez-Vidal},
  {Gosset}, {Guy}, {Halbwachs}, {Hambly}, {Harrison}, {Hern{\'a}ndez},
  {Hestroffer}, {Hodgkin}, {Hutton}, {Jasniewicz}, {Jean-Antoine-Piccolo},
  {Jordan}, {Korn}, {Krone-Martins}, {Lanzafame}, {Lebzelter}, {L{\"o}ffler},
  {Manteiga}, {Marrese}, {Mart{\'\i}n-Fleitas}, {Moitinho}, {Mora}, {Muinonen},
  {Osinde}, {Pancino}, {Pauwels}, {Petit}, {Recio-Blanco}, {Richards},
  {Rimoldini}, {Sarro}, {Siopis}, {Smith}, {Sozzetti}, {S{\"u}veges}, {Torra},
  {van Reeven}, {Abbas}, {Abreu Aramburu}, {Accart}, {Aerts}, {Altavilla},
  {{\'A}lvarez}, {Alvarez}, {Alves}, {Anderson}, {Andrei}, {Anglada Varela},
  {Antiche}, {Arcay}, {Astraatmadja}, {Bach}, {Baker},
  {Balaguer-N{\'u}{\~n}ez}, {Balm}, {Barache}, {Barata}, {Barbato}, {Barblan},
  {Barklem}, {Barrado}, {Barros}, {Barstow}, {Bartholom{\'e} Mu{\~n}oz},
  {Bassilana}, {Becciani}, {Bellazzini}, {Berihuete}, {Bertone}, {Bianchi},
  {Bienaym{\'e}}, {Blanco-Cuaresma}, {Boch}, {Boeche}, {Bombrun}, {Borrachero},
  {Bossini}, {Bouquillon}, {Bourda}, {Bragaglia}, {Bramante}, {Breddels},
  {Bressan}, {Brouillet}, {Br{\"u}semeister}, {Brugaletta}, {Bucciarelli},
  {Burlacu}, {Busonero}, {Butkevich}, {Buzzi}, {Caffau}, {Cancelliere},
  {Cannizzaro}, {Cantat-Gaudin}, {Carballo}, {Carlucci}, {Carrasco},
  {Casamiquela}, {Castellani}, {Castro-Ginard}, {Charlot}, {Chemin},
  {Chiavassa}, {Cocozza}, {Costigan}, {Cowell}, {Crifo}, {Crosta}, {Crowley},
  {Cuypers}, {Dafonte}, {Damerdji}, {Dapergolas}, {David}, {David}, {de
  Laverny}, {De Luise}, {De March}, {de Souza}, {de Torres}, {Debosscher}, {del
  Pozo}, {Delbo}, {Delgado}, {Delgado}, {Diakite}, {Diener}, {Distefano},
  {Dolding}, {Drazinos}, {Dur{\'a}n}, {Edvardsson}, {Enke}, {Eriksson},
  {Esquej}, {Eynard Bontemps}, {Fabre}, {Fabrizio}, {Faigler}, {Falc a},
  {Farr{\`a}s Casas}, {Federici}, {Fedorets}, {Fernique}, {Filippi},
  {Findeisen}, {Fonti}, {Fraile}, {Fraser}, {Fr{\'e}zouls}, {Gai}, {Galleti},
  {Garabato}, {Garc{\'\i}a-Sedano}, {Garofalo}, {Garralda}, {Gavel}, {Gavras},
  {Gerssen}, {Geyer}, {Giacobbe}, {Gilmore}, {Girona}, {Giuffrida}, {Glass},
  {Gomes}, {Granvik}, {Gueguen}, {Guerrier}, {Guiraud}, {Guti{\'e}}, {Haigron},
  {Hatzidimitriou}, {Hauser}, {Haywood}, {Heiter}, {Helmi}, {Heu}, {Hilger},
  {Hobbs}, {Hofmann}, {Holland}, {Huckle}, {Hypki}, {Icardi}, {Jan{\ss}en},
  {Jevardat de Fombelle}, {Jonker}, {Juh{\'a}sz}, {Julbe}, {Karampelas},
  {Kewley}, {Klar}, {Kochoska}, {Kohley}, {Kolenberg}, {Kontizas}, {Kontizas},
  {Koposov}, {Kordopatis}, {Kostrzewa-Rutkowska}, {Koubsky}, {Lambert},
  {Lanza}, {Lasne}, {Lavigne}, {Le Fustec}, {Le Poncin-Lafitte}, {Lebreton},
  {Leccia}, {Leclerc}, {Lecoeur-Taibi}, {Lenhardt}, {Leroux}, {Liao}, {Licata},
  {Lindstr{\o}m}, {Lister}, {Livanou}, {Lobel}, {L{\'o}pez}, {Managau}, {Mann},
  {Mantelet}, {Marchal}, {Marchant}, {Marconi}, {Marinoni}, {Marschalk{\'o}},
  {Marshall}, {Martino}, {Marton}, {Mary}, {Massari}, {Matijevi{\v{c}}},
  {Mazeh}, {McMillan}, {Messina}, {Michalik}, {Millar}, {Molina}, {Molinaro},
  {Moln{\'a}r}, {Montegriffo}, {Mor}, {Morbidelli}, {Morel}, {Morris},
  {Mulone}, {Muraveva}, {Musella}, {Nelemans}, {Nicastro}, {Noval},
  {O'Mullane}, {Ord{\'e}novic}, {Ord{\'o}{\~n}ez-Blanco}, {Osborne}, {Pagani},
  {Pagano}, {Pailler}, {Palacin}, {Palaversa}, {Panahi}, {Pawlak},
  {Piersimoni}, {Pineau}, {Plachy}, {Plum}, {Poujoulet}, {Pr{\v{s}}a},
  {Pulone}, {Racero}, {Ragaini}, {Rambaux}, {Ramos-Lerate}, {Regibo}, {Riclet},
  {Ripepi}, {Riva}, {Rivard}, {Rixon}, {Roegiers}, {Roelens}, {Rowell},
  {Royer}, {Ruiz-Dern}, {Sadowski}, {Sagrist{\`a} Sell{\'e}s}, {Sahlmann},
  {Salgado}, {Salguero}, {Sanna}, {Santana-Ros}, {Sarasso}, {Savietto},
  {Schultheis}, {Sciacca}, {Segol}, {Segovia}, {S{\'e}gransan}, {Shih},
  {Siltala}, {Silva}, {Smart}, {Smith}, {Solano}, {Solitro}, {Sordo}, {Soria
  Nieto}, {Souchay}, {Spagna}, {Spoto}, {Stampa}, {Steele},
  {Steidelm{\"u}ller}, {Stephenson}, {Stoev}, {Suess}, {Surdej}, {Szabados},
  {Szegedi-Elek}, {Tapiador}, {Taris}, {Tauran}, {Taylor}, {Teixeira},
  {Terrett}, {Teyssandier}, {Thuillot}, {Titarenko}, {Torra Clotet}, {Turon},
  {Ulla}, {Utrilla}, {Uzzi}, {Vaillant}, {Valentini}, {Valette}, {van Elteren},
  {Van Hemelryck}, {van Leeuwen}, {Vaschetto}, {Vecchiato}, {Veljanoski},
  {Viala}, {Vicente}, {Vogt}, {von Essen}, {Voss}, {Votruba}, {Voutsinas},
  {Walmsley}, {Weiler}, {Wertz}, {Wevers}, {Wyrzykowski}, {Yoldas},
  {{\v{Z}}erjal}, {Ziaeepour}, {Zorec}, {Zschocke}, {Zucker}, {Zurbach}, \&
  {Zwitter}}]{Gaia2018}
{Gaia Collaboration}, {Katz}, D., {Antoja}, T., {et~al.} 2018, \aap, 616, A11

\bibitem[{{Gaia Collaboration} {et~al.}(2023){Gaia Collaboration},
  {Recio-Blanco}, {Kordopatis}, {de Laverny}, {Palicio}, {Spagna}, {Spina},
  {Katz}, {Re Fiorentin}, {Poggio}, {McMillan}, {Vallenari}, {Lattanzi},
  {Seabroke}, {Casamiquela}, {Bragaglia}, {Antoja}, {Bailer-Jones},
  {Schultheis}, {Andrae}, {Fouesneau}, {Cropper}, {Cantat-Gaudin}, {Bijaoui},
  {Heiter}, {Brown}, {Prusti}, {de Bruijne}, {Arenou}, {Babusiaux}, {Biermann},
  {Creevey}, {Ducourant}, {Evans}, {Eyer}, {Guerra}, {Hutton}, {Jordi},
  {Klioner}, {Lammers}, {Lindegren}, {Luri}, {Mignard}, {Panem}, {Pourbaix},
  {Randich}, {Sartoretti}, {Soubiran}, {Tanga}, {Walton}, {Bastian}, {Drimmel},
  {Jansen}, {van Leeuwen}, {Bakker}, {Cacciari}, {Casta{\~n}eda}, {De Angeli},
  {Fabricius}, {Fr{\'e}mat}, {Galluccio}, {Guerrier}, {Masana}, {Messineo},
  {Mowlavi}, {Nicolas}, {Nienartowicz}, {Pailler}, {Panuzzo}, {Riclet}, {Roux},
  {Sordo}, {Th{\'e}venin}, {Gracia-Abril}, {Portell}, {Teyssier}, {Altmann},
  {Audard}, {Bellas-Velidis}, {Benson}, {Berthier}, {Blomme}, {Burgess},
  {Busonero}, {Busso}, {C{\'a}novas}, {Carry}, {Cellino}, {Cheek},
  {Clementini}, {Damerdji}, {Davidson}, {de Teodoro}, {Nu{\~n}ez Campos},
  {Delchambre}, {Dell'Oro}, {Esquej}, {Fern{\'a}ndez-Hern{\'a}ndez}, {Fraile},
  {Garabato}, {Garc{\'\i}a-Lario}, {Gosset}, {Haigron}, {Halbwachs}, {Hambly},
  {Harrison}, {Hern{\'a}ndez}, {Hestroffer}, {Hodgkin}, {Holl}, {Jan{\ss}en},
  {Jevardat de Fombelle}, {Jordan}, {Krone-Martins}, {Lanzafame},
  {L{\"o}ffler}, {Marchal}, {Marrese}, {Moitinho}, {Muinonen}, {Osborne},
  {Pancino}, {Pauwels}, {Reyl{\'e}}, {Riello}, {Rimoldini}, {Roegiers},
  {Rybizki}, {Sarro}, {Siopis}, {Smith}, {Sozzetti}, {Utrilla}, {van Leeuwen},
  {Abbas}, {{\'A}brah{\'a}m}, {Abreu Aramburu}, {Aerts}, {Aguado}, {Ajaj},
  {Aldea-Montero}, {Altavilla}, {{\'A}lvarez}, {Alves}, {Anders}, {Anderson},
  {Anglada Varela}, {Baines}, {Baker}, {Balaguer-N{\'u}{\~n}ez}, {Balbinot},
  {Balog}, {Barache}, {Barbato}, {Barros}, {Barstow}, {Bartolom{\'e}},
  {Bassilana}, {Bauchet}, {Becciani}, {Bellazzini}, {Berihuete}, {Bernet},
  {Bertone}, {Bianchi}, {Binnenfeld}, {Blanco-Cuaresma}, {Boch}, {Bombrun},
  {Bossini}, {Bouquillon}, {Bramante}, {Breedt}, {Bressan}, {Brouillet},
  {Brugaletta}, {Bucciarelli}, {Burlacu}, {Butkevich}, {Buzzi}, {Caffau},
  {Cancelliere}, {Carballo}, {Carlucci}, {Carnerero}, {Carrasco}, {Castellani},
  {Castro-Ginard}, {Chaoul}, {Charlot}, {Chemin}, {Chiaramida}, {Chiavassa},
  {Chornay}, {Comoretto}, {Contursi}, {Cooper}, {Cornez}, {Cowell}, {Crifo},
  {Crosta}, {Crowley}, {Dafonte}, {Dapergolas}, {David}, {De Luise}, {De
  March}, {De Ridder}, {de Souza}, {de Torres}, {del Peloso}, {del Pozo},
  {Delbo}, {Delgado}, {Delisle}, {Demouchy}, {Dharmawardena}, {Di Matteo},
  {Diakite}, {Diener}, {Distefano}, {Dolding}, {Edvardsson}, {Enke}, {Fabre},
  {Fabrizio}, {Faigler}, {Fedorets}, {Fernique}, {Figueras}, {Fournier},
  {Fouron}, {Fragkoudi}, {Gai}, {Garcia-Gutierrez}, {Garcia-Reinaldos},
  {Garc{\'\i}a-Torres}, {Garofalo}, {Gavel}, {Gavras}, {Gerlach}, {Geyer},
  {Giacobbe}, {Gilmore}, {Girona}, {Giuffrida}, {Gomel}, {Gomez},
  {Gonz{\'a}lez-N{\'u}{\~n}ez}, {Gonz{\'a}lez-Santamar{\'\i}a},
  {Gonz{\'a}lez-Vidal}, {Granvik}, {Guillout}, {Guiraud},
  {Guti{\'e}rrez-S{\'a}nchez}, {Guy}, {Hatzidimitriou}, {Hauser}, {Haywood},
  {Helmer}, {Helmi}, {Sarmiento}, {Hidalgo}, {H{\l}adczuk}, {Hobbs}, {Holland},
  {Huckle}, {Jardine}, {Jasniewicz}, {Jean-Antoine Piccolo},
  {Jim{\'e}nez-Arranz}, {Juaristi Campillo}, {Julbe}, {Karbevska}, {Kervella},
  {Khanna}, {Korn}, {K{\'o}sp{\'a}l}, {Kostrzewa-Rutkowska}, {Kruszy{\'n}ska},
  {Kun}, {Laizeau}, {Lambert}, {Lanza}, {Lasne}, {Le Campion}, {Lebreton},
  {Lebzelter}, {Leccia}, {Leclerc}, {Lecoeur-Taibi}, {Liao}, {Licata},
  {Lindstr{\o}m}, {Lister}, {Livanou}, {Lobel}, {Lorca}, {Loup}, {Madrero
  Pardo}, {Magdaleno Romeo}, {Managau}, {Mann}, {Manteiga}, {Marchant},
  {Marconi}, {Marcos}, {Marcos Santos}, {Mar{\'\i}n Pina}, {Marinoni},
  {Marocco}, {Marshall}, {Martin Polo}, {Mart{\'\i}n-Fleitas}, {Marton},
  {Mary}, {Masip}, {Massari}, {Mastrobuono-Battisti}, {Mazeh}, {Messina},
  {Michalik}, {Millar}, {Mints}, {Molina}, {Molinaro}, {Moln{\'a}r}, {Monari},
  {Mongui{\'o}}, {Montegriffo}, {Montero}, {Mor}, {Mora}, {Morbidelli},
  {Morel}, {Morris}, {Muraveva}, {Murphy}, {Musella}, {Nagy}, {Noval},
  {Oca{\~n}a}, {Ogden}, {Ordenovic}, {Osinde}, {Pagani}, {Pagano}, {Palaversa},
  {Pallas-Quintela}, {Panahi}, {Payne-Wardenaar}, {Pe{\~n}alosa Esteller},
  {Penttil{\"a}}, {Pichon}, {Piersimoni}, {Pineau}, {Plachy}, {Plum},
  {Pr{\v{s}}a}, {Pulone}, {Racero}, {Ragaini}, {Rainer}, {Raiteri}, {Ramos},
  {Ramos-Lerate}, {Regibo}, {Richards}, {Rios Diaz}, {Ripepi}, {Riva}, {Rix},
  {Rixon}, {Robichon}, {Robin}, {Robin}, {Roelens}, {Rogues}, {Rohrbasser},
  {Romero-G{\'o}mez}, {Rowell}, {Royer}, {Ruz Mieres}, {Rybicki}, {Sadowski},
  {S{\'a}ez N{\'u}{\~n}ez}, {Sagrist{\`a} Sell{\'e}s}, {Sahlmann}, {Salguero},
  {Samaras}, {Sanchez Gimenez}, {Sanna}, {Santove{\~n}a}, {Sarasso}, {Sciacca},
  {Segol}, {Segovia}, {S{\'e}gransan}, {Semeux}, {Shahaf}, {Siddiqui},
  {Siebert}, {Siltala}, {Silvelo}, {Slezak}, {Slezak}, {Smart}, {Snaith},
  {Solano}, {Solitro}, {Souami}, {Souchay}, {Spoto}, {Steele},
  {Steidelm{\"u}ller}, {Stephenson}, {S{\"u}veges}, {Surdej}, {Szabados},
  {Szegedi-Elek}, {Taris}, {Taylor}, {Teixeira}, {Tolomei}, {Tonello}, {Torra},
  {Torra}, {Torralba Elipe}, {Trabucchi}, {Tsounis}, {Turon}, {Ulla}, {Unger},
  {Vaillant}, {van Dillen}, {van Reeven}, {Vanel}, {Vecchiato}, {Viala},
  {Vicente}, {Voutsinas}, {Weiler}, {Wevers}, {Wyrzykowski}, {Yoldas}, {Yvard},
  {Zhao}, {Zorec}, {Zucker}, \& {Zwitter}}]{Gaia2023}
{Gaia Collaboration}, {Recio-Blanco}, A., {Kordopatis}, G., {et~al.} 2023,
  \aap, 674, A38

\bibitem[{{Gazak} {et~al.}(2015){Gazak}, {Kudritzki}, {Evans}, {Patrick},
  {Davies}, {Bergemann}, {Plez}, {Bresolin}, {Bender}, {Wegner}, {Bonanos}, \&
  {Williams}}]{Gazaketal2015}
{Gazak}, J.~Z., {Kudritzki}, R., {Evans}, C., {et~al.} 2015, \apj, 805, 182

\bibitem[{{Genovali} {et~al.}(2014){Genovali}, {Lemasle}, {Bono}, {Romaniello},
  {Fabrizio}, {Ferraro}, {Iannicola}, {Laney}, {Nonino}, {Bergemann},
  {Buonanno}, {Fran{\c{c}}ois}, {Inno}, {Kudritzki}, {Matsunaga}, {Pedicelli},
  {Primas}, \& {Th{\'e}venin}}]{Genovalietal2014}
{Genovali}, K., {Lemasle}, B., {Bono}, G., {et~al.} 2014, \aap, 566, A37

\bibitem[{{Genovali} {et~al.}(2015){Genovali}, {Lemasle}, {da Silva}, {Bono},
  {Fabrizio}, {Bergemann}, {Buonanno}, {Ferraro}, {Fran{\c c}ois}, {Iannicola},
  {Inno}, {Laney}, {Kudritzki}, {Matsunaga}, {Nonino}, {Primas}, {Romaniello},
  {Urbaneja}, \& {Th{\'e}venin}}]{Genovalietal2015}
{Genovali}, K., {Lemasle}, B., {da Silva}, R., {et~al.} 2015, \aap, 580, A17

\bibitem[{{Gioannini} {et~al.}(2017){Gioannini}, {Matteucci}, {Vladilo}, \&
  {Calura}}]{Gioanninietal2017}
{Gioannini}, L., {Matteucci}, F., {Vladilo}, G., \& {Calura}, F. 2017, \mnras,
  464, 985

\bibitem[{{GRAVITY Collaboration} {et~al.}(2018){GRAVITY Collaboration},
  {Abuter}, {Amorim}, {Anugu}, {Baub{\"o}ck}, {Benisty}, {Berger}, {Blind},
  {Bonnet}, {Brandner}, {Buron}, {Collin}, {Chapron}, {Cl{\'e}net}, {Coud{\'e}
  Du Foresto}, {de Zeeuw}, {Deen}, {Delplancke-Str{\"o}bele}, {Dembet},
  {Dexter}, {Duvert}, {Eckart}, {Eisenhauer}, {Finger}, {F{\"o}rster
  Schreiber}, {F{\'e}dou}, {Garcia}, {Garcia Lopez}, {Gao}, {Gendron},
  {Genzel}, {Gillessen}, {Gordo}, {Habibi}, {Haubois}, {Haug}, {Hau{\ss}mann},
  {Henning}, {Hippler}, {Horrobin}, {Hubert}, {Hubin}, {Jimenez Rosales},
  {Jochum}, {Jocou}, {Kaufer}, {Kellner}, {Kendrew}, {Kervella}, {Kok},
  {Kulas}, {Lacour}, {Lapeyr{\`e}re}, {Lazareff}, {Le Bouquin}, {L{\'e}na},
  {Lippa}, {Lenzen}, {M{\'e}rand}, {M{\"u}ler}, {Neumann}, {Ott}, {Palanca},
  {Paumard}, {Pasquini}, {Perraut}, {Perrin}, {Pfuhl}, {Plewa}, {Rabien},
  {Ram{\'\i}rez}, {Ramos}, {Rau}, {Rodr{\'\i}guez-Coira}, {Rohloff}, {Rousset},
  {Sanchez-Bermudez}, {Scheithauer}, {Sch{\"o}ller}, {Schuler}, {Spyromilio},
  {Straub}, {Straubmeier}, {Sturm}, {Tacconi}, {Tristram}, {Vincent}, {von
  Fellenberg}, {Wank}, {Waisberg}, {Widmann}, {Wieprecht}, {Wiest},
  {Wiezorrek}, {Woillez}, {Yazici}, {Ziegler}, \& {Zins}}]{Gravity2018}
{GRAVITY Collaboration}, {Abuter}, R., {Amorim}, A., {et~al.} 2018, \aap, 615,
  L15

\bibitem[{{Green}(2014)}]{Green2014}
{Green}, D.~A. 2014, in IAU Symposium, Vol. 296, Supernova Environmental
  Impacts, ed. A.~{Ray} \& R.~A. {McCray}, 188--196

\bibitem[{{Grisoni} {et~al.}(2018){Grisoni}, {Spitoni}, \&
  {Matteucci}}]{Grisonietal2018}
{Grisoni}, V., {Spitoni}, E., \& {Matteucci}, F. 2018, \mnras, 481, 2570

\bibitem[{{Gullikson} {et~al.}(2014){Gullikson}, {Dodson-Robinson}, \&
  {Kraus}}]{Gulliksonetal2014}
{Gullikson}, K., {Dodson-Robinson}, S., \& {Kraus}, A. 2014, \aj, 148, 53

\bibitem[{{Henry} {et~al.}(2004){Henry}, {Kwitter}, \&
  {Balick}}]{Henryetal2004}
{Henry}, R.~B.~C., {Kwitter}, K.~B., \& {Balick}, B. 2004, \aj, 127, 2284

\bibitem[{{Iwamoto} {et~al.}(1999){Iwamoto}, {Brachwitz}, {Nomoto},
  {Kishimoto}, {Umeda}, {Hix}, \& {Thielemann}}]{Iwamotoetal1999}
{Iwamoto}, K., {Brachwitz}, F., {Nomoto}, K., {et~al.} 1999, \apjs, 125, 439

\bibitem[{{Jayasinghe} {et~al.}(2021){Jayasinghe}, {Kochanek}, {Stanek},
  {Shappee}, {Holoien}, {Thompson}, {Prieto}, {Dong}, {Pawlak}, {Pejcha},
  {Pojmanski}, {Otero}, {Hurst}, \& {Will}}]{Jayasingheetal2021}
{Jayasinghe}, T., {Kochanek}, C.~S., {Stanek}, K.~Z., {et~al.} 2021, \mnras,
  503, 200

\bibitem[{{Kaufer} {et~al.}(1999){Kaufer}, {Stahl}, {Tubbesing},
  {N{\o}rregaard}, {Avila}, {Francois}, {Pasquini}, \&
  {Pizzella}}]{Kauferetal1999}
{Kaufer}, A., {Stahl}, O., {Tubbesing}, S., {et~al.} 1999, The Messenger, 95, 8

\bibitem[{{Kennicutt}(1998)}]{Kennicutt1998}
{Kennicutt}, Robert~C., J. 1998, \araa, 36, 189

\bibitem[{{Klagyivik} \& {Szabados}(2009)}]{KlagyivikSzabados2009}
{Klagyivik}, P. \& {Szabados}, L. 2009, \aap, 504, 959

\bibitem[{{Kobayashi} {et~al.}(2020){Kobayashi}, {Karakas}, \&
  {Lugaro}}]{Kobayashietal2020}
{Kobayashi}, C., {Karakas}, A.~I., \& {Lugaro}, M. 2020, \apj, 900, 179

\bibitem[{{Kobayashi} {et~al.}(2011){Kobayashi}, {Karakas}, \&
  {Umeda}}]{Kobayashietal2011}
{Kobayashi}, C., {Karakas}, A.~I., \& {Umeda}, H. 2011, \mnras, 414, 3231

\bibitem[{{Kobayashi} {et~al.}(2006){Kobayashi}, {Umeda}, {Nomoto}, {Tominaga},
  \& {Ohkubo}}]{Kobayashietal2006}
{Kobayashi}, C., {Umeda}, H., {Nomoto}, K., {Tominaga}, N., \& {Ohkubo}, T.
  2006, \apj, 653, 1145

\bibitem[{{Korotin}(2009)}]{Korotin2009}
{Korotin}, S.~A. 2009, Astronomy Reports, 53, 651

\bibitem[{{Kovtyukh}(2007)}]{Kovtyukh2007}
{Kovtyukh}, V.~V. 2007, \mnras, 378, 617

\bibitem[{{Kraft}(1966)}]{Kraft1966}
{Kraft}, R.~P. 1966, \aj, 71, 166

\bibitem[{{Kudritzki} \& {Urbaneja}(2018)}]{KudritzkiUrbaneja2018}
{Kudritzki}, R. \& {Urbaneja}, M.~A. 2018, arXiv e-prints, arXiv:1810.01102

\bibitem[{{Kudritzki} {et~al.}(2014){Kudritzki}, {Urbaneja}, {Bresolin},
  {Hosek}, \& {Przybilla}}]{Kudritzkietal2014}
{Kudritzki}, R.-P., {Urbaneja}, M.~A., {Bresolin}, F., {Hosek}, Matthew~W., J.,
  \& {Przybilla}, N. 2014, \apj, 788, 56

\bibitem[{{Lemasle} {et~al.}(2013){Lemasle}, {Fran{\c c}ois}, {Genovali},
  {Kovtyukh}, {Bono}, {Inno}, {Laney}, {Kaper}, {Bergemann}, {Fabrizio},
  {Matsunaga}, {Pedicelli}, {Primas}, \& {Romaniello}}]{Lemasleetal2013}
{Lemasle}, B., {Fran{\c c}ois}, P., {Genovali}, K., {et~al.} 2013, \aap, 558,
  A31

\bibitem[{{Lemasle} {et~al.}(2018){Lemasle}, {Hajdu}, {Kovtyukh}, {Inno},
  {Grebel}, {Catelan}, {Bono}, {Fran{\c{c}}ois}, {Kniazev}, {da Silva}, \&
  {Storm}}]{Lemasleetal2018}
{Lemasle}, B., {Hajdu}, G., {Kovtyukh}, V., {et~al.} 2018, \aap, 618, A160

\bibitem[{{Lemasle} {et~al.}(2022){Lemasle}, {Lala}, {Kovtyukh}, {Hanke},
  {Prudil}, {Bono}, {Braga}, {da Silva}, {Fabrizio}, {Fiorentino},
  {Fran{\c{c}}ois}, {Grebel}, \& {Kniazev}}]{Lemasleetal2022}
{Lemasle}, B., {Lala}, H.~N., {Kovtyukh}, V., {et~al.} 2022, \aap, 668, A40

\bibitem[{{L{\'e}pine} {et~al.}(2017){L{\'e}pine}, {Michtchenko}, {Barros}, \&
  {Vieira}}]{Lepineetal2017}
{L{\'e}pine}, J. R.~D., {Michtchenko}, T.~A., {Barros}, D.~A., \& {Vieira}, R.
  S.~S. 2017, \apj, 843, 48

\bibitem[{{L{\'e}pine} {et~al.}(2011){L{\'e}pine}, {Roman-Lopes}, {Abraham},
  {Junqueira}, \& {Mishurov}}]{Lepineetal2011}
{L{\'e}pine}, J.~R.~D., {Roman-Lopes}, A., {Abraham}, Z., {Junqueira}, T.~C.,
  \& {Mishurov}, Y.~N. 2011, \mnras, 414, 1607

\bibitem[{{Leung} \& {Nomoto}(2018)}]{LeungNomoto2018}
{Leung}, S.-C. \& {Nomoto}, K. 2018, \apj, 861, 143

\bibitem[{{Leung} \& {Nomoto}(2020)}]{LeungNomoto2020}
{Leung}, S.-C. \& {Nomoto}, K. 2020, \apj, 888, 80

\bibitem[{{Limongi} \& {Chieffi}(2003)}]{LimongiChieffi2003}
{Limongi}, M. \& {Chieffi}, A. 2003, \apj, 592, 404

\bibitem[{{Limongi} \& {Chieffi}(2018)}]{LimongiChieffi2018}
{Limongi}, M. \& {Chieffi}, A. 2018, \apjs, 237, 13

\bibitem[{{Liu} {et~al.}(2022){Liu}, {Kudritzki}, {Zhao}, {Urbaneja}, {Huang},
  {Zhang}, \& {Zhao}}]{Liuetal2022}
{Liu}, C., {Kudritzki}, R.-P., {Zhao}, G., {et~al.} 2022, \apj, 932, 29

\bibitem[{{Liu} {et~al.}(2015){Liu}, {Arav}, \& {Rupke}}]{Liuetal2015}
{Liu}, G., {Arav}, N., \& {Rupke}, D. S.~N. 2015, \apjs, 221, 9

\bibitem[{{L{\'o}pez-Corredoira} {et~al.}(2019){L{\'o}pez-Corredoira}, {Sylos
  Labini}, {Kalberla}, \& {Allende Prieto}}]{LopezCorredoiraetal2019}
{L{\'o}pez-Corredoira}, M., {Sylos Labini}, F., {Kalberla}, P.~M.~W., \&
  {Allende Prieto}, C. 2019, \aj, 157, 26

\bibitem[{{Lucertini} {et~al.}(2022){Lucertini}, {Monaco}, {Caffau},
  {Bonifacio}, \& {Mucciarelli}}]{Lucertinietal2022a}
{Lucertini}, F., {Monaco}, L., {Caffau}, E., {Bonifacio}, P., \& {Mucciarelli},
  A. 2022, \aap, 657, A29

\bibitem[{{Lucertini} {et~al.}(2023){Lucertini}, {Monaco}, {Caffau},
  {Mucciarelli}, {Villanova}, {Bonifacio}, \& {Sbordone}}]{Lucertinietal2022b}
{Lucertini}, F., {Monaco}, L., {Caffau}, E., {et~al.} 2023, \aap, 671, A137

\bibitem[{{Luck}(2018)}]{Luck2018}
{Luck}, R.~E. 2018, \aj, 156, 171

\bibitem[{{Luck} {et~al.}(2011){Luck}, {Andrievsky}, {Kovtyukh}, {Gieren}, \&
  {Graczyk}}]{Lucketal2011}
{Luck}, R.~E., {Andrievsky}, S.~M., {Kovtyukh}, V.~V., {Gieren}, W., \&
  {Graczyk}, D. 2011, \aj, 142, 51

\bibitem[{{Luck} \& {Lambert}(2011)}]{LuckLambert2011}
{Luck}, R.~E. \& {Lambert}, D.~L. 2011, \aj, 142, 136

\bibitem[{{Madore} {et~al.}(2013){Madore}, {Hoffman}, {Freedman}, {Kollmeier},
  {Monson}, {Persson}, {Rich}, {Scowcroft}, \& {Seibert}}]{Madoreetal2013}
{Madore}, B.~F., {Hoffman}, D., {Freedman}, W.~L., {et~al.} 2013, \apj, 776,
  135

\bibitem[{{Magrini} {et~al.}(2017{\natexlab{a}}){Magrini}, {Randich},
  {Kordopatis}, {Prantzos}, {Romano}, {Chieffi}, {Limongi}, {Fran{\c{c}}ois},
  {Pancino}, {Friel}, {Bragaglia}, {Tautvai{\v{s}}ien{\.{e}}}, {Spina},
  {Overbeek}, {Cantat-Gaudin}, {Donati}, {Vallenari}, {Sordo},
  {Jim{\'e}nez-Esteban}, {Tang}, {Drazdauskas}, {Sousa}, {Duffau}, {Jofr{\'e}},
  {Gilmore}, {Feltzing}, {Alfaro}, {Bensby}, {Flaccomio}, {Koposov},
  {Lanzafame}, {Smiljanic}, {Bayo}, {Carraro}, {Casey}, {Costado}, {Damiani},
  {Franciosini}, {Hourihane}, {Lardo}, {Lewis}, {Monaco}, {Morbidelli},
  {Sacco}, {Sbordone}, {Worley}, \& {Zaggia}}]{Magrinietal2017}
{Magrini}, L., {Randich}, S., {Kordopatis}, G., {et~al.} 2017{\natexlab{a}},
  \aap, 603, A2

\bibitem[{{Magrini} {et~al.}(2017{\natexlab{b}}){Magrini}, {Randich},
  {Kordopatis}, {Prantzos}, {Romano}, {Chieffi}, {Limongi}, {Fran{\c{c}}ois},
  {Pancino}, {Friel}, {Bragaglia}, {Tautvai{\v{s}}ien{\.{e}}}, {Spina},
  {Overbeek}, {Cantat-Gaudin}, {Donati}, {Vallenari}, {Sordo},
  {Jim{\'e}nez-Esteban}, {Tang}, {Drazdauskas}, {Sousa}, {Duffau}, {Jofr{\'e}},
  {Gilmore}, {Feltzing}, {Alfaro}, {Bensby}, {Flaccomio}, {Koposov},
  {Lanzafame}, {Smiljanic}, {Bayo}, {Carraro}, {Casey}, {Costado}, {Damiani},
  {Franciosini}, {Hourihane}, {Lardo}, {Lewis}, {Monaco}, {Morbidelli},
  {Sacco}, {Sbordone}, {Worley}, \& {Zaggia}}]{Magrinietal2018}
{Magrini}, L., {Randich}, S., {Kordopatis}, G., {et~al.} 2017{\natexlab{b}},
  \aap, 603, A2

\bibitem[{{Magrini} {et~al.}(2023){Magrini}, {Viscasillas V{\'a}zquez},
  {Spina}, {Randich}, {Romano}, {Franciosini}, {Recio-Blanco}, {Nordlander},
  {D'Orazi}, {Baratella}, {Smiljanic}, {Dantas}, {Pasquini}, {Spitoni},
  {Casali}, {Van der Swaelmen}, {Bensby}, {Stonkute}, {Feltzing}, {Sacco},
  {Bragaglia}, {Pancino}, {Heiter}, {Biazzo}, {Gilmore}, {Bergemann},
  {Tautvai{\v{s}}ien{\.{e}}}, {Worley}, {Hourihane}, {Gonneau}, \&
  {Morbidelli}}]{Magrinietal2023}
{Magrini}, L., {Viscasillas V{\'a}zquez}, C., {Spina}, L., {et~al.} 2023, \aap,
  669, A119

\bibitem[{{Matsunaga} {et~al.}(2018){Matsunaga}, {Bono}, {Chen}, {de Grijs},
  {Inno}, \& {Nishiyama}}]{Matsunagaetal2018}
{Matsunaga}, N., {Bono}, G., {Chen}, X., {et~al.} 2018, \ssr, 214, 74

\bibitem[{{Matteucci}(2021)}]{Matteucci2021}
{Matteucci}, F. 2021, \aapr, 29, 5

\bibitem[{{Matteucci} {et~al.}(2020){Matteucci}, {Vasini}, {Grisoni}, \&
  {Schultheis}}]{Matteuccietal2020}
{Matteucci}, F., {Vasini}, A., {Grisoni}, V., \& {Schultheis}, M. 2020, \mnras,
  494, 5534

\bibitem[{{Mayor} {et~al.}(2003){Mayor}, {Pepe}, {Queloz}, {Bouchy},
  {Rupprecht}, {Lo Curto}, {Avila}, {Benz}, {Bertaux}, {Bonfils}, {Dall},
  {Dekker}, {Delabre}, {Eckert}, {Fleury}, {Gilliotte}, {Gojak}, {Guzman},
  {Kohler}, {Lizon}, {Longinotti}, {Lovis}, {Megevand}, {Pasquini}, {Reyes},
  {Sivan}, {Sosnowska}, {Soto}, {Udry}, {van Kesteren}, {Weber}, \&
  {Weilenmann}}]{Mayoretal2003}
{Mayor}, M., {Pepe}, F., {Queloz}, D., {et~al.} 2003, The Messenger, 114, 20

\bibitem[{{McWilliam} {et~al.}(2008){McWilliam}, {Matteucci}, {Ballero},
  {Rich}, {Fulbright}, \& {Cescutti}}]{McWilliametal2008}
{McWilliam}, A., {Matteucci}, F., {Ballero}, S., {et~al.} 2008, \aj, 136, 367

\bibitem[{{Minchev} {et~al.}(2011){Minchev}, {Famaey}, {Combes}, {Di Matteo},
  {Mouhcine}, \& {Wozniak}}]{Minchevetal2011}
{Minchev}, I., {Famaey}, B., {Combes}, F., {et~al.} 2011, \aap, 527, A147

\bibitem[{{Miyamoto} \& {Nagai}(1975)}]{MiyamotoNagai1975}
{Miyamoto}, M. \& {Nagai}, R. 1975, \pasj, 27, 533

\bibitem[{{Mizumoto} {et~al.}(2023){Mizumoto}, {Sameshima}, {Kobayashi}, {},
  {}, \& {}}]{Mizumotoetal2023}
{Mizumoto}, M., {Sameshima}, H., {Kobayashi}, N., {et~al.} 2023, \apj,
  submitted

\bibitem[{{Nakanishi} \& {Sofue}(2003)}]{NakanishiSofue2003}
{Nakanishi}, H. \& {Sofue}, Y. 2003, \pasj, 55, 191

\bibitem[{{Nakanishi} \& {Sofue}(2006)}]{NakanishiSofue2006}
{Nakanishi}, H. \& {Sofue}, Y. 2006, \pasj, 58, 847

\bibitem[{{Navarro} {et~al.}(1996){Navarro}, {Frenk}, \&
  {White}}]{Navarroetal1996}
{Navarro}, J.~F., {Frenk}, C.~S., \& {White}, S. D.~M. 1996, \apj, 462, 563

\bibitem[{{Ngeow} {et~al.}(2003){Ngeow}, {Kanbur}, {Nikolaev}, {Tanvir}, \&
  {Hendry}}]{Ngeowetal2003}
{Ngeow}, C.-C., {Kanbur}, S.~M., {Nikolaev}, S., {Tanvir}, N.~R., \& {Hendry},
  M.~A. 2003, \apj, 586, 959

\bibitem[{{Noll} {et~al.}(2012){Noll}, {Kausch}, {Barden}, {Jones}, {Szyszka},
  {Kimeswenger}, \& {Vinther}}]{Nolletal2012}
{Noll}, S., {Kausch}, W., {Barden}, M., {et~al.} 2012, \aap, 543, A92

\bibitem[{{Nomoto} {et~al.}(2013){Nomoto}, {Kobayashi}, \&
  {Tominaga}}]{Nomotoetal2013}
{Nomoto}, K., {Kobayashi}, C., \& {Tominaga}, N. 2013, \araa, 51, 457

\bibitem[{{Oppenheimer} \& {Dav{\'e}}(2008)}]{OppenheimerDave2008}
{Oppenheimer}, B.~D. \& {Dav{\'e}}, R. 2008, \mnras, 387, 577

\bibitem[{{Palla}(2021)}]{Palla2021}
{Palla}, M. 2021, \mnras, 503, 3216

\bibitem[{{Palla} {et~al.}(2020){Palla}, {Matteucci}, {Spitoni}, {Vincenzo}, \&
  {Grisoni}}]{Pallaetal2020}
{Palla}, M., {Matteucci}, F., {Spitoni}, E., {Vincenzo}, F., \& {Grisoni}, V.
  2020, \mnras, 498, 1710

\bibitem[{{Palla} {et~al.}(2022){Palla}, {Santos-Peral}, {Recio-Blanco}, \&
  {Matteucci}}]{Pallaetal2022}
{Palla}, M., {Santos-Peral}, P., {Recio-Blanco}, A., \& {Matteucci}, F. 2022,
  \aap, 663, A125

\bibitem[{{Perdigon} {et~al.}(2021){Perdigon}, {de Laverny}, {Recio-Blanco},
  {Fernandez-Alvar}, {Santos-Peral}, {Kordopatis}, \&
  {{\'A}lvarez}}]{Perdigonetal2021}
{Perdigon}, J., {de Laverny}, P., {Recio-Blanco}, A., {et~al.} 2021, \aap, 647,
  A162

\bibitem[{{Pietrukowicz} {et~al.}(2021){Pietrukowicz}, {Soszy{\'n}ski}, \&
  {Udalski}}]{Pietrukowiczetal2021}
{Pietrukowicz}, P., {Soszy{\'n}ski}, I., \& {Udalski}, A. 2021, \actaa, 71, 205

\bibitem[{{Piskunov} \& {Valenti}(2017)}]{PiskunovValenti2017}
{Piskunov}, N. \& {Valenti}, J.~A. 2017, \aap, 597, A16

\bibitem[{{Prantzos} {et~al.}(2018){Prantzos}, {Abia}, {Limongi}, {Chieffi}, \&
  {Cristallo}}]{Prantzosetal2018}
{Prantzos}, N., {Abia}, C., {Limongi}, M., {Chieffi}, A., \& {Cristallo}, S.
  2018, \mnras, 476, 3432

\bibitem[{{Proxauf} {et~al.}(2018){Proxauf}, {da Silva}, {Kovtyukh}, {Bono},
  {Inno}, {Lemasle}, {Pritchard}, {Przybilla}, {Storm}, {Urbaneja}, {Valenti},
  {Bergemann}, {Buonanno}, {D'Orazi}, {Fabrizio}, {Ferraro}, {Fiorentino},
  {Fran{\c{c}}ois}, {Iannicola}, {Laney}, {Kudritzki}, {Matsunaga}, {Nonino},
  {Primas}, {Romaniello}, \& {Th{\'e}venin}}]{Proxaufetal2018}
{Proxauf}, B., {da Silva}, R., {Kovtyukh}, V.~V., {et~al.} 2018, \aap, 616, A82

\bibitem[{{Queiroz} {et~al.}(2020){Queiroz}, {Anders}, {Chiappini},
  {Khalatyan}, {Santiago}, {Steinmetz}, {Valentini}, {Miglio}, {Bossini},
  {Barbuy}, {Minchev}, {Minniti}, {Garc{\'\i}a Hern{\'a}ndez}, {Schultheis},
  {Beaton}, {Beers}, {Bizyaev}, {Brownstein}, {Cunha},
  {Fern{\'a}ndez-Trincado}, {Frinchaboy}, {Lane}, {Majewski}, {Nataf},
  {Nitschelm}, {Pan}, {Roman-Lopes}, {Sobeck}, {Stringfellow}, \&
  {Zamora}}]{Queirozetal2020}
{Queiroz}, A.~B.~A., {Anders}, F., {Chiappini}, C., {et~al.} 2020, \aap, 638,
  A76

\bibitem[{{Quillen} {et~al.}(2009){Quillen}, {Minchev}, {Bland-Hawthorn}, \&
  {Haywood}}]{Quillenetal2009}
{Quillen}, A.~C., {Minchev}, I., {Bland-Hawthorn}, J., \& {Haywood}, M. 2009,
  \mnras, 397, 1599

\bibitem[{{Randich} {et~al.}(2022){Randich}, {Gilmore}, {Magrini}, {Sacco},
  {Jackson}, {Jeffries}, {Worley}, {Hourihane}, {Gonneau}, {Viscasillas
  Vazquez}, {Franciosini}, {Lewis}, {Alfaro}, {Allende Prieto}, {Bensby},
  {Blomme}, {Bragaglia}, {Flaccomio}, {Fran{\c{c}}ois}, {Irwin}, {Koposov},
  {Korn}, {Lanzafame}, {Pancino}, {Recio-Blanco}, {Smiljanic}, {Van Eck},
  {Zwitter}, {Asplund}, {Bonifacio}, {Feltzing}, {Binney}, {Drew}, {Ferguson},
  {Micela}, {Negueruela}, {Prusti}, {Rix}, {Vallenari}, {Bayo}, {Bergemann},
  {Biazzo}, {Carraro}, {Casey}, {Damiani}, {Frasca}, {Heiter}, {Hill},
  {Jofr{\'e}}, {de Laverny}, {Lind}, {Marconi}, {Martayan}, {Masseron},
  {Monaco}, {Morbidelli}, {Prisinzano}, {Sbordone}, {Sousa}, {Zaggia},
  {Adibekyan}, {Bonito}, {Caffau}, {Daflon}, {Feuillet}, {Gebran}, {Gonzalez
  Hernandez}, {Guiglion}, {Herrero}, {Lobel}, {Maiz Apellaniz}, {Merle},
  {Mikolaitis}, {Montes}, {Morel}, {Soubiran}, {Spina}, {Tabernero},
  {Tautvai{\v{s}}iene}, {Traven}, {Valentini}, {Van der Swaelmen}, {Villanova},
  {Wright}, {Abbas}, {Aguirre B{\o}rsen-Koch}, {Alves}, {Balaguer-Nunez},
  {Barklem}, {Barrado}, {Berlanas}, {Binks}, {Bressan}, {Capuzzo-Dolcetta},
  {Casagrande}, {Casamiquela}, {Collins}, {D'Orazi}, {Dantas}, {Debattista},
  {Delgado-Mena}, {Di Marcantonio}, {Drazdauskas}, {Evans}, {Famaey},
  {Franchini}, {Fr{\'e}mat}, {Friel}, {Fu}, {Geisler}, {Gerhard}, {Gonzalez
  Solares}, {Grebel}, {Gutierrez Albarran}, {Hatzidimitriou}, {Held},
  {Jim{\'e}nez-Esteban}, {J{\"o}nsson}, {Jordi}, {Khachaturyants},
  {Kordopatis}, {Kos}, {Lagarde}, {Mahy}, {Mapelli}, {Marfil}, {Martell},
  {Messina}, {Miglio}, {Minchev}, {Moitinho}, {Montalban}, {Monteiro},
  {Morossi}, {Mowlavi}, {Mucciarelli}, {Murphy}, {Nardetto}, {Ortolani},
  {Paletou}, {Palou{\v{s}}}, {Paunzen}, {Pickering}, {Quirrenbach}, {Re
  Fiorentin}, {Read}, {Romano}, {Ryde}, {Sanna}, {Santos}, {Seabroke},
  {Spagna}, {Steinmetz}, {Stonkut{\'e}}, {Sutorius}, {Th{\'e}venin}, {Tosi},
  {Tsantaki}, {Vink}, {Wright}, {Wyse}, {Zoccali}, {Zorec}, {Zucker}, \&
  {Walton}}]{Randichetal2022}
{Randich}, S., {Gilmore}, G., {Magrini}, L., {et~al.} 2022, \aap, 666, A121

\bibitem[{{Reid} {et~al.}(2019){Reid}, {Menten}, {Brunthaler}, {Zheng}, {Dame},
  {Xu}, {Li}, {Sakai}, {Wu}, {Immer}, {Zhang}, {Sanna}, {Moscadelli}, {Rygl},
  {Bartkiewicz}, {Hu}, {Quiroga-Nuñez}, \& {van Langevelde}}]{Reidetal2019}
{Reid}, M.~J., {Menten}, K.~M., {Brunthaler}, A., {et~al.} 2019, ApJ, 885, 131

\bibitem[{{Reid} {et~al.}(2014){Reid}, {Menten}, {Brunthaler}, {Zheng}, {Dame},
  {Xu}, {Wu}, {Zhang}, {Sanna}, {Sato}, {Hachisuka}, {Choi}, {Immer},
  {Moscadelli}, {Rygl}, \& {Bartkiewicz}}]{Reidetal2014}
{Reid}, M.~J., {Menten}, K.~M., {Brunthaler}, A., {et~al.} 2014, \apj, 783, 130

\bibitem[{{Ripepi} {et~al.}(2020){Ripepi}, {Catanzaro}, {Molinaro}, {Marconi},
  {Clementini}, {Cusano}, {De Somma}, {Leccia}, {Musella}, \&
  {Testa}}]{Ripepietal2020}
{Ripepi}, V., {Catanzaro}, G., {Molinaro}, R., {et~al.} 2020, \aap, 642, A230

\bibitem[{{Ripepi} {et~al.}(2022){Ripepi}, {Chemin}, {Molinaro}, {Cioni},
  {Bekki}, {Clementini}, {de Grijs}, {De Somma}, {El Youssoufi}, {Girardi},
  {Groenewegen}, {Ivanov}, {Marconi}, {McMillan}, \& {van
  Loon}}]{Ripepietal2022}
{Ripepi}, V., {Chemin}, L., {Molinaro}, R., {et~al.} 2022, \mnras, 512, 563

\bibitem[{{Romaniello} {et~al.}(2022){Romaniello}, {Riess}, {Mancino},
  {Anderson}, {Freudling}, {Kudritzki}, {Macr{\`\i}}, {Mucciarelli}, \&
  {Yuan}}]{Romanielloetal2022}
{Romaniello}, M., {Riess}, A., {Mancino}, S., {et~al.} 2022, \aap, 658, A29

\bibitem[{{Romano} {et~al.}(2010){Romano}, {Karakas}, {Tosi}, \&
  {Matteucci}}]{Romanoetal2010}
{Romano}, D., {Karakas}, A.~I., {Tosi}, M., \& {Matteucci}, F. 2010, \aap, 522,
  A32

\bibitem[{{S{\'a}nchez-Bl{\'a}zquez} {et~al.}(2014){S{\'a}nchez-Bl{\'a}zquez},
  {Rosales-Ortega}, {M{\'e}ndez-Abreu}, {P{\'e}rez}, {S{\'a}nchez}, {Zibetti},
  {Aguerri}, {Bland-Hawthorn}, {Catal{\'a}n-Torrecilla}, {Cid Fernandes}, {de
  Amorim}, {de Lorenzo-Caceres}, {Falc{\'o}n-Barroso}, {Galazzi}, {Garc{\'\i}a
  Benito}, {Gil de Paz}, {Gonz{\'a}lez Delgado}, {Husemann},
  {Iglesias-P{\'a}ramo}, {Jungwiert}, {Marino}, {M{\'a}rquez}, {Mast},
  {Mendoza}, {Moll{\'a}}, {Papaderos}, {Ruiz-Lara}, {van de Ven}, {Walcher}, \&
  {Wisotzki}}]{SanchezBlazquezetal2014}
{S{\'a}nchez-Bl{\'a}zquez}, P., {Rosales-Ortega}, F.~F., {M{\'e}ndez-Abreu},
  J., {et~al.} 2014, \aap, 570, A6

\bibitem[{{S{\'a}nchez-Menguiano} {et~al.}(2016){S{\'a}nchez-Menguiano},
  {S{\'a}nchez}, {P{\'e}rez}, {Garc{\'\i}a-Benito}, {Husemann}, {Mast},
  {Mendoza}, {Ruiz-Lara}, {Ascasibar}, {Bland-Hawthorn}, {Cavichia},
  {D{\'\i}az}, {Florido}, {Galbany}, {G{\'o}nzalez Delgado}, {Kehrig},
  {Marino}, {M{\'a}rquez}, {Masegosa}, {M{\'e}ndez-Abreu}, {Moll{\'a}}, {Del
  Olmo}, {P{\'e}rez}, {S{\'a}nchez-Bl{\'a}zquez}, {Stanishev}, {Walcher},
  {L{\'o}pez-S{\'a}nchez}, \& {CALIFA
  Collaboration}}]{SanchezMenguianoetal2016}
{S{\'a}nchez-Menguiano}, L., {S{\'a}nchez}, S.~F., {P{\'e}rez}, I., {et~al.}
  2016, \aap, 587, A70

\bibitem[{{Santos-Peral} {et~al.}(2021){Santos-Peral}, {Recio-Blanco},
  {Kordopatis}, {Fern{\'a}ndez-Alvar}, \& {de Laverny}}]{SantosPeral2021}
{Santos-Peral}, P., {Recio-Blanco}, A., {Kordopatis}, G.,
  {Fern{\'a}ndez-Alvar}, E., \& {de Laverny}, P. 2021, \aap, 653, A85

\bibitem[{{Santucci} {et~al.}(2023){Santucci}, {Brough}, {van de Sande},
  {McDermid}, {Barsanti}, {Bland-Hawthorn}, {Bryant}, {Croom}, {Lagos},
  {Lawrence}, {Owers}, {van de Ven}, {Vaughan}, \& {Yi}}]{Santuccietal2023}
{Santucci}, G., {Brough}, S., {van de Sande}, J., {et~al.} 2023, \mnras, 521,
  2671

\bibitem[{{Schlafly} \& {Finkbeiner}(2011)}]{SchlaflyFinkbeiner2011}
{Schlafly}, E.~F. \& {Finkbeiner}, D.~P. 2011, \apj, 737, 103

\bibitem[{{Sch{\"o}nrich} {et~al.}(2010){Sch{\"o}nrich}, {Binney}, \&
  {Dehnen}}]{Schonrichetal2010}
{Sch{\"o}nrich}, R., {Binney}, J., \& {Dehnen}, W. 2010, \mnras, 403, 1829

\bibitem[{{Sch{\"o}nrich} \& {McMillan}(2017)}]{SchonrichMcMillan2017}
{Sch{\"o}nrich}, R. \& {McMillan}, P.~J. 2017, \mnras, 467, 1154

\bibitem[{{Skowron} {et~al.}(2019){Skowron}, {Skowron}, {Mr{\'o}z}, {Udalski},
  {Pietrukowicz}, {Soszy{\'n}ski}, {Szyma{\'n}ski}, {Poleski}, {Koz{\l}owski},
  {Ulaczyk}, {Rybicki}, \& {Iwanek}}]{Skowronetal2019}
{Skowron}, D.~M., {Skowron}, J., {Mr{\'o}z}, P., {et~al.} 2019, Science, 365,
  478

\bibitem[{Sneden(2002)}]{Sneden2002}
Sneden, C. 2002, The MOOG code, \url{http://www.as.utexas.edu/~chris/moog.html}

\bibitem[{{Sofue}(2013)}]{Sofue2013}
{Sofue}, Y. 2013, \pasj, 65, 118

\bibitem[{{Soszy{\'n}ski} {et~al.}(2017){Soszy{\'n}ski}, {Udalski},
  {Szyma{\'n}ski}, {Wyrzykowski}, {Ulaczyk}, {Poleski}, {Pietrukowicz},
  {Koz{\l}owski}, {Skowron}, {Skowron}, {Mr{\'o}z}, \&
  {Pawlak}}]{Soszynskietal2017}
{Soszy{\'n}ski}, I., {Udalski}, A., {Szyma{\'n}ski}, M.~K., {et~al.} 2017,
  \actaa, 67, 103

\bibitem[{{Sousa} {et~al.}(2015){Sousa}, {Santos}, {Adibekyan}, {Delgado-Mena},
  \& {Israelian}}]{Sousaetal2015}
{Sousa}, S.~G., {Santos}, N.~C., {Adibekyan}, V., {Delgado-Mena}, E., \&
  {Israelian}, G. 2015, \aap, 577, A67

\bibitem[{{Sousa} {et~al.}(2007){Sousa}, {Santos}, {Israelian}, {Mayor}, \&
  {Monteiro}}]{Sousaetal2007}
{Sousa}, S.~G., {Santos}, N.~C., {Israelian}, G., {Mayor}, M., \& {Monteiro},
  M.~J.~P.~F.~G. 2007, \aap, 469, 783

\bibitem[{{Spitoni} {et~al.}(2022){Spitoni}, {Aguirre B{\o}rsen-Koch}, {Verma},
  \& {Stokholm}}]{Spitonietal2022}
{Spitoni}, E., {Aguirre B{\o}rsen-Koch}, V., {Verma}, K., \& {Stokholm}, A.
  2022, \aap, 663, A174

\bibitem[{{Spitoni} {et~al.}(2019){Spitoni}, {Silva Aguirre}, {Matteucci},
  {Calura}, \& {Grisoni}}]{Spitonietal2019}
{Spitoni}, E., {Silva Aguirre}, V., {Matteucci}, F., {Calura}, F., \&
  {Grisoni}, V. 2019, \aap, 623, A60

\bibitem[{{Spitoni} {et~al.}(2021){Spitoni}, {Verma}, {Silva Aguirre},
  {Vincenzo}, {Matteucci}, {Vai{\v{c}}ekauskait{\.{e}}}, {Palla}, {Grisoni}, \&
  {Calura}}]{Spitonietal2021}
{Spitoni}, E., {Verma}, K., {Silva Aguirre}, V., {et~al.} 2021, \aap, 647, A73

\bibitem[{{Stahler} \& {Palla}(2005)}]{StahlerPalla2005}
{Stahler}, S.~W. \& {Palla}, F. 2005, {The Formation of Stars}

\bibitem[{{Strassmeier} {et~al.}(2004){Strassmeier}, {Granzer}, {Weber},
  {Woche}, {Andersen}, {Bartus}, {Bauer}, {Dionies}, {Popow}, {Fechner},
  {Hildebrandt}, {Washuettl}, {Ritter}, {Schwope}, {Staude}, {Paschke},
  {Stolz}, {Serre-Ricart}, {de la Rosa}, \& {Arnay}}]{Strassmeieretal2004}
{Strassmeier}, K.~G., {Granzer}, T., {Weber}, M., {et~al.} 2004, Astronomische
  Nachrichten, 325, 527

\bibitem[{{Strassmeier} {et~al.}(2010){Strassmeier}, {Granzer}, {Weber},
  {Woche}, {Popow}, {J{\"a}rvinen}, {Bartus}, {Bauer}, {Dionies}, {Fechner},
  {Bittner}, \& {Paschke}}]{Strassmeieretal2010}
{Strassmeier}, K.~G., {Granzer}, T., {Weber}, M., {et~al.} 2010, Advances in
  Astronomy, 2010, 970306

\bibitem[{{Sun} {et~al.}(2015){Sun}, {Xu}, {Yang}, {Li}, {Du}, {Zhang}, \&
  {Zhou}}]{Sunetal2015}
{Sun}, Y., {Xu}, Y., {Yang}, J., {et~al.} 2015, \apjl, 798, L27

\bibitem[{{Takeda} {et~al.}(2005){Takeda}, {Hashimoto}, {Taguchi}, {Yoshioka},
  {Takada-Hidai}, {Saito}, \& {Honda}}]{Takedaetal2005}
{Takeda}, Y., {Hashimoto}, O., {Taguchi}, H., {et~al.} 2005, \pasj, 57, 751

\bibitem[{{Tautvai{\v{s}}ien{\.{e}}} {et~al.}(2015){Tautvai{\v{s}}ien{\.{e}}},
  {Drazdauskas}, {Mikolaitis}, {Barisevi{\v{c}}ius}, {Puzeras},
  {Stonkut{\.{e}}}, {Chorniy}, {Magrini}, {Romano}, {Smiljanic}, {Bragaglia},
  {Carraro}, {Friel}, {Morel}, {Pancino}, {Donati}, {Jim{\'e}nez-Esteban},
  {Gilmore}, {Randich}, {Jeffries}, {Vallenari}, {Bensby}, {Flaccomio},
  {Recio-Blanco}, {Costado}, {Hill}, {Jofr{\'e}}, {Lardo}, {de Laverny},
  {Masseron}, {Moribelli}, {Sousa}, \& {Zaggia}}]{Tautvaivsieneetal2015}
{Tautvai{\v{s}}ien{\.{e}}}, G., {Drazdauskas}, A., {Mikolaitis}, {\v{S}}.,
  {et~al.} 2015, \aap, 573, A55

\bibitem[{{Tian} {et~al.}(2017){Tian}, {Liu}, {Wan}, {Wang}, {Wang}, {Deng},
  {Cao}, {Hou}, {Wang}, {Wu}, \& {Zhao}}]{Tianetal2017}
{Tian}, H.-J., {Liu}, C., {Wan}, J.-C., {et~al.} 2017, Research in Astronomy
  and Astrophysics, 17, 114

\bibitem[{{Trentin} {et~al.}(2023){Trentin}, {Ripepi}, {Catanzaro}, {Storm},
  {Marconi}, {De Somma}, {Testa}, \& {Musella}}]{Trentinetal2023}
{Trentin}, E., {Ripepi}, V., {Catanzaro}, G., {et~al.} 2023, \mnras, 519, 2331

\bibitem[{{Tsujimoto}(2023)}]{Tsujimoto2023}
{Tsujimoto}, T. 2023, \mnras, 518, 3475

\bibitem[{{Twarog} {et~al.}(1997){Twarog}, {Ashman}, \&
  {Anthony-Twarog}}]{Twarogetal1997}
{Twarog}, B.~A., {Ashman}, K.~M., \& {Anthony-Twarog}, B.~J. 1997, \aj, 114,
  2556

\bibitem[{{Urbaneja} {et~al.}(2005){Urbaneja}, {Herrero}, {Bresolin},
  {Kudritzki}, {Gieren}, {Puls}, {Przybilla}, {Najarro}, \&
  {Pietrzy{\'n}ski}}]{Urbanejaetal2005}
{Urbaneja}, M.~A., {Herrero}, A., {Bresolin}, F., {et~al.} 2005, \apj, 622, 862

\bibitem[{{Viscasillas V{\'a}zquez} {et~al.}(2022){Viscasillas V{\'a}zquez},
  {Magrini}, {Casali}, {Tautvai{\v{s}}ien{\.{e}}}, {Spina}, {Van der Swaelmen},
  {Randich}, {Bensby}, {Bragaglia}, {Friel}, {Feltzing}, {Sacco}, {Turchi},
  {Jim{\'e}nez-Esteban}, {D'Orazi}, {Delgado-Mena}, {Mikolaitis},
  {Drazdauskas}, {Minkevi{\v{c}}i{\={u}}t{\.{e}}}, {Stonkut{\.{e}}},
  {Bagdonas}, {Montes}, {Guiglion}, {Baratella}, {Tabernero}, {Gilmore},
  {Alfaro}, {Francois}, {Korn}, {Smiljanic}, {Bergemann}, {Franciosini},
  {Gonneau}, {Hourihane}, {Worley}, \& {Zaggia}}]{Viscasillasetal2022}
{Viscasillas V{\'a}zquez}, C., {Magrini}, L., {Casali}, G., {et~al.} 2022,
  \aap, 660, A135

\bibitem[{{Vladilo} {et~al.}(2018){Vladilo}, {Gioannini}, {Matteucci}, \&
  {Palla}}]{Vladiloetal2018}
{Vladilo}, G., {Gioannini}, L., {Matteucci}, F., \& {Palla}, M. 2018, \apj,
  868, 127

\bibitem[{{Wang} {et~al.}(2018){Wang}, {Chen}, {de Grijs}, \&
  {Deng}}]{Wangetal2018}
{Wang}, S., {Chen}, X., {de Grijs}, R., \& {Deng}, L. 2018, \apj, 852, 78

\bibitem[{{Wegg} {et~al.}(2015){Wegg}, {Gerhard}, \& {Portail}}]{Weggetal2015}
{Wegg}, C., {Gerhard}, O., \& {Portail}, M. 2015, MNRAS, 450, 4050

\bibitem[{{Wehrhahn}(2021)}]{Wehrhahn2021}
{Wehrhahn}, A. 2021, in The 20.5th Cambridge Workshop on Cool Stars, Stellar
  Systems, and the Sun (CS20.5), Cambridge Workshop on Cool Stars, Stellar
  Systems, and the Sun, 1

\bibitem[{{Woosley} \& {Weaver}(1995)}]{WoosleyWeaver1995}
{Woosley}, S.~E. \& {Weaver}, T.~A. 1995, \apjs, 101, 181

\bibitem[{{Yong} {et~al.}(2012){Yong}, {Carney}, \& {Friel}}]{Yongetal2012}
{Yong}, D., {Carney}, B.~W., \& {Friel}, E.~D. 2012, \aj, 144, 95

\bibitem[{{Zaritsky} {et~al.}(1994){Zaritsky}, {Kennicutt}, \&
  {Huchra}}]{Zaritskyetal1994}
{Zaritsky}, D., {Kennicutt}, Robert~C., J., \& {Huchra}, J.~P. 1994, \apj, 420,
  87

\bibitem[{{Zhu} {et~al.}(2018){Zhu}, {van den Bosch}, {van de Ven},
  {Lyubenova}, {Falc{\'o}n-Barroso}, {Meidt}, {Martig}, {Shen}, {Li},
  {Yildirim}, {Walcher}, \& {Sanchez}}]{Zhuetal2018}
{Zhu}, L., {van den Bosch}, R., {van de Ven}, G., {et~al.} 2018, \mnras, 473,
  3000

\end{thebibliography}


\begin{appendix}

\section{Pulsating type classification}
\label{appendix:bailey_fourier}

\begin{figure}[h]
\centering
\resizebox{\hsize}{!}{\includegraphics{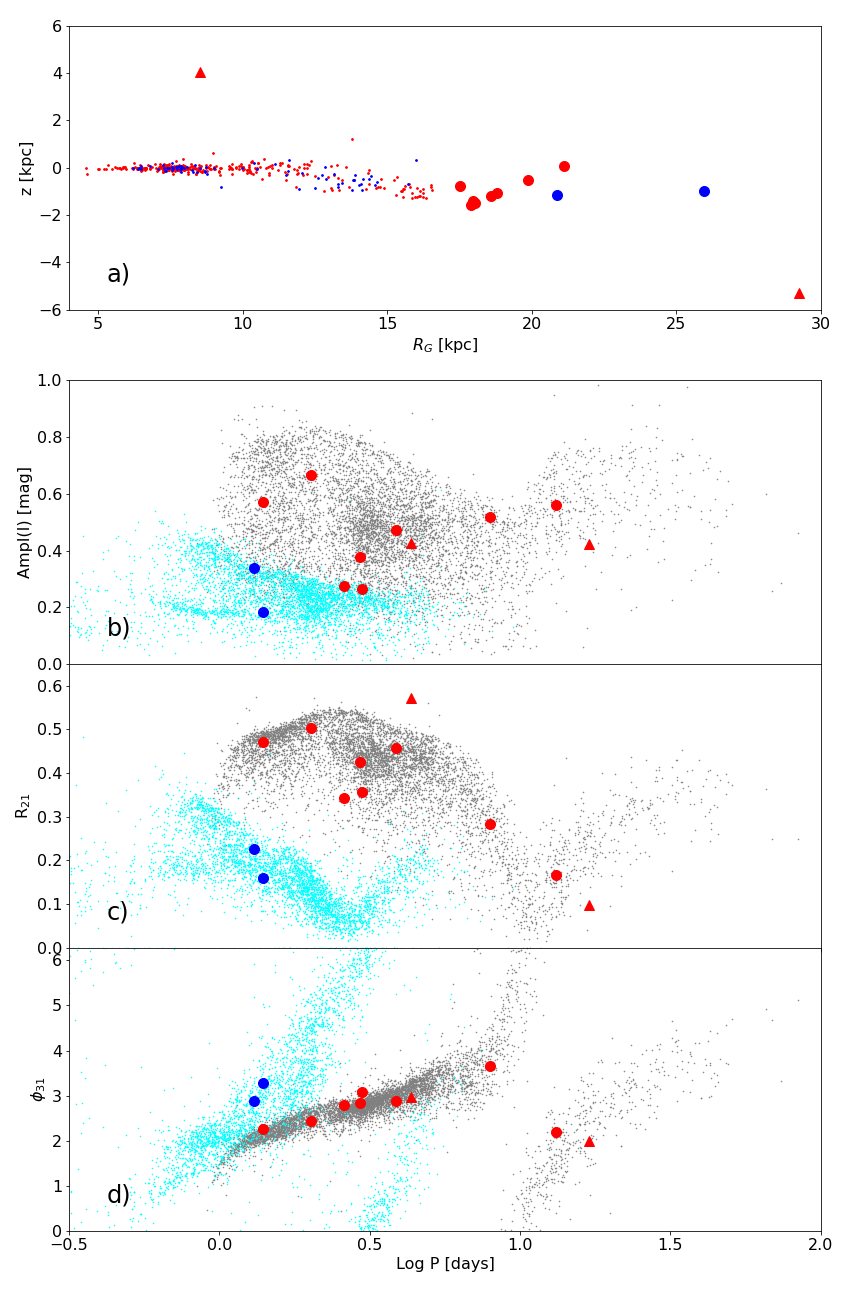}}
\caption{Height from the Galactic plane, Bailey diagram, and Fourier parameters.
{\it Panel a)} -- Distance from the Galactic plane as a function of the Galactocentric distance for the current sample of 379 classical Cepheids.
{\it Panel b)} -- Bailey diagram: $I$-band luminosity amplitude as a function of the logarithmic pulsation period. Grey and cyan dots display, respectively, fundamental (FU) and first overtone (FO) Galactic CCs collected by OGLE~IV. The red triangles indicate two of our Cepheids for which only the $V$-band light curves are available \citep[][ASAS-SN V]{Jayasingheetal2021} and that were transformed into the $I$-band using an amplitude ratio of 1.61 \citep[see][]{KlagyivikSzabados2009}. Red/blue circles display FU/FO Cepheids located in the outermost disk region.
{\it Panels c) and d)} -- Same as in panel (b), but showing the R$_{21}$ and the $\phi_{31}$ Fourier parameters. For the two CCs with only $V$-band light curves, the Fourier parameters were transformed into the $I$-band using empirical relations from \citet{Ngeowetal2003}.
}
\label{figure:check_cepheids}
\end{figure}

By having a look at the position of our CCs in the MW, we found 12 objects that needed further investigation in order to validate our results. More specifically, we found eleven CCs having a distance from the Galactic center larger than 17~kpc and two with height from the Galactic plane larger than 3~kpc (see Fig.~\ref{figure:check_cepheids}, panel a). Therefore, with the aim at separating different types of pulsating variables, we checked their positions in diagrams that are commonly used for variable star classification, namely, the Bailey diagram (panel b of the same figure) and the Fourier parameters plotted as a function of the logarithmic pulsation period (panels c and d).

Comparing our sample of CCs with bona-fide CCs from the OGLE survey (including both the LMC, SMC, disk, and Bulge sample), we found that only the two fundamental-mode CCs with a large $Z$ (i.e., the two triangles), occupy anomalous positions in some of these diagram. More specifically, the star \object{ASAS\,J062939-1840.5} (the one at $R_{\rm G}$$\sim$29~kpc and with pulsation period $\sim$17~days), is slightly outside of the main sequence of fundamental CCs in both the Bailey and the R$_{21}$ plots. \object{V1253\,Cen} ($R_{\rm G}$$\sim$8.5~kpc and $P$$\sim$4~days), instead, is outside the locus of fundamental CCs only in the R$_{21}$ versus $\log{P}$ diagram. All the other CCs in our sample are well within the loci of both fundamental and first overtone CCs.

\newpage
\clearpage

\section{Sulfur spectral synthesis examples}
\label{appendix:spectral_synthesis_ex}
\begin{figure}[h]
\centering
\resizebox{\hsize}{!}{\includegraphics{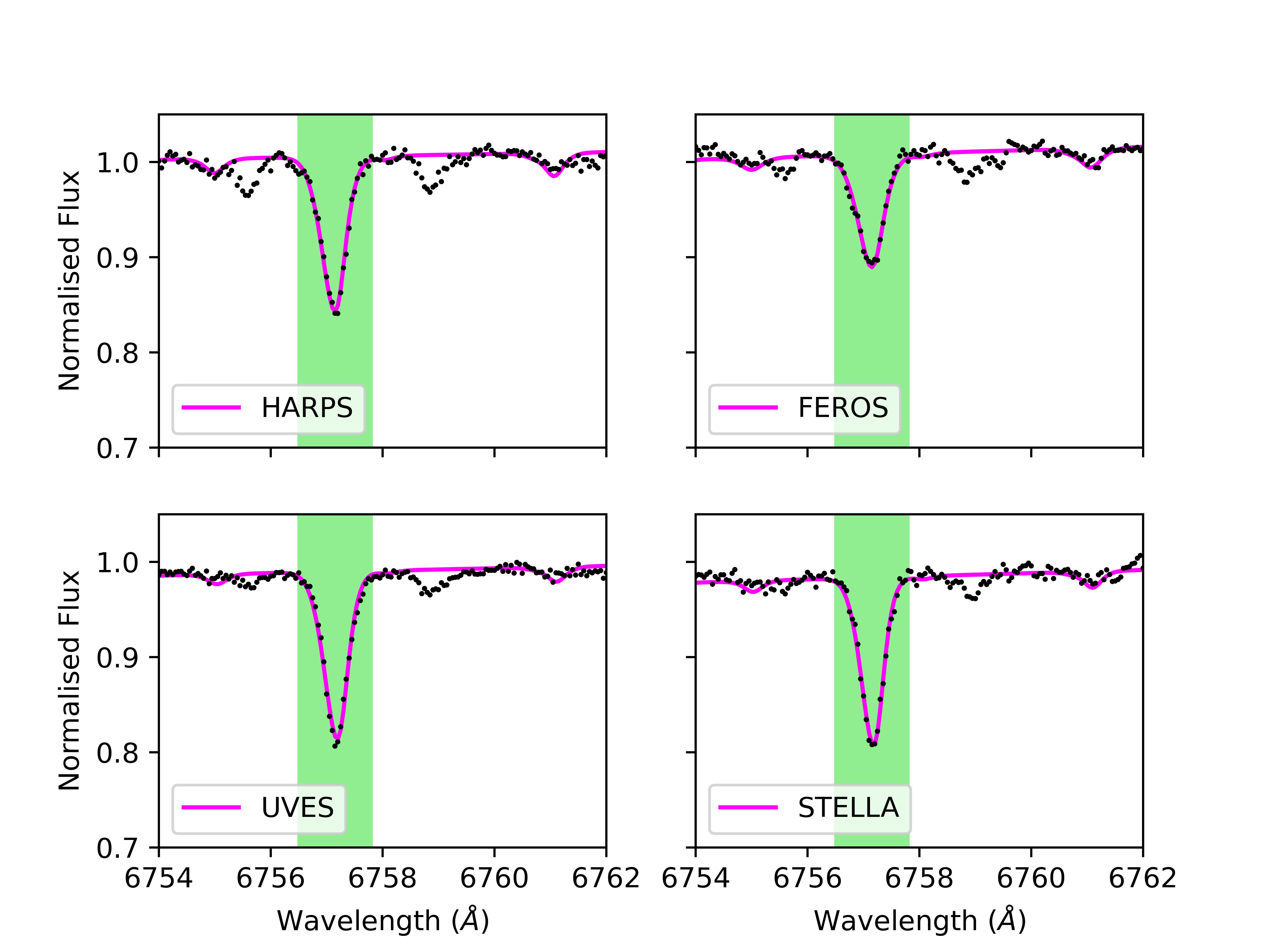}}
\caption[]{Examples of spectral synthesis for the S~{\sc i} line at 6757~\AA. The observed (HARPS, FEROS, UVES, and STELLA) and best-fit synthetic spectra are plotted in black points and magenta solid lines, respectively. The line of interest is marked as the green area.}
\label{figure:examples_synth}
\end{figure}

Figure~\ref{figure:examples_synth} shows the comparison between observed spectra and the best-fit synthetic models for a sample of four stars, acquired with different spectrographs, in the sulfur line region at 6757~\AA. Specifically: the HARPS spectrum (R$\sim$115\,000) is for the star \object{$\zeta$\,Gem} (\teff\ = 5617~K, \logg\ = 1.2~dex, [Fe/H] = 0.02~dex, \vmic\ = 3.4~\kms, A(S) = 7.13\,$\pm$\,0.03~dex); the FEROS spectrum (R$\sim$48\,000) is for \object{T\,Vel} (\teff\ = 5721~K, \logg\ = 1.6~dex, [Fe/H] = $-$0.06~dex, \vmic\ = 3.3~\kms, A(S) = 7.02\,$\pm$\,0.05~dex); the UVES spectrum (R$\sim$40\,000) is for \object{XX\,Sgr} (\teff\ = 6259~K, \logg\ = 1.3~dex, [Fe/H] = $-$0.06~dex, \vmic\ = 2.9~\kms, A(S) = 7.05\,$\pm$\,0.05~dex); and the STELLA spectrum (R$\sim$55\,000) is for the star \object{VZ\,Cyg} (\teff\ = 6169~K, \logg\ = 1.3~dex, [Fe/H] = $-$0.02~dex, \vmic\ = 2.9~\kms, A(S) = 7.03\,$\pm$\,0.04~dex).

\section{Comparison with stars in common}
\label{appendix:paratm_ab_literature}

\begin{figure}[h]
\centering
\resizebox{\hsize}{!}{\includegraphics{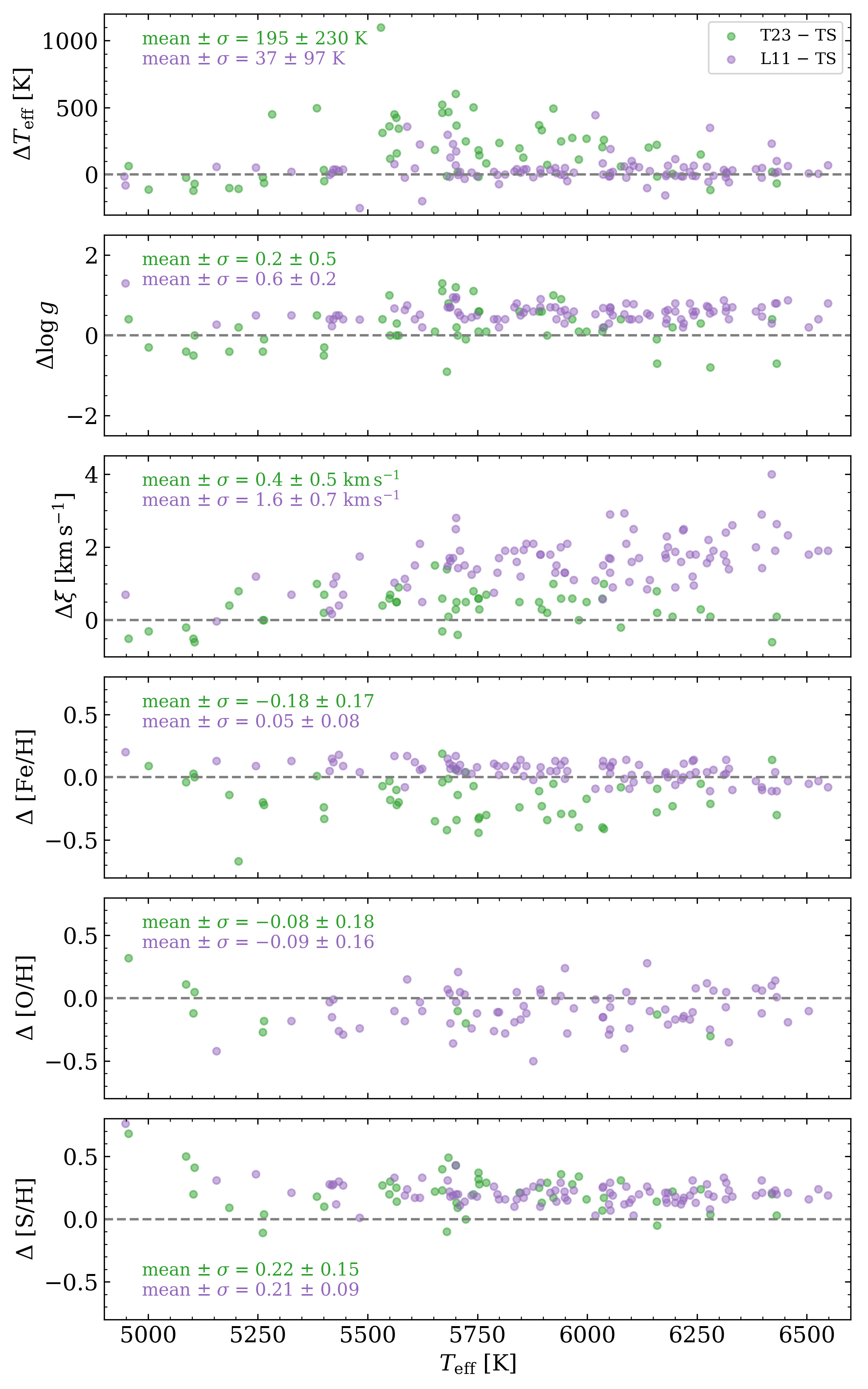}}
\caption{Comparison with literature of the atmospheric parameters and abundances estimated in the current work. The panels shows the differences as a function of the effective temperature for stars in common between the current study (TS) and data from \citet[][L11]{Lucketal2011} and \citet[][T23]{Trentinetal2023}.}
\label{figure:paratm_ab_literature}
\end{figure}

Figure~\ref{figure:paratm_ab_literature} shows a comparison of the atmospheric parameters and the O and S abundances for Cepheids in common between our sample and the literature \citep{Lucketal2011,Trentinetal2023}. The differences are, within the errors, close to zero for most of the comparisons. The surface gravity and the microturbulent velocity are systematically higher in \citet{Lucketal2011}, which might be due to differences in the line lists used. Sulfur abundances are systematically overabundant in both literature works compared to our estimates. Again, line list differences are likely playing an important role. Our measurements are based on one single line of sulfur whereas those authors used additional S lines in their estimates.



\newpage
\clearpage

\section{Abundances along the pulsation cycle}
\label{appendix:ab_pulsation_cycle}

Figure~\ref{figure:xh_phase} shows the O and S abundances 
plotted as a function of the pulsation phase for the 20 calibrating 
CCs provided by \citet{daSilvaetal2022}. This is a selection of 
Cepheids from our sample for which high-resolution spectra 
cover either the entire cycle or a significant portion of 
the pulsation cycle.

The phases ($\phi$) along the pulsation cycle were calculated 
by using the following equation and the ephemerides that we 
previously published in \citet{daSilvaetal2022}:
\begin{equation*}
\phi =
\begin{cases}
    x - int(x) - 1 & \text{if $x - int(x) > 0$} \\
    x - int(x) + 1 & \text{if $x - int(x) < 0$} \\
    x - int(x)     & \text{otherwise}
\end{cases}
\end{equation*}
\begin{equation}
x = \frac{{\rm MJD} - (T_0 - 2\,400\,000) + 0.5}{P}
\end{equation}
where MJD is the Modified Julian Date from their Table~4, while $T_0$ (as defined in that paper) and the pulsation period ($P$) are from their Table~5. For Cepheids for which more than one value of $T_0$ are available, we adopted the first one (Ephemerides source = 0). We notice that a shift was applied to the phases of \object{VY\,Sgr} ($-$0.05), \object{RZ\,Vel} (+0.03), and \object{WZ\,Sgr} ($-$0.03) for the reasons mentioned in the Notes to Table~5. Finally, we mention that the oxygen abundance for the variable \object{Y\,Oph} was not included in our current study. Although the S/N of the spectra available for this variable is quite high (they range from $\sim$70 to more than 180), there are several emission features close to the oxygen line at 6300.3~\AA.~This means that the position of the continuum could not be accurately assessed. This oxygen line is relatively weak, which makes the abundance measurements very sensible to the continuum definition. The observed features are possibly related to the presence of airglow emission lines \citep[see][]{Nolletal2012}, which may affect some of the spectra adopted in our analysis.

\begin{figure*}
\centering
\begin{minipage}[t]{0.49\textwidth}
\centering
\resizebox{\hsize}{!}{\includegraphics{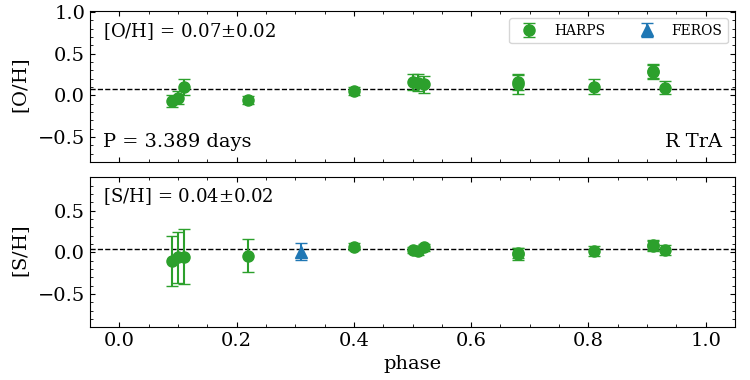}}
\end{minipage}
\begin{minipage}[t]{0.49\textwidth}
\centering
\resizebox{\hsize}{!}{\includegraphics{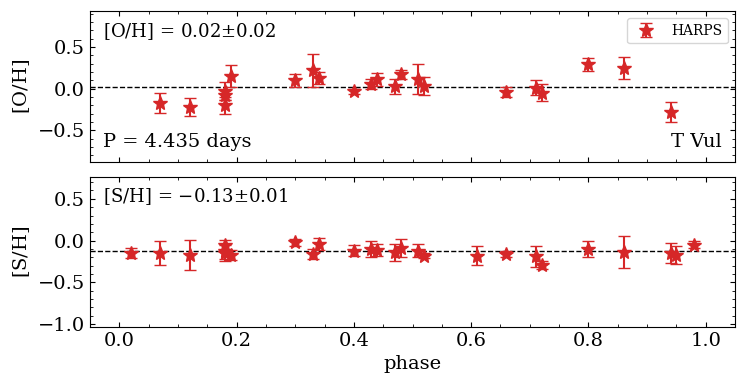}}
\end{minipage} \\
\begin{minipage}[t]{0.49\textwidth}
\centering
\resizebox{\hsize}{!}{\includegraphics{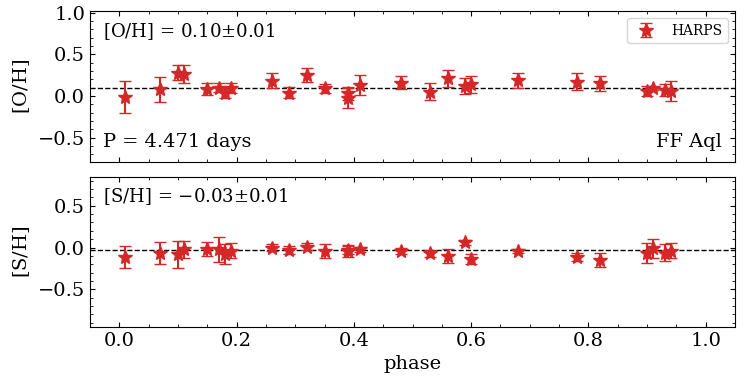}}
\end{minipage}
\begin{minipage}[t]{0.49\textwidth}
\centering
\resizebox{\hsize}{!}{\includegraphics{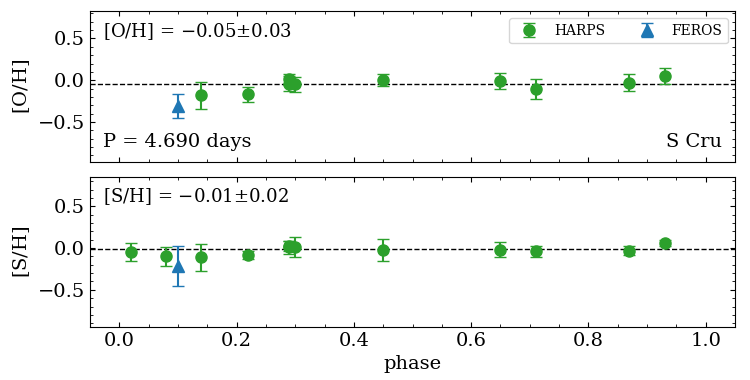}}
\end{minipage} \\
\begin{minipage}[t]{0.49\textwidth}
\centering
\resizebox{\hsize}{!}{\includegraphics{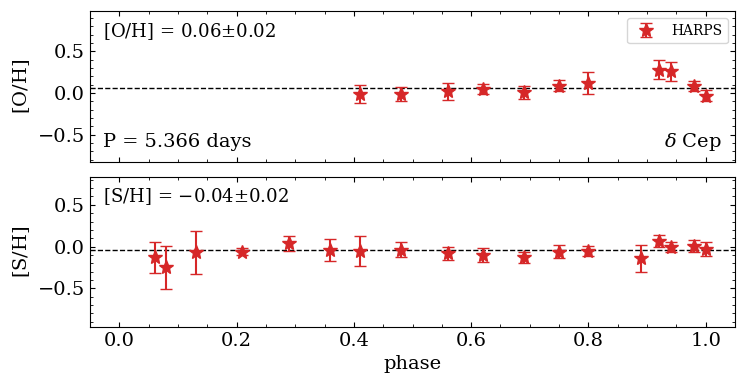}}
\end{minipage}
\begin{minipage}[t]{0.49\textwidth}
\centering
\resizebox{\hsize}{!}{\includegraphics{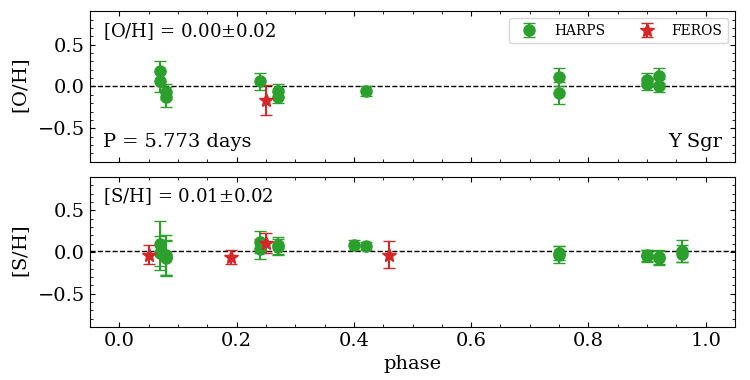}}
\end{minipage} \\
\begin{minipage}[t]{0.49\textwidth}
\centering
\resizebox{\hsize}{!}{\includegraphics{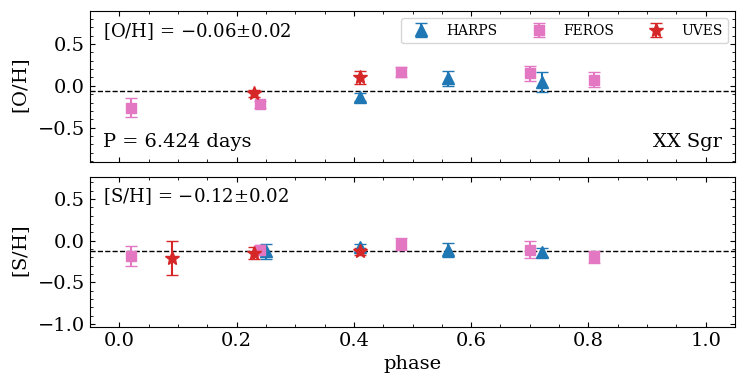}}
\end{minipage}
\begin{minipage}[t]{0.49\textwidth}
\centering
\resizebox{\hsize}{!}{\includegraphics{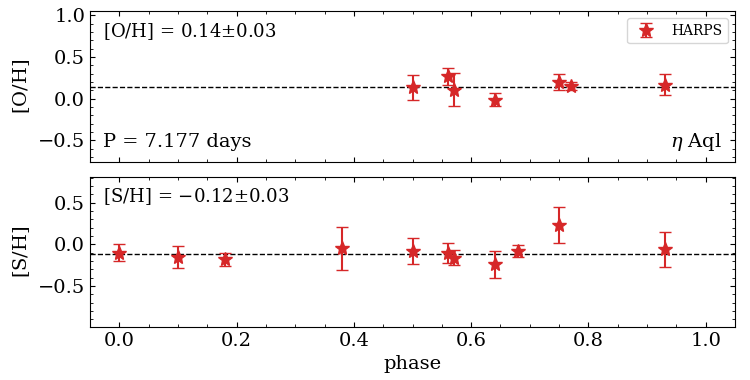}}
\end{minipage} \\
\begin{minipage}[t]{0.49\textwidth}
\centering
\resizebox{\hsize}{!}{\includegraphics{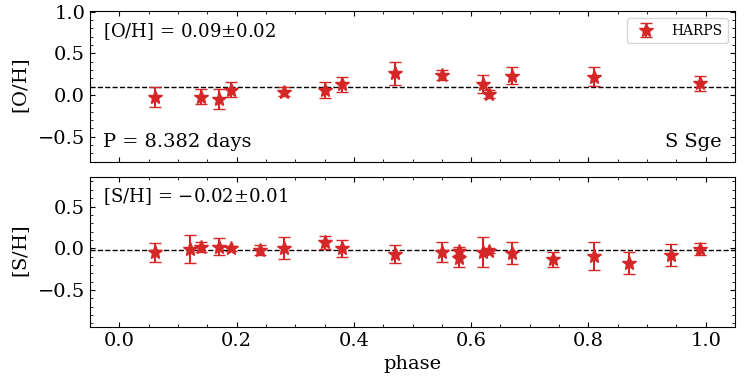}}
\end{minipage}
\begin{minipage}[t]{0.49\textwidth}
\centering
\resizebox{\hsize}{!}{\includegraphics{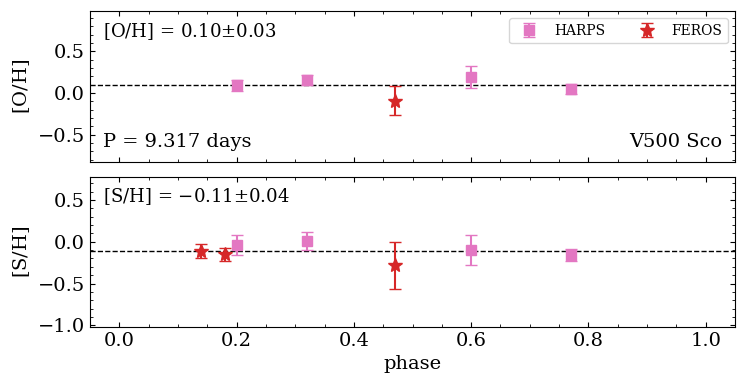}}
\end{minipage}
\caption{Abundance ratios as a function of the pulsation phase. Measurements from different spectrographs are indicated with different colors and symbols. The error bars are in some cases smaller than the symbol size. The weighted mean (dashed line) and the standard error of the abundances are also shown together with the pulsation period.}
\label{figure:xh_phase}
\end{figure*}
\addtocounter{figure}{-1}
\begin{figure*}
\centering
\begin{minipage}[t]{0.49\textwidth}
\centering
\resizebox{\hsize}{!}{\includegraphics{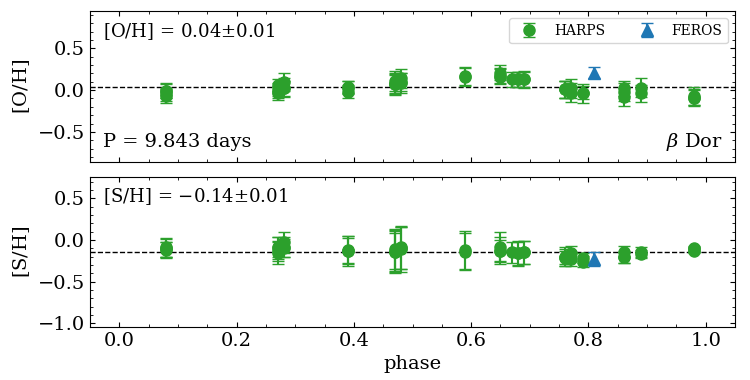}}
\end{minipage}
\begin{minipage}[t]{0.49\textwidth}
\centering
\resizebox{\hsize}{!}{\includegraphics{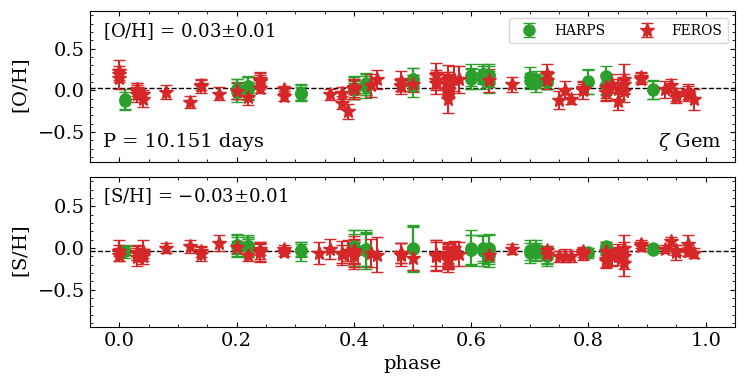}}
\end{minipage} \\
\begin{minipage}[t]{0.49\textwidth}
\centering
\resizebox{\hsize}{!}{\includegraphics{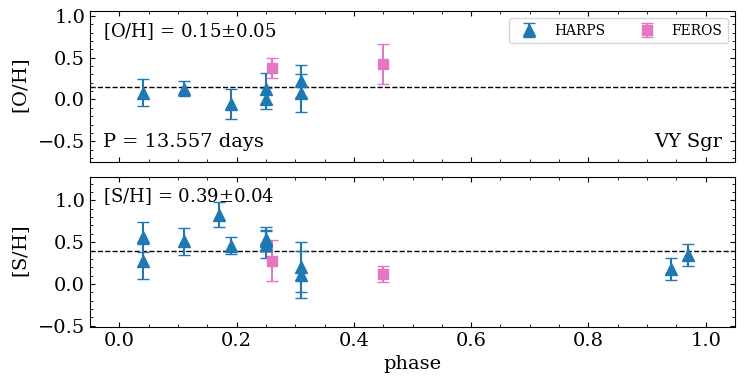}}
\end{minipage}
\begin{minipage}[t]{0.49\textwidth}
\centering
\resizebox{\hsize}{!}{\includegraphics{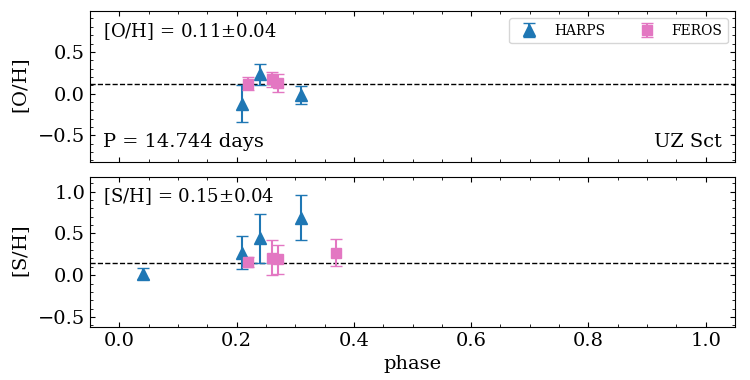}}
\end{minipage} \\
\begin{minipage}[t]{0.49\textwidth}
\centering
\resizebox{\hsize}{!}{\includegraphics{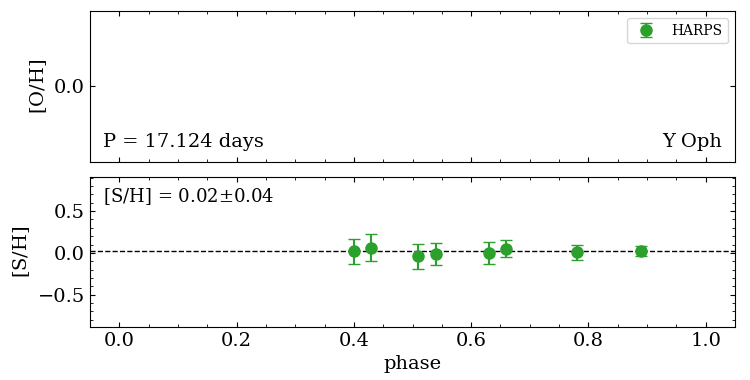}}
\end{minipage}
\begin{minipage}[t]{0.49\textwidth}
\centering
\resizebox{\hsize}{!}{\includegraphics{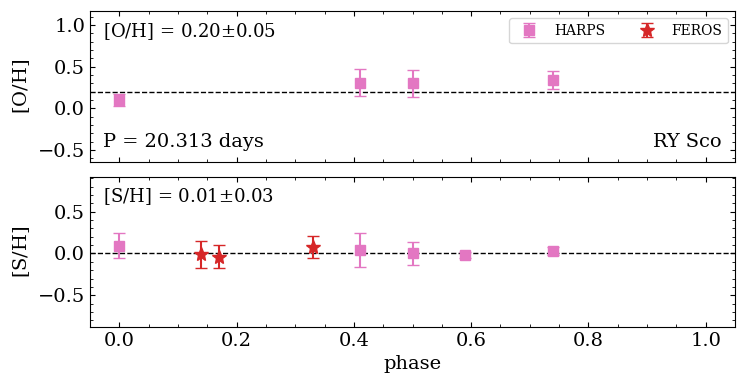}}
\end{minipage} \\
\begin{minipage}[t]{0.49\textwidth}
\centering
\resizebox{\hsize}{!}{\includegraphics{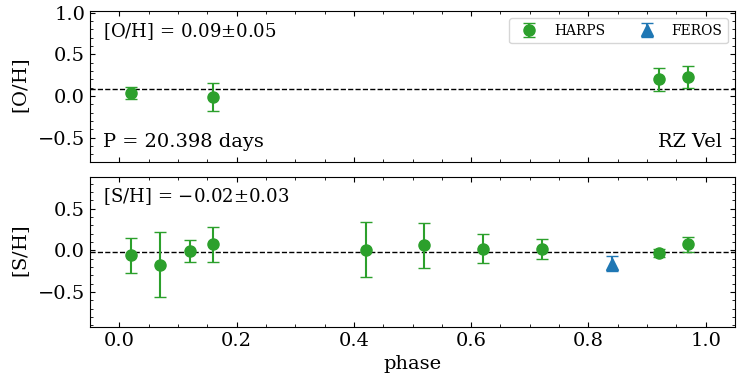}}
\end{minipage}
\begin{minipage}[t]{0.49\textwidth}
\centering
\resizebox{\hsize}{!}{\includegraphics{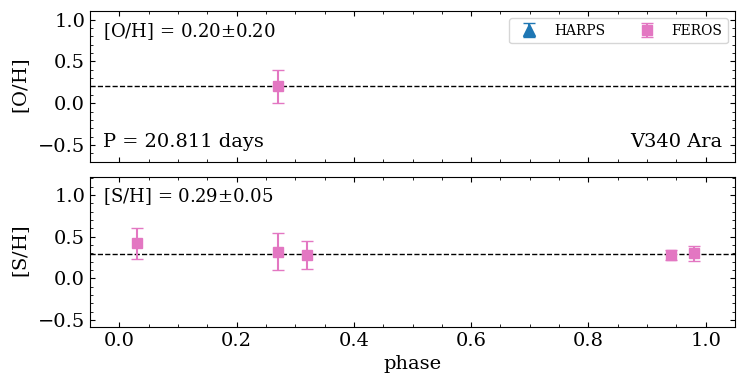}}
\end{minipage} \\
\begin{minipage}[t]{0.49\textwidth}
\centering
\resizebox{\hsize}{!}{\includegraphics{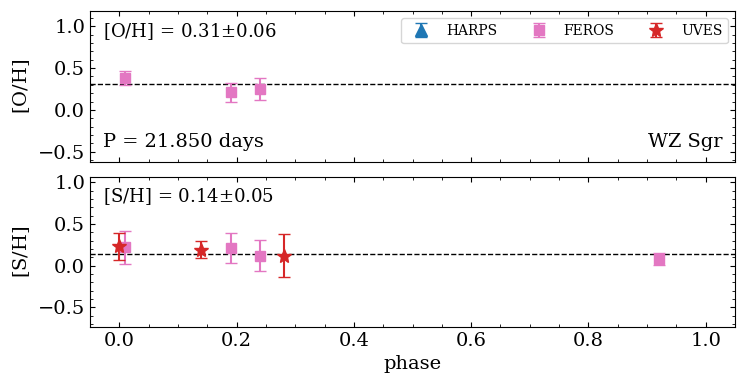}}
\end{minipage}
\begin{minipage}[t]{0.49\textwidth}
\centering
\resizebox{\hsize}{!}{\includegraphics{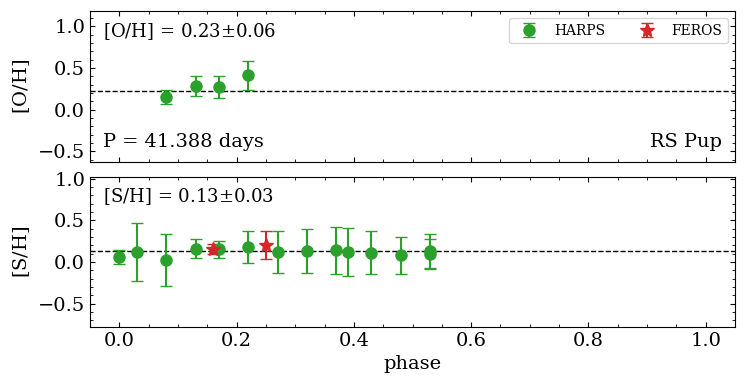}}
\end{minipage}
\caption[]{continued.}
\end{figure*}

\end{appendix}


\end{document}